\documentclass[prl,twocolumn,floats,aps,epsfig,nofootinbib,axodraw,amssymb]{revtex4}
\usepackage{amsmath,amsfonts}
\usepackage{subfigure}
\def\beq{\begin{equation}}
\def\eeq{\end{equation}}
\def\bea{\begin{eqnarray}}
\def\eea{\end{eqnarray}}

\def\leqn#1{(\ref{#1})}

\evensidemargin=\oddsidemargin \textwidth=172mm \textheight=230mm

\def\pslash{\not{\hbox{\kern-4pt $p$}}}
\def\qslash{\not{\hbox{\kern-4pt $q$}}}
\def\lv{\not{\hbox{\kern-4pt $L$}}}
\def\lsim{\mathrel{\raise.3ex\hbox{$<$\kern-.75em\lower1ex\hbox{$\sim$}}}}
\def\gsim{\mathrel{\raise.3ex\hbox{$>$\kern-.75em\lower1ex\hbox{$\sim$}}}}
\def\ifmath#1{\relax\ifmmode #1\else $#1$\fi}

\usepackage{graphicx}
\usepackage{bm}

\begin{document}

\title{Unitarity Constraints on Higgs Portals}
\bigskip
\author{Devin~G.~E.~Walker}
\address{SLAC National Accelerator Laboratory, 2575 Sand Hill Road, Menlo Park, CA 94025, U.S.A.}

\begin{abstract}
\noindent
Dark matter that was once in thermal equilibrium with the Standard Model is generally prohibited from obtaining all of its mass from the electroweak phase transition.  This implies a new scale of physics and mediator particles to facilitate dark matter annihilation.  In this work, we focus on dark matter that annihilates through a generic Higgs portal.  We show how partial wave unitarity places an upper bound on the mass of the mediator (or dark) Higgs when its mass is increased to be the largest scale in the effective theory.  For models where the dark matter annihilates via fermion exchange, an upper bound is generated when unitarity breaks down around 8.5 TeV.  Models where the dark matter annihilates via fermion and higgs boson exchange push the bound to 45.5 TeV.  We also show that if dark matter obtains all of its mass from a new symmetry breaking scale that scale is also constrained.  We improve these constraints by requiring perturbativity in the Higgs sector up to each unitarity bound.  In this limit, the bounds on the dark symmetry breaking vev and the dark Higgs mass are now 2.4 and 3 TeV, respectively, when the dark matter annihilates via fermion exchange.  When dark matter annihilates via fermion and higgs boson exchange, the bounds are now 12 and 14.2 TeV, respectively.  Given the unitarity bounds, the available parameter space for Higgs portal dark matter annihilation is outlined.  We also show how the various bounds are dramatically improved if Higgs portal dark matter is only a fraction of the observed relic abundance.  Finally, we discuss how to apply these arguments to other dark matter scenarios and discuss prospects for direct detection and future collider searches.  Notably, if the Higgs portal is responsible for dark matter annihilation, direct detection experiments planned within the next decade will cover almost all the parameter space.  The ILC and/or VLHC, however, is needed to establish the Higgs portal mechanism.
\end{abstract}
\maketitle

\section{I. Introduction}
\label{sec:intro}
\noindent
Understanding the nature of dark matter is one of the most pressing, unresolved problems in particle physics.  Dark matter is needed to understand structure formation, the measured galactic rotation curves~\cite{Kowalski:2008ez,Ahn:2013gms,Beringer:1900zz} and the acoustic peaks in the cosmic microwave background~\cite{Ade:2013lta}.  Moreover, the dark matter relic abundance is measured to be~\cite{Ade:2013lta}  
\begin{equation}
h^2\, \Omega_c = 0.1199 \pm 0.0027.  \label{eq:relic}
\end{equation}
A compelling argument for the origin of this abundance is to assume dark matter was once in thermal contact with the baryon-photon plasma during the early universe.  Since all known forms of matter in the universe arise from thermal equilibrium, this type dark matter is theoretically persuasive\footnote{Unless otherwise specified, we always consider dark matter that is in thermal equilibrium during the early universe.}.  In this scenario, the measured relic abundance is generated by dark matter annihilations into Standard Model (SM) particles.  Because of constraints from the observed large scale structure in the universe, dark matter must be stable and non-relativistic when leaving thermal equilibrium~\cite{Ahn:2013gms}.  
\newline
\newline
The SM alone cannot account for the missing matter in the universe~\cite{Bertone:2004pz}.  Experimental constraints, however, provide guidance on structure of the underlying theory.  For example, the lack of large missing energy signatures at the LHC and other colliders suggest~\cite{ATLAS-CONF-2012-147,ATLAS-CONF-2013-073,CMS-PAS-EXO-12-048} that dark matter is either heavy or has such a small coupling that it evades detection.  Additionally, direct detection experiments~\cite{Aprile:2012nq,Ahmed:2011gh}, updated precision electroweak constraints~\cite{devinjoannetim,Baak:2012kk} and precision Z-pole experiments~\cite{devinjoannetim,ALEPH:2005ab} all severely constrain the direct coupling of dark matter to the SM Higgs or Z~boson.  As emphasized in~\cite{devinjoannetim}, these constraints also imply dark matter cannot obtain all of it's mass from the SM Higgs alone.  Thus, we are led to scenarios where (1) a new mediator particles mix with SM bosons in order to facilitate interactions between the dark matter and the SM and (2) a new fundamental scale of physics is needed that is (or partly) responsible for the dark matter mass.  The mixing helps to evade current experimental constraints by decoupling the dark matter from the SM.  Should these scenarios be realized in nature, the discovery of mediator particles provides an important piece to understand the nature of dark matter.  It is therefore crucial to place bounds on mass and couplings of these mediators.  Moreover, defining a new scale of physics is essential for understanding and characterizing new physics beyond the SM.
\newline
\newline
The most popular ways for dark matter to annihilate via a mediator particle are through the Higgs~\cite{Patt:2006fw} and/or a new, heavy gauge boson.  We refer to this as the Higgs and gauge portals, respectively.  Well-studied and motivated supersymmetric models have pure Higgsinos or winos as viable dark matter candidates~\cite{CahillRowley:2012kx}.  They annihilate via Higgses and/or sfermon mediated processes.  In this work we focus on generating constraints Higgs portal dark matter annihilation.  We study gauge portal constraints in~\cite{devinwill}.  
\newline
\newline
The Higgs portal~\cite{Patt:2006fw} requires a new hidden symmetry breaking sector in addition to the SM Higgs sector.  The dark matter is charged and rendered stable by the symmetries in the hidden sector.  After all the symmetries are broken, the SM and ``dark" Higgs mix and thereby facilitating tree-level dark matter annihilation.  This mixing is subject to constraints from unitarity.  For example, the mixing between the dark Higgs and the SM Higgs forces the SM Higgs to incompletely cancel the gauge contribution to $WW$ scattering.  Since the SM Higgs mass is now known~\cite{atlasHiggsresults,CMSHiggsresults}, an upper bound on the dark Higgs mass can be given for a value of the mixing angle.  Requiring the dark matter annihilations to satisfy the relic abundance fixes the mixing angle to generate absolute upper bound on the dark Higgs' mass.
\newline
\newline
In the next section, we sketch the basic unitarity argument for the Higgs portal in more detail.  Section~III. introduces a generic model to place our unitary bounds.  In Section IV., we characterize the Higgs portal by a few parameters and describe how these parameters can be constrained by a variety of theoretical constraints.  Section V.~computes the dark matter relic abundance and direct detection cross sections.  Section VI.~describes existing experimental constraints on the SM Higgs.  Section VII.~introduces our unitarity constraints.  We implement the unitarity constraints with a parameter scan in Section VIII.  The basic unitarity bounds appear here.  In Section IX.,~we improve the unitarity bounds by requiring perturbativity on the Higgs sector.  Future experimental signatures are in Section X.  As discussed in the next section, we have assumed a mass hierarchy where the dark Higgs is the heaviest particle in the effective theory.  In Section XI., we discuss alternative hierarchies and potential constraints.  Conclusions and Appendices follow.

\section{II.  The Basic Unitarity Argument}  
\noindent
Dark matter annihilating via the Higgs portal requires a new scalar particle (a ``dark" Higgs) which couples to the dark matter and mixes with the SM Higgs~\cite{Patt:2006fw}.  Because of this mass mixing, the coupling of the SM Higgs to SM particles is proportional to $\cos\theta$.  Conversely, the coupling of the dark Higgs to SM particles is proportional to $\sin\theta$.  Here $\theta$ is the mass mixing angle which is roughly the ratio of the SM and dark Higgs masses times the coupling(s) in the Higgs potential that mixes the different Higgs sectors,
\begin{equation}
\sin\theta \sim \lambda_\mathrm{mix}\, m_h \bigl/ m_\rho.  \label{eq:roughmix} 
\end{equation}
We define $v$ and $u$ to be the electroweak and dark symmetry breaking veversus  For most of this work, we assume the dark Higgs mass is heavier than the SM Higgs mass\footnote{ LEP and the Tevatron did not discover light Higgses with nontrivial couplings to the SM.  Thus this assumption is somewhat justified.  We relax this condition in Section X.}.  The reason for this brings us to our first important point:
\begin{enumerate}
\newcounter{enum_saved2}
\item Higgs portal unitarity constraints are triggered when the dark Higgs mass is much larger than the SM Higgs and dark matter masses.  
\setcounter{enum_saved2}{\value{enumi}}
\end{enumerate}
This is analogous to the electroweak case when the SM Higgs mass is assumed to be large relative to the $W$ and $Z$ masses.  That calculation produced an upper bound on the SM Higgs mass.  To see how an upper bound on the dark Higgs mass is generated, first consider the dark matter relic abundance.  When the dark Higgs mass is larger than the dark matter and SM Higgs masses, the low-velocity scattering amplitude(s) must have the form
\begin{align}
\langle \sigma v \rangle \sim {\sin^4\theta \over m_\chi^2} &&\mathrm{or}&&  \langle \sigma v \rangle \sim {\sin^2\theta\cos^2\theta \over m_\chi^2}. \label{eq:scatteringamp}
\end{align}
The relation on the left is for dark matter annihilation into SM Higgses.  The right is for annihilation into SM fermions or weak gauge boson in the final state.  This brings us to our second important point.
\begin{enumerate}
\setcounter{enumi}{\value{enum_saved2}}
\item The measured relic abundance in equation~\leqn{eq:relic} requires $\sin\theta$ to be nonzero.  
\setcounter{enum_saved2}{\value{enumi}}
\end{enumerate}
Now consider the unitarity bounds from, e.g., high-energy $WW$ scattering.  The most important contributions to the tree-level scattering amplitudes are
\begin{eqnarray}
\mathcal{M}_\mathrm{gauge} &=& {g^2 \over 4\,m_W^2}(s + t) \label{eq:gaugecontrib} \\
\mathcal{M}_\mathrm{SM\,\,Higgs} &=& -{g^2 \over 4\,m_W^2}(s + t)\, \cos^2\theta  \label{eq:smhcontrib} \\
\mathcal{M}_\mathrm{dark\,\,Higgs} &=&  -{g^2 \over 4\,m_W^2}(s + t) \,\sin^2\theta  \label{eq:dhcontrib}
\end{eqnarray}  
Both the dark Higgs and SM Higgs exchange diagrams are needed to unitarize $WW$ scattering!  Equation~\leqn{eq:dhcontrib} assumes the dark Higgs mass is much smaller than $\sqrt{s}$.  In practice, the dark Higgs mass can be raised to be much larger than any other scale of interest while keeping the mixing angle fixed and non-zero\footnote{We demonstrate this explicitly in Section III.}.  In this limit, the SM Higgs amplitude can only partially cancel out the gauge contribution in equation~\leqn{eq:gaugecontrib}.  Consequently, partial wave unitarity has the potential to place an upper bound on the dark Higgs mass.  This brings us to our final point.
\begin{enumerate}
\setcounter{enumi}{\value{enum_saved2}}
\item In the limit of small $\sin\theta$, the SM Higgs amplitude in equation~\leqn{eq:smhcontrib} cancels most of the gauge contribution and therefore accommodates a heavier dark Higgs mass.  However, because of the relic abundance constraint, there is a lower bound on the mixing angle and therefore an absolute upper bound on how heavy the dark Higgs mass can be.
\setcounter{enum_saved2}{\value{enumi}}
\end{enumerate}
Beyond the dark Higgs mass, we can also place a bound on the dark symmetry breaking scale.  
If the mixing between the SM and dark Higgses is large (e.g.,~$\sin\theta \sim 1/\sqrt{2}$), the unitary constraint from $WW$ scattering alone gives
\begin{equation}
m_\mathrm{dark\,\,Higgs} \lesssim 1.4\,\,\,\mathrm{TeV}.
\end{equation}
This estimate follows from equations~\leqn{eq:gaugecontrib},~\leqn{eq:smhcontrib} and~\leqn{eq:unitconstraint}.  The LHC14/VLHC at high luminosity likely will be able probe these scales.  Using equation~\leqn{eq:roughmix} and assuming the couplings in the Higgs sector are $\mathcal{O}(1)$, the scale of new physics is roughly,
\begin{equation}
u \leq \mathcal{O}(3-4\,\,\mathrm{TeV}).
\end{equation}
which may be possible to directly probe with the next generation of colliders.
\newline
\newline
In the coming sections, we make these basic arguments explicit and constrain the basic Higgs portal parameters.  Even though there is an interplay between constraints from unitarity and the dark matter relic abundance, we refer to the derived bounds as unitarity constraints.  Before moving on, we note that unitarity bounds~\cite{Dicus:1992vj,Lee:1977eg} were the essential argument for why the SM Higgs boson was expected to be discovered at the LHC.  Our desire is for these simple arguments to motivate new searches for physics at scales which may be obtainable for present/near-term experiments. 

\section{III. A Representative Model}
\noindent
Without loss of generality, we consider a model in which a chiral $Z_2$ is broken to the diagonal,
\begin{equation}
Z_{2} \times Z_{2} \to Z_{2}.
\end{equation}
The resulting $Z_2$ stabilizes the dark matter candidates.  The dark matter ($\psi$, $\xi$) and dark Higgs ($\phi$) transform under the chiral symmetry as, 
\begin{eqnarray}
\phi &\to& (-,-)\,\,\phi \label{phisym} \\
\xi   &\to& (-,+)\,\,\xi,\label{eq:xisym} \\
\psi &\to& (+,-)\,\,\psi, \label{eq:psisym} 
\end{eqnarray}
where the entries in parenthesis notate if the particle is even or odd under the first or second $Z_2$.  Without loss of generality we focus on fermonic dark matter and comment in the Appendix on any differences when considering bosonic dark matter.
%

\subsection{III.1. A Generic Higgs Sector}
\noindent
The SM Higgs ($h$) is neutral under the discrete chiral symmetry.  The Higgs potential is,
\begin{eqnarray}
V &=&  \lambda_1\,\biggl(h^\dagger h - {v^2 \over 2} \biggr)^2 +  \lambda_2\,\biggl(\phi^2 - {u^2 \over 2} \biggr)^2\label{eq:Higgspotential} \\
 &+& \lambda_3\,\biggl(h^\dagger h - {v^2 \over 2} \biggr)\biggl(\phi^2 - {u^2 \over 2} \biggr), \nonumber
\end{eqnarray}
where $v$ and $u$ are the electroweak and dark vevs, respectively.  We parametrize the Higgs and Goldstone boson as 
\begin{equation}
\phi = (u + \rho)/\sqrt{2}
\end{equation}
where 
$\rho$ is the dark Higgs.  It is clear the dark vev is even under the diagonal $Z_2$.  The resulting mass matrix is
\begin{equation}
M^2 = \begin{pmatrix} 2 \lambda_1 v^2 & \lambda_3\, u\, v \\ \lambda_3 \,u \,v & 2 \lambda_2 \,u^2 \end{pmatrix}.
\end{equation}
The Higgs masses are
\begin{eqnarray}
m_h^2 &=& 2\,\lambda_1 v^2\biggl(1 - {\lambda_3^2 \over 4\,\lambda_1\lambda_2} + \ldots\biggr)  \label{eq:smHiggsmass}\\
m_\rho^2 &=& 2\,\lambda_2\,u^2\biggl(1  +{\lambda_3^2 \over 4\,\lambda_2^2} {v^2\over u^2}+ \ldots\biggr) \label{eq:darkHiggsmass}
\end{eqnarray}
where $m_h$ is the SM Higgs mass and is fixed to 125.5 GeV.  $m_\rho$ is the dark Higgs mass.  The Higgses mix in the mass matrix, 
\begin{equation}
\begin{pmatrix} h' \\ \rho' \end{pmatrix} = \begin{pmatrix} \cos\theta & - \sin\theta \\ \sin\theta & \cos\theta \end{pmatrix}\, \begin{pmatrix} h \\ \rho \end{pmatrix}, \label{eq:mixing}  
\end{equation}
where the primes are the mass eigenstates.  For brevity going forward, we refer to both the mass eigenstates without primes.  In the limit of $u \gg v$, 
\begin{align}
\cos\theta \sim 1-\frac{\lambda _3^2 \,v^2}{8 \lambda _2^2\,u^2}  && \sin\theta \sim \frac{\lambda _3\,v}{2\lambda _2\,u}. \label{eq:cossin}
\end{align}  
As expected, the decoupling limit requires $\lambda_3 \to 0$ and/or sending the dark vev, $u$, to infinity.  
\newline
\newline
In the introduction, we asserted that the mass of the dark Higgs could be raised while keeping the mixing angle constant.   We can now make this explicit.  
The dark Higgs mass and $\sin\theta$ have a different parametric dependence on $\lambda_2$,
\begin{align}
m_\rho \sim \sqrt{2\,\lambda_2}\,\,u && \sin\theta \sim \frac{\lambda _3\,v}{2\lambda _2\,u}.
\end{align}
Because of this dependence as $u$ is increased to infinity, $\lambda_2$ can be reduced to keep $\sin\theta$ fixed.
Before moving on, we note the dark Higgs mass does not have to be the result of spontaneously broken symmetry.  Although this is not exactly a Higgs portal, one can simply mix a real, massive scalar with the SM Higgs to generate a potential analogous to equation~\leqn{eq:Higgspotential}.   We comment on this case in the Appendix.  

\subsection{III.2. Dark Matter Sector}
\noindent
We focus on fermonic dark matter\footnote{Scalar dark matter requires a mechanism to stabilize it's mass from large quantum corrections.  We comment on potential differences when one considers bosonic dark matter in the Appendix}.  For simplicity, we assume the dark Higgs is solely responsible for the dark matter mass.  The discrete symmetries in equations~\leqn{phisym}-\leqn{eq:psisym} forbid tree-level majorana mass terms generated from the dark symmetry breaking.  Having majorana mass terms just introduces more parameters to constrain.  The dark matter sector now has the following Yukawa terms,
\begin{equation}
\mathcal{L} = \overline{\chi} \,\bigl( \lambda_{\chi_V}  + i\,\lambda_{\chi_A} \gamma_5  \bigr)\,\Phi\, \chi. \label{eq:darkYukawa}
\end{equation}
We have included both with scalar and psuedo-scalar couplings.  In equation~\leqn{eq:darkYukawa} we defined
\begin{align}
\chi = \begin{pmatrix} \psi \\ \xi \end{pmatrix} &&  \Phi = \begin{pmatrix} \phi &  \\  & \phi \end{pmatrix}.
\end{align}
The dark Higgs gives the dark matter candidates the following mass
\begin{eqnarray}
m_{\chi} &=& \Bigl(\sqrt{\lambda_{\chi_V}^2 + \lambda_{\chi_A}^2 }\,u\Bigr)\bigl/\sqrt{2} \equiv \lambda_\chi u\bigl/\sqrt{2}.
\label{eq:dmfmass}
\end{eqnarray}
We often use $\lambda_\chi$ to represent both the scalar and psuedoscalar couplings.  We are focusing on a model where the dark matter obtains all of its mass from a new dark symmetry breaking scale.  The Higgs portal is defined by this requirement~\cite{Patt:2006fw}.  However in~\cite{devinjoannetim}, we weaken this requirement and consider the associated bounds.
\newline
\newline
It has been shown~\cite{LopezHonorez:2012kv} that the scalar and psuedo-scalar couplings in equation~\leqn{eq:darkYukawa} are needed to generate the most unconstrained Higgs portal scenario.  We therefore explore two scenarios,
\newline
\newline
\textbf{Model 1:} $\lambda_{\chi\,A} = 0$,\\ 
\newline
\textbf{Model 2:}  $\lambda_{\chi\,A}$ and  $\lambda_{\chi\,V}$ are non-zero,
%
\newline
\newline
and provide unitarity bounds for each.  As we will see in Section V., the dark matter in Model 1 can only annihilate only through t-channel fermion exchange.  The dark matter in Model 2 can annihilate through both s-channel and t-channel diagrams.  This difference forces the Higgs mixing angle, on average, to be larger for Model 1 in comparison to Model 2 for a given dark matter mass.  This difference leads to stronger bounds on Model 1.

\subsection{III.3. Couplings}
\noindent
As in Section I, here we emphasize that the Higgs mixing modifies the SM  and dark Higgs couplings by sines and cosines.  For example, \begin{eqnarray}
\Gamma^{\mu\nu}(WWh) &=& i \,g\,m_W \cos\theta \,g^{\mu\nu}  \label{eq:WWh} \\
   \Gamma(\bar{\chi} \chi h) &=& - i \,\bigl(\lambda_{\chi_V} + i\,\lambda_{\chi_A}\gamma_5 \bigr)\, \sin\theta/\sqrt{2}  \,\,\,\label{eq:chichih}
\end{eqnarray}
Additionally through mixing, all couplings in the Higgs potential are a function of both the dark and electroweak vev as well as sines and cosines in relatively complex ways. The Appendix lists all the couplings in the Higgs potential after mixing for reference.  For the above, it is clear in the decoupling limit that the SM Higgs couples to the SM particles with SM values.  

\section{IV. Generic Higgs Portal}
\noindent
We are most familiar with unitarity constraints in the electroweak sector.  There the only unknown parameter is the quartic coupling, $\lambda_\mathrm{SM}$, in the SM Higgs potential,
\begin{equation}
V_\mathrm{electroweak} =  \lambda_\mathrm{SM}\,\biggl(h^\dagger h - {v^2 \over 2} \biggr)^2 \label{eq:ewHiggspotential}
\end{equation}
The standard elecroweak unitarity bound simply constrains this one parameter to give an upper bound on the SM Higgs mass.  For the Higgs portal however, there are five generic parameters 
\begin{equation}
\{m_h, m_\rho, m_\chi,\sin\theta, u\}. \label{eq:parameterspace2}
\end{equation}
Without loss of generality, we can trade these parameters for
\begin{equation}
\{\lambda_1, \lambda_2, \lambda_3, \lambda_\chi,u\}.\label{eq:parameterspace}
\end{equation}
In Section VII., we compute the unitarity constraints on these couplings using the Goldstone boson equivalence theorem.  Specifically, 
\begin{enumerate}
\newcounter{enum_saved}
\item  Goldstone-Goldstone, Goldstone-Higgs and Higgs-Higgs scattering directly constrains $\lambda_1$, $\lambda_2$ and~$\lambda_3$.
\item Dark matter self-scattering mediated by the dark Higgs directly constrains $\lambda_\chi$ (or $\lambda_{\chi_V}$ and  $\lambda_{\chi_A}$ in concert).
\setcounter{enum_saved}{\value{enumi}}
\end{enumerate}
This leaves one unknown parameter.
\begin{enumerate}
\setcounter{enumi}{\value{enum_saved}}
\item Through the mixing angles, the relic abundance efficiently constrains the dark symmetry breaking scale, $u$.\setcounter{enum_saved}{\value{enumi}}
\end{enumerate}
As described in Section II, the relic abundance is able to constrain the symmetry breaking scale in the limit where
\begin{equation}
m_\rho > m_h, m_\chi. \label{eq:primaryhierarchy}
\end{equation}
This is shown explicitly as the relic abundance is calculated in the next section.  As we will see, the abundance also directly constrains the coupling, $\lambda_{\chi_A}$.  Direct detection searches directly constrain both $\lambda_{\chi_V}$ and $\lambda_{\chi_A}$.  Also, requiring $m_h = 125.5$~GeV provides an additional constraint the parameter set.  The mass hierarchy in equation~\leqn{eq:primaryhierarchy} is relaxed in Section XI.
\newline
\newline
The relic abundance is also very important because it is a function of all the parameters listed in equations~\leqn{eq:parameterspace2} and \leqn{eq:parameterspace}.  Thus, the unitarity constraints generated from particle scattering  (in Points 1 and 2 above) are disjointed and have no relationship to each other without these ``unifying" sets of equations.  This is a key to this analysis and why this approach can be applied to many scenarios in a model independent fashion.  If there are mediator(s) involved in dark matter annihilation, bounds on it's coupling and mass can be given.  Section VIII.~details how unitarity constraints listed in Section VII.~and relic abundance in Section V.~work in concert to place bounds on the dark Higgs mass and the dark symmetry breaking vev.
 \newline
 \newline
 \section{V. Dark Matter Constraints}
 
\subsection{V.1. Dark Matter Relic Abundance}
\noindent
We compute the dark matter relic abundance to reduce the unknowns in equation~\leqn{eq:parameterspace}.  In order to generate unitarity constraints on the dark Higgs mass, we raise the dark Higgs mass so that 
\begin{equation}
m_\rho > m_h, \,m_\chi.
\end{equation}  
For completeness, in Section VIII we consider constraints on the dark Higgs mass without this limit.  
\newline
\newline
If the dark matter relic abundance is established by thermal freeze-out, the abundance is given by
\begin{equation}
\Omega_\chi\, h^2  =  {1 \times 10^9 \over M_\mathrm{pl}} \,{x_F \over \sqrt{g_*}}\,{1 \over \langle \sigma v \rangle}\,{1 \over \mathrm{GeV}},
 \label{eq:relicabundance}
\end{equation}
where $x_F \equiv m_\chi/T_F$ and $T_F$ is the freeze out temperature.   $g_*$ is calculable depending on the dark matter mass.  It is intriguing to note that the correct thermal relic abundance is generated for annihilation cross sections that are typical for the weak scale interactions,
\begin{equation} 
\Omega_\chi\, h^2  \sim {0.1\,\, \mathrm{pb} \over \langle \sigma v \rangle}.
\end{equation}
This suggests a connection between electroweak and dark matter physics.  For simplicity, we do not consider dark matter annihilation into slightly heavier new physics.  The basic arguments contained in this paper remain unaltered in coannhilation scenarios.  

 \subsubsection{V.1.1. t-channel annihilation}
 \noindent
If the dark matter has a mass $m_\chi > m_h$
, then t-channel annihilation of fermonic dark matter into the Higgses,
 \begin{align}
\chi + \chi \to h +  h, 
\end{align}
dominates.  As discussed in Section III.2., this is the dominant channel for the Higgs portal dark matter to annihilate for Model 1.  The other channels are velocity suppressed.  thermally averaged cross section (in the low velocity limit) is
\begin{widetext}
\begin{eqnarray}
\langle \sigma |v| \rangle &=& \frac{\sin^4\theta}{4\pi\,\left(2\, m_\chi^2- m_h^2\right)^2}  \,\sqrt{1 - {m_h^2 \over m_\chi^2} }\,\biggl( m_\chi^2 \left(\lambda_{\chi_A}^4+6 \,\lambda_{\chi_A}^2 \lambda_{\chi_V}^2+\lambda_{\chi_V}^4\right) - m_h^2 \left(\lambda_{\chi_A}^2+\lambda_{\chi_V}^2\right)^2   \biggr) + \ldots \label{eq:tchannelannhil} 
\end{eqnarray}
\end{widetext}  
Here $\sin\theta$ is the mass mixing angle defined in equation~\leqn{eq:cossin}.  We wrote out this equation to reemphasize a key point from Section II.  The cross section is multiplied by $\sin^4\theta$ which forces $\theta$ to remain nontrivial.  Moreover, $\sin\theta$ must be relatively large to generate enough dark matter annihilation.   %
Because of the larger values for $\sin^4\theta$, Model 1 has lower unitarity bounds and is much more constrained.

\subsubsection{V.1.2. s-channel annihilation}
\noindent
Depending on the dark matter mass, the annihilation processes are,
\begin{align}
\chi + \chi &\to \bar{q} + q &  \chi + \chi &\to W +  W \\
\chi + \chi &\to \bar{l} +  l &  \chi + \chi &\to Z +  Z,
\end{align}
in addition to, 
\begin{align}
\chi + \chi \to h +  h, \label{eq:Higgsschannel}
\end{align}
where $q = u, d, c, s, t, b$ and $l = e, \mu, \tau$.  
thermally averaged cross section is
\begin{eqnarray}
\langle \sigma |v| \rangle &=& \langle \sigma |v| \rangle_{\bar{f}f}  + \langle \sigma |v| \rangle_{VV} +  \langle \sigma |v| \rangle_{hh}  \nonumber
\end{eqnarray}
where
\begin{widetext}
\begin{eqnarray}
\langle \sigma |v| \rangle_{\bar{f}f} &=&\frac{\lambda_{\chi_A}^2 \sin^2\theta \cos^2\theta}{4 \pi} \sum_{f=u,d,c,s,t,b,e,\mu,\tau} \sqrt{1 - {m_f^2 \over m_\chi^2}}\,\Biggl({g\,m_f \over m_W}\Biggr)^2 \,\Biggl( {m_\chi^2 - m_f^2 \over \left(4\,m_\chi^2 - m_h^2\right)^2} \Biggr) + \ldots \label{eq:schannelff} \\
\langle \sigma |v| \rangle_{VV} &=&   {\lambda^2_{\chi_A} \,m_W^2\,\sin^2\theta \cos^2\theta \over 8\,\pi}\, \sum_{V = W,Z} 
\sqrt{1 - {m_V^2 \over m_\chi^2}} \,\Biggl({g_{Vh}^2 \over m_V^4 \left(4 m_\chi^2- m_h^2\right)^2} \Biggr)\Biggl(3 \,m_V^4-4 m_V^2 \,m_\chi^2+4\, m_\chi^4\Biggr)+ \ldots \,\,\,\,\,\,\,\,\,\,\\ \nonumber \\
\langle \sigma |v| \rangle_{hh} &=& {\lambda_{h^3}^2 \,\lambda^2_{\chi_A}\,\sin^2\theta \over 2\,\pi}\,\sqrt{1 - {m_h^2 \over m_\chi^2}}  
{9\,u^2 \over \left( 4\,m_\chi^2 - m_h^2\right)^2}  + \ldots \label{eq:schannelhh}
\end{eqnarray}
\end{widetext}
Here $g_{Wh} = g$ and $g_{Zh} = g\,(\cos\theta_W)^{-2}$.  Again, we have taken $x_F = m_\chi/T_F$ and $T_F$ is the freeze-out temperature.  See equation~\leqn{eq:lambdah3} for the definition of $\lambda_{h^3}$.  For brevity, we did not list the velocity suppressed terms.  We did, however, include them in our analysis.  The pseudoscalar coupling, $\lambda_{\chi_A}$, is responsible for these velocity unsuppressed s-channel annihilation terms.  This allows Model 2 (see Section II.2) to have more annihilation channels than Model 1.  Notice, the s-channel terms have a $\sin^2\theta\,\cos^2\theta$ prefactor which forces $\theta$ to be nontrivial and on average smaller than the t-channel prefactor, $\sin^4\theta$.  The annihilation channels and the s-channel prefactor leads to weaker bounds for Model 2.
\newline
\newline
The pseudoscalar coupling, $\lambda_{\chi_A}$, is basically responsible for the  annihilation strength in these channels.  This allows Model 2 (see Section II.2 for the definition) to have weaker unitarity bounds than Model 1.  Also, to emphasize a point made in Section III, the above equations are multiplied by $\sin\theta \cos\theta$ which forces $\theta$ to be nonzero.

\subsection{V.2. Direct Detection}
\noindent
XENON100 provides the strongest constraints on the Higgs portal parameter space.  In this section we compute the spin-independent dark matter-nucleon scattering cross section.  To write down the effective dark matter-nucleon interaction, we follow the derivation in~\cite{Drees:1993bu,Ellis:2000ds}.  Consider the dark matter-quark effective interaction, 
 \begin{equation}
 \mathcal{L}_{q\chi} = f_q\,\chi_L\chi_R\,\bar{q}\,q
 \end{equation}
 where we have defined,
 \begin{equation}
 f_q = \lambda_\chi\lambda_q\,\sin\theta\cos\theta\,\biggl(   {1 \over 2\,m_h^2} -  {1 \over 2\,m_\rho^2}\biggr)
 \end{equation}
Both Higgses couple to the heavier quarks more significantly than the light quarks.  Thus the coupling to nucleons is least suppressed through a heavy quark loop.  Thus, the effective nucleon-dark matter interaction is
\begin{equation}
 \mathcal{L}_{N\chi} = f_N\,\chi_L\chi_R\,\bar{\Psi}_N \Psi_N,
\end{equation}
where,
 \begin{eqnarray}
 f_N = m_N\,\Biggl[ \,\,\sum_{q = u,d,s} {f_q \over m_q}\,f_{Tq} + {2 \over 27}\,f_{TG} \sum_{Q = c,b,t} {f_Q \over m_Q} \,\,\Biggr].\,\,\,\,
 \end{eqnarray}
$f_{TG} = 1 - \sum_{q= u,d,s}f_{Tq}$ and $N = $ proton or neutron.  $f_{Tq}$ is the fraction of the nucleon mass that is due to the light quark, $q$.
\begin{equation}
\langle N |\, m_q\, \bar{q} q \,| N \rangle = m_N f_{Tq}.
\end{equation}
From~\cite{Ellis:2000ds}, we take
\begin{align} 
f^{(p)}_{Tu}  = 0.020 \pm 0.004 && f^{(n)}_{Tu}  = 0.014 \pm 0.003 \\
f^{(p)}_{Td}  = 0.026 \pm 0.005 && f^{(n)}_{Td}  = 0.036 \pm 0.008  \\ 
f^{(p)}_{Ts} = 0.118 \pm 0.062 && f^{(n)}_{Ts} = f^{(p)}_{Ts} \hspace{1.4cm}
\end{align}
The elastic, spin-indpendent dark matter-nucleon cross section for the idealized case of a point-like nucleon is given by
\begin{equation}
\sigma = {4\,m_\chi^2 \,m_A^2 \over \pi\,(m_\chi + m_A)^2}\,f_p^2
\end{equation}
and $m_A$ is the proton mass.  The above is in the limit where $f_p = f_n$.  

 \section{VI. Higgs Constraints}
 \label{sec:smHiggsconstraints}
 \noindent
%
The LHC has discovered what likely is the SM Higgs boson~\cite{atlasHiggsresults,CMSHiggsresults}.  We find constraints on the Higgs mixing angle is weak at best and do not lead to a narrowing of the available parameter space.  In this Section, we review these conclusions.

\subsection{VI.1. Higgs Mixing Constraints}
 \label{sec:Higgsmix}
\noindent
It is not well known how much the measured Higgs signal cross section deviates from SM expectations in a statistically significant way.  The strength of the signal cross section places constraints on the mass mixing angle, $\cos\theta$.  We take the range of $\cos\theta$ to be
\begin{equation}
\cos\theta \in [1/\sqrt{2},\,1]. \label{eq:cthetarange}
\end{equation} 
In the Appendix, we use current ATLAS and CMS results to argue for this range.  Many points in parameter space approach the upper limit.  With more data, however, Higgs mixing can severely constrain Higgs portal annihilation.  The potential of new experiments to probe and thereby constrain Higgs portals is discussed in Section X. 
 \subsection{VI.2. Previous Higgs Searches}
 \label{sec:prevHiggs}
 \noindent
We primarily focus on the limit where $m_\rho > m_h$.  Previous searches for Higgses at LEP~\cite{Barate:2003sz} and the Tevatron~\cite{Group:2012zca} did not discover Higgs-like particles  therefore motivate this choice.  Recall, the dark Higgs has a coupling to SM particles proportional to the SM Higgs' coupling modulo a factor of $\sin\theta$ (see Section II.3).  This implies if  $m_\rho < m_h$ then the mixing angle is small.  We discuss this in detail in Section VIII.

 \subsection{VI.3. Precision Electroweak Constraints}
 \label{sec:precision}
 \noindent
Our scenario adds a minimum amount of new content to the SM.  However, the dark Higgs does generate logarithmic enhanced corrections to the $S$ and $T$ precision electroweak parameters~\cite{Peskin:1991sw}.  We checked that the scans over the Higgs portal parameter space (in Section VII) are well within the most recent 95\% c.l. precision electroweak constraint ellipse~\cite{Baak:2012kk}.  Of course, this can change if one departs from this minimal scenario.  For reference, the $S$ and $T$ parameter corrections are listed in the Appendix.

\section{VII. Unitarity Constraints}
\noindent
There are five unknown Higgs portal parameters, see equation~\leqn{eq:parameterspace}, that are constrained to four by the measured SM Higgs mass.  In this section, we derive the high-energy scattering amplitudes needed to place the unitarity constraints.  These amplitudes directly constrain the couplings in equation~\leqn{eq:parameterspace}.   We use the partial wave unitary constraint~\cite{Marciano:1989ns,Luscher:1988gc},
\begin{equation}
\bigl| \mathrm{Re}\, \mathcal{M}^{(j)}\bigr| \leq {1\over 2},
\label{eq:unitconstraint}
\end{equation}
for all of our computations.  
Notably, $M^{(j)}$ is a matrix and the condition in equation~\leqn{eq:unitconstraint} must be applied to every eigenvalue.  Finally, as described before, the relic abundance constraint in concert with the unitarity bounds derived below sets the scale of new physics when $m_\rho > m_\chi, m_h$.  We show this in the next section.  Section XI~relaxes this mass hierarchy.

\subsection{VII.1. Goldstone-Higgs Boson Scattering Diagrams}
\noindent
To generate the unitarity constraints on the Higgs sector, we employ the Goldstone boson equivalence theorem and focus on Higgs-Higgs, Goldstone-Higgs and Goldstone-Goldstone scattering.  The uneaten Goldstone bosons are the longitudinal components of the electroweak gauge bosons.  Because of the equivalence theorem, at high energies the standard $W$ and $Z$ scattering can be replaced by  Goldstone-Higgs scattering amplitudes.  The uneaten Goldstone bosons are,
\begin{align}
h = {1 \over \sqrt{2}}\begin{pmatrix} w^+ \\ v + h_0 + i\,z \end{pmatrix} && \rho = {1 \over \sqrt{2}}\,(u + \rho),
\end{align}
where we defined $w^\pm = w_1 \mp i\,w_2$.  Going forward, we relabel $h_0 \to h$ for consistency.  The scalar potential,  equation~\leqn{eq:Higgspotential}, is now
\begin{eqnarray}
V &=& \lambda_1 \,\biggl( v^2 \,h^2 + v\, h\,\left( 2 \,w^+w^- + h^2 + z^2 \right) \label{eq:Higgspotentialext} \\
&+& {1 \over 4}\left(2\,w^+w^- + h^2 + z^2 \right)^2\biggr) \nonumber \\
&+&  \lambda_2 \,\biggl( u^2\,\rho^2 + u\,\rho^3 + {1 \over 4}\,\rho^4  \biggr) \nonumber \\
&+& \lambda_3\,\biggl(u\,v\,\rho\,h + {1 \over 2}\,v\,h\,\rho^2   \nonumber \\
&+& {1 \over 2} \,u\,\rho\,\left(2\,w^+w^- + h^2 + z^2 \right) \nonumber \\
&+& {1 \over 4}\,\rho^2\left(2\, w^+w^- + h^2 + z^2 \right)\biggr).   \nonumber
\end{eqnarray}
To simplify matters, we focus on the charge neutral scattering processes,
\begin{eqnarray}
V + V^* &\to& V^* + V   \label{eq:VVVV} \\ 
V + V^* &\leftrightarrow& H + H \label{eq:VVHH} \\
V + V^* &\leftrightarrow& H + z \label{eq:WWHZ} \\
H + H &\to& H + H \label{eq:HHHH} \\
H + H &\leftrightarrow& H + z  \label{eq:HHHZ} \\
H + z &\to&  H + z. \label{eq:HZHZ} 
\end{eqnarray}
We denote $V = w^+, z$ and $H = h, \rho$.  The Goldstone-Higgs scattering replaces the standard scattering between the longitudinal states of the $W$ and $Z$ bosons.  The above processes account for all possible scattering combinations.  In this section we generate unitary constraints on these processes by essentially generalizing the analysis in~\cite{Lee:1977eg}.  In the Appendix, we list the amplitudes for Higgs-Higgs, Goldstone boson-Higgs and Goldstone boson-Goldstone boson scattering for reference.
\newline
\newline
The scattering amplitudes in the Appendix have two unknown scales, $u$,  $m_\rho$.  This is unlike the analogous electroweak computation which had only the SM Higgs mass as an unknown.  We compute the unitarity bound in the traditional limit $s \gg u^2, m_\rho^2$ as well as the limit where $u^2, m_\rho^2 \gg s \gg m_h^2$.  We argue the other hierarchy, $u^2 \gg s \gg m_\rho^2, m_h^2$, is similar to $s \gg u^2, m_\rho^2$.
\newline
\newline
Because of the number of Higgses and Goldstone bosons, we have a seven channel system (equations~\leqn{eq:VVVV}-\leqn{eq:HZHZ}) onto which we must apply the unitarity bounds.  The system can be represented by the vector,
\begin{equation}
\left(w^+ w^-, \,{z z \over \sqrt{2}}\,, \,{h h \over\sqrt{2}}, \,{\rho \rho \over \sqrt{2}}, \,h \rho, \,h z, \,\rho z \right),
\end{equation}
which describes initial and final states for different interactions.  The eigenvalues of the resulting Higgs-Higgs, Goldstone-Higgs and Goldstone-Goldstone scattering matrix must satisfy equation~\leqn{eq:unitconstraint}.

\subsubsection{VII.1.1. Scattering Matrix for  $s \gg u^2, m_\rho^2, m_h^2$}
\noindent
When $s \gg u^2, m_\rho^2, m_h^2$, 
the zeroth partial wave of coupled seven channel system has the form,
\begin{widetext}
\begin{eqnarray}
\mathcal{M}^{(0)}_I &=& \label{eq:firstunitconstr}  
-{\lambda_1\over 4\pi}\,\begin{pmatrix}  
1 				& {1\over \sqrt{8}} 		& {c^2\over \sqrt{8}} 	&  {s^2 \over \sqrt{8}} 	& {s c \over 2}  			& 0 & 0 \\ \\
{1\over \sqrt{8}}		& {3 \over 4} 			& {c^2 \over 4} 		& {s^2 \over 4}  			&  {s c \over \sqrt{8}}  	& 0 & 0 \\ \\
{c^2 \over \sqrt{8}} 	& {c^2 \over 4} 			& {3 c^4 \over 4}	 & {3 s^2 c^2 \over 4}	&  {3\, s c^3 \over \sqrt{8}} 	& 0 & 0 \\ \\
{s^2 \over \sqrt{8}}  	& {s^2 \over 4} 			& {3\, s^2 c^2 \over 4}			& {3 s^4 \over 4}		& {3\, c s^3 \over \sqrt{8}}	& 0 & 0 \\ \\
{s c \over 2} 		& {s c \over \sqrt{8}} 		 & {3\, s c^3 \over \sqrt{8}}	&  {3\, c s^3 \over \sqrt{8}} 	& {3\, c^2 s^2 \over 2}		& 0 & 0 \\ \\
0 & 0 & 0 & 0 & 0 	& {c^2 \over 2} 	& {s c \over 2}  \\ \\
0 & 0 & 0 & 0 & 0 	& {s c \over 2} 		& {s^2 \over 2} 
\end{pmatrix}   
- {\lambda_3\over 4\pi}\,
\begin{pmatrix}  
0 				& 0					& {s^2\over \sqrt{32}} 			&  {c^2 \over \sqrt{32}}			& -{s c \over 4}  			& 0 & 0 \\ \\
0				& 0 					& {s^2 \over 8} 					& {c^2 \over 8}  					&  {-s c \over \sqrt{32}}  	& 0 & 0 \\ \\
{s^2\over \sqrt{32}} 	& {s^2 \over 8} 			& \kappa						& \delta						& \xi 				& 0 & 0 \\ \\
{c^2	 \over \sqrt{32}}	& {c^2 \over 8} 			& \delta						&  \alpha						& \beta					& 0 & 0 \\ \\
-{s c \over 4}  		& -{s c \over \sqrt{32}} 	& \xi							& \beta						& \eta		& 0 & 0 \\ \\
0 				& 0 					& 0 							& 0 & 0 						& {s^2 \over 4} 	& -{s c \over 4}  \\ \\
0 				& 0 					& 0 							& 0 & 0 						& -{s c \over 4}	& {c^2 \over 4} 
\end{pmatrix}, \,\,\,\,\,\,\,\,
\end{eqnarray}
\end{widetext}
where $c = \cos\theta$, $s = \sin\theta $ and 
\begin{subequations}
\begin{align}
c = 1-\frac{\lambda _3^2 \,v^2}{8 \lambda _2^2\,u^2} && s  = \frac{\lambda _3\,v}{2\lambda _2\,u}.
\end{align}
Also,
\begin{align}
\kappa &= {3 \over 4}s^2\biggl(c^2  + {\lambda_2 \over \lambda_3} s^2\biggr) \\
\delta &= \frac{1}{8} \biggl(s^4+c^4\biggr)-\frac{1}{2} s^2c^2 \biggl(1 - {3 \over 2}\frac{\lambda_2}{\lambda_3}\biggr)  \\ 
\xi &= {3 \,c s \over \sqrt{8}} \,\left(-\frac{1}{2} \left(c^2-s^2\right)-\frac{\lambda_2 }{\lambda_3}s^2\right) \\
\alpha &= \kappa(s\leftrightarrow c) \\
\beta	&= \xi(s\leftrightarrow c) \\
\eta  	&= 2\,\delta.
\end{align}
\end{subequations}
In the decoupling limit, this seven by seven matrix reduces to the four-channel Higgs system given in~\cite{Lee:1977eg} plus a disassociated dark Higgs.  
\newline
\newline
The unitarity arguments in~\cite{Lee:1977eg} place bounds only on the SM Higgs' quartic coupling; the matrix therein can 
be analytically diagonalized.  The above matrix is 
most efficiently solved numerically thereby necessitating a parameter scan.  We do this in the next section.  For example, if $\lambda_1 = \lambda_2 = \lambda_3 = 1$ and $u = 1$ TeV, four of the eigenvalues %
%
violate the unitarity condition (equation~\leqn{eq:unitconstraint}) thereby invalidating this parameter point.  

\subsubsection{VII.1.2. Scattering Matrix for $u^2, m_\rho^2 \gg s \gg m_h^2$}
\noindent
In the limit where $u, m_\rho$ is much larger than $\sqrt{s}$ another  unitarity bound can be generated.  
Here the dark Higgs is integrated out.  A four channel system remains consisting of 
\begin{equation}
\left(w^+ w^-, \,{z z \over \sqrt{2}}\,, \,{h h \over\sqrt{2}}, \,h z\right)
\end{equation}
remains.  This computation is very similar to the famous SM Higgs computation in~\cite{Lee:1977eg} modulo additional contributions from operators resulting from the integrated out dark Higgs.  The resulting zeroth partial wave is
\begin{widetext}
\begin{eqnarray}
\label{eq:secondunitconstr} 
\mathcal{M}^{(0)} =  -{\lambda_1\over 4\pi}\,\begin{pmatrix}  
1 				& {1\over \sqrt{8}} 		& {1 \over \sqrt{8}} 	& 0 \\ \\
{1\over \sqrt{8}}		& {3 \over 4} 			& {1 \over 4} 		& 0  \\ \\
{1 \over \sqrt{8}} 	& {1 \over 4} 			& {3 \over 2}		& 0 \\ \\
0 & 0 & 0 	& {1 \over 2}  
\end{pmatrix} 
+ {1\over 16\, \pi}\,{\lambda_3^2 \over \lambda_2}\,
\begin{pmatrix}  
1 				& {1 \over \sqrt{8}}			& {3\over \sqrt{8}} 	& 0  \\ \\
{1 \over \sqrt{8}}	& {3 \over 4}				& {3 \over 4} 		 	& 0\\ \\
{3 \over \sqrt{8}} 	& {3 \over 4} 				& {3 \over 2}				& 0 \\ \\
0 				& 0 					& 0 					& 0 	
\end{pmatrix}. 
\end{eqnarray}  
\end{widetext}
These eigenvalues can be found analytically.  

\subsubsection{VII.1.3. Scattering Matrix for $u^2 \gg s \gg m_\rho^2, m_h^2$}
\noindent
In the limit of $u^2 \gg s \gg m_\rho^2, m_h^2$, the partial wave amplitudes are sum of equation~\leqn{eq:firstunitconstr} plus additional contributions that go as $\epsilon$ or $\epsilon \,\log \epsilon$, where $\epsilon = m^2/s$ and $m$ are the light masses, $m_\rho$, $m_h$, $m_Z$ or $m_W$.  The $\log \epsilon$ factor originates from those terms with $t$- and/or $u$-channel propagators after integrating over the scattering angle in the center-of-mass fame.
%
The $\epsilon \,\log \epsilon$ and $\epsilon$ terms are well behaved and in the limit of $\epsilon \to 0$.  This leads us back to  equation~\leqn{eq:firstunitconstr} to apply unitarity bounds.
%

\subsection{VII.2. Dark Matter Scattering Diagrams}
\noindent
In the addition to the Higgs-Higgs, Goldstone-Higgs and Goldstone-Goldstone scattering we also consider dark matter scattering diagrams.  The dark matter-dark matter self-scattering amplitudes do not grow as energy; however, they have do a prefactor of $\lambda_\chi^2$ which can be constrained by equation~\leqn{eq:unitconstraint}.  We place the explicit scattering amplitudes in the Appendix.  We use the unitarity bounds derived in this section along with the relic abundance to place constraints on this Higgs portal parameter space when $m_\rho > m_h, m_\chi$.
\subsubsection{VII.2.1. Scattering Matrix for  $s \gg m^2_\chi, m_\rho^2, m_h^2$}
\noindent
In order to directly place constraints on the dark matter Yukawa coupling, $\lambda_\chi$, we also consider the process
\begin{eqnarray}
 \chi + \overline{\chi} &\to& \chi + \overline{\chi}.  
\end{eqnarray}
An analogous process was considered in~\cite{Chanowitz:1978uj,Chanowitz:1978mv} to place upper bounds on new fermion masses resulting from electroweak symmetry breaking.  Considering a dark photon in theory that directly couples to the dark matter generates a nontrivial scattering matrix analogous to equation~\leqn{eq:firstunitconstr}.  We consider this in~\cite{devinwill}.
\newline
\newline
We list the self-scattering amplitudes in the Appendix.  In the limit where $s \gg m_\chi^2, m_\rho^2, m_h^2$  (and in the order $\chi+ \overline{\chi} \to \chi + \overline{\chi}$), the amplitudes have the form, 
\begin{widetext}
\begin{eqnarray}
\mathcal{M} = \,- {1 \over 2} (\lambda_{\chi_V}^2 + \lambda_{\chi_A}^2) \hspace{0.75cm} && \begin{cases} (+ \,\, + \,\to \,+ \,\, +)  \\  (- \,\, - \,\to \,- \,\, -) \end{cases} \hspace{-0.2cm}\mathrm{s\!-\!channel}\hspace{0.1cm} \nonumber \\
\mathcal{M} =  {1 \over 2} (\lambda_{\chi_V} + i\,\lambda_{\chi_A})^2 \hspace{0.75cm} && \begin{cases} (+ \,\, + \,\to \,- \,\, -) \end{cases} \hspace{-0.2cm}\mathrm{s\!-\!channel} \hspace{0.1cm}  \nonumber\\
\mathcal{M} =  {1 \over 2} (\lambda_{\chi_V} - i\,\lambda_{\chi_A})^2 \hspace{0.75cm} && \begin{cases} (- \,\, - \,\to \,+ \,\, +) \end{cases} \hspace{-0.2cm}\mathrm{s\!-\!channel} \hspace{0.1cm}  \nonumber\\ \nonumber \\
 \mathcal{M} =  -{1 \over 2} (\lambda_{\chi_V}^2 + \lambda_{\chi_A}^2) \hspace{0.82cm} && \begin{cases} (+ \,\, - \,\to \,- \,\, +)  \\  (- \,\, + \,\to \,+ \,\, -) \end{cases} \hspace{-0.2cm}\mathrm{t\!-\!channel} \hspace{0.01cm} \nonumber \\
 \mathcal{M} =  {1 \over 2} (\lambda_{\chi_V} + i\,\lambda_{\chi_A})^2  \hspace{0.75cm}&& \begin{cases} (+ \,\, + \,\to \,- \,\, -)  \end{cases} \hspace{-0.2cm}\mathrm{t\!-\!channel}\hspace{0.01cm} \nonumber \\
  \mathcal{M} =  {1 \over 2} (\lambda_{\chi_V} - i\,\lambda_{\chi_A})^2. \hspace{0.75cm}&& \begin{cases} (- \,\, - \,\to \,+ \,\, +) \end{cases} \hspace{-0.2cm}\mathrm{t\!-\!channel}\hspace{0.01cm} \nonumber 
\end{eqnarray}
\end{widetext}
These are equivalent to the amplitudes found in~\cite{Chanowitz:1978mv} in the limit where $\lambda_{\chi_A} \to 0$.  Only the following helicities contribute to the $j = 0$ partial wave amplitude,
\begin{eqnarray}
(+ \,\, + \,\to \,+ \,\, +), \nonumber\\ 
(- \,\, - \,\to \,- \,\, -),\nonumber\\
(+ \,\, +\, \,\leftrightarrow \,-\,\, -).\nonumber
\end{eqnarray}
Given that the vector
\begin{equation}
\left(\chi_+ \overline{\chi}_+, \,\chi_- \overline{\chi}_- \right),
\end{equation}
the coupled channels of the zeroth partial wave amplitudes are
\begin{eqnarray}
\label{eq:thirdunitarityconst}
\mathcal{M}^{(0)} &=&  -{1 \over 32\pi}\,\begin{pmatrix}  
\lambda^*\lambda				& -2\,\lambda^2	 \\ \\
-2\,\lambda^{*\,2}				& \lambda^*\lambda		
\end{pmatrix}. 
\end{eqnarray}
Here $\lambda =  \lambda_{\chi_V}  + i\, \lambda_{\chi_A}$.  The unitarity constraint is,
\begin{equation}
\lambda_{\chi} = \sqrt{\lambda_{\chi_V}^2 + \lambda_{\chi_A}^2 } \lesssim 4,
\end{equation}
for the dark matter Yukawa coupling.
\newline
\newline
In addition to the $j = 0$ partial wave amplitudes for dark matter scattering, we also considered processes that contribute to the $j = 1$ amplitudes,
\begin{eqnarray}
\chi + \overline{\chi} &\leftrightarrow& w^+ + w^- \label{eq:chichiw+w-} \\
\chi + \overline{\chi} &\leftrightarrow& z + z  \label{eq:chichizz} \\
\chi + \overline{\chi} &\leftrightarrow& H + z  \label{eq:chichihz} \\ 
\chi + \overline{\chi} &\leftrightarrow& H + H \label{eq:chichihh}.
\end{eqnarray}
Here we defined $V = w^+, z$ and $H = h, \rho$.  The amplitudes in equations~\leqn{eq:chichiw+w-}-\leqn{eq:chichihz} are s-channel.  At large $\sqrt{s}$, these diagrams vanish.  The amplitude in equation~\leqn{eq:chichihh} decouples.  
See~\cite{Chanowitz:1978mv} for an explanation of the how similar diagrams decouple.  
%
%
%
%
%
%
%
%

\section{VIII.  Higgs Portal Parameter Scans}
\noindent
In this section, we take the unitarity constraints from Sections VI.~as well as the dark matter and Higgs constraints from Section IV.~and V.~and perform a general parameter scan over the Higgs portal parameter space.  As noted in Section III., each of these constraints has a special role in directly restrict the strength of the dark vev or the Higgs portal couplings. The combination of all the constraints yield the bound on the different particle masses.

\subsection{VIII.1. Understanding the Parameter Scans}
\noindent
In addition to the constraints discussed in the previous sections, we also require the cold dark matter is cold.  This requires the dark matter freeze out temperature, $T_f$, to satisfy $T_f  < m_\chi$.  
Also, our initial scan requires the Higgs portal parameter space to satisfy the relic abundance constraint at $\pm \,3\sigma$ level.  We assumed a perturbative expansion in equation~\leqn{eq:cossin}.  Thus, we require
\begin{equation}
{\lambda_3 \,v \over 2 \lambda_2\, u} < 1.
\end{equation}
The scan operates in the following fashion:  We take the couplings in equation~\leqn{eq:parameterspace} and vary them from 0 to 15.  For $\lambda_\chi$ we vary both $\lambda_{\chi_V}$ and $\lambda_{\chi_A}$ independently.  The unitarity constraint from dark matter scattering generates a bound on $\lambda_\chi$.  The dark symmetry vev, $u$, initially varied from 250 GeV to 150 TeV.  The upper value was steadily reduced when it became clear there were no points satisfying the all the constraints above the given bound.  Each plot starts with at least 400 million points in the parameter space before the various constraints are applied.  As emphasized throughout, the unitarity bounds are possible in the limit where $m_\rho \gtrsim m_\chi$.  The 400 million points have this limit already applied.
\newline
\newline
We organize the parameter scan plots in the following way:  
We first consider Model 1 and plot the parameter space which satisfies the measured relic abundance in equation~\leqn{eq:relic}.  This parameter space is plotted with and without the Xenon100 constraints.  We repeat this exercise for the parameter space that  satisfies only \textit{half} the measured relic abundance.  This is to illustrate how the bounds change in the limit where the Higgs portal dark matter is a fraction of the overall measured relic abundance.  We conclude this section after generating the same plots for Model 2.  The Xenon100 constraints for Model 2  are sufficiently weak that they do not visibly deform the parameter space.  In order not to repeat a plot that is visibly the same, instead we describe in words where most of the excluded parameter space points reside.
\newline
\newline 
The plots in this section have the dark Higgs mass versus the dark symmetry breaking vev.  To guide the eye, the plots have a dotted diagonal line with a slope that is roughly 0.25.  The parameter space points near this dotted line are near the decoupling limit, $\cos\theta \to 1$.  Thus for these points, equation~\leqn{eq:firstunitconstr} becomes the SM Higgs scattering amplitude~\cite{Lee:1977eg} along with a decoupled dark Higgs.  The largest eigenvalue comes from the SM Higgs amplitude and therefore generates the unitarity bound.  This line reproduces the result in~\cite{Lee:1977eg} when $u \to 246$ GeV.  We also include a solid, horizontal line in these plots which gives the bound on the dark symmetry breaking vev for each scenario.  This bound (and the fraying of the plot near the solid line) primarily comes from the relic abundance constraint and the unitarity bound generated from equation~\leqn{eq:secondunitconstr}.  Finally, the solid vertical line is the bound on the dark Higgs mass.  It is the result of the confluence of all the unitarity constraints listed in the previous section as well as the relic abundance.  \textit{From now on, we will often refer to the upper bound on the dark Higgs mass as the ``unitarity" bound even though multiple unitarity bounds and the relic abundance contributed to the limit.}  This follows the tradition of calling bounds on the SM Higgs mass the unitarity bound.  In the Appendix, we also plot the dark matter mass versus the Higgs mass mixing angle, $\cos\theta$.  In those plots, the solid vertical line is the constraint on the dark matter mass.  This bound is a result of equation~\leqn{eq:thirdunitarityconst} as well as the relic abundance and the other unitarity constraints. We refer to the bound on the dark matter mass as the  \textit{``dark matter unitarity bound."}
\begin{figure*}[t]
\centering
{\label{fig:parameterspace3}	
	\includegraphics[width=10truecm,height=7.0truecm,clip=true]{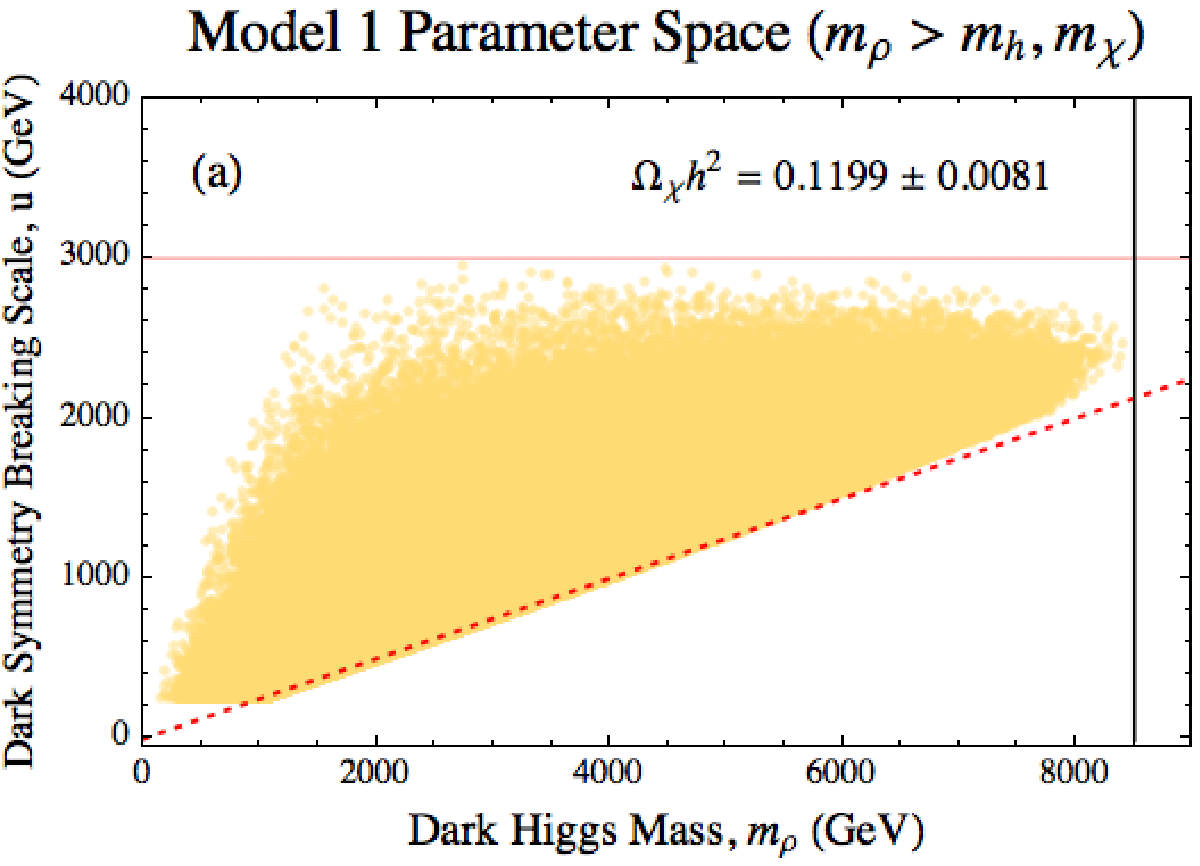} \vspace{1.5cm}  } 
	{\includegraphics[width=10truecm,height=7.0truecm,clip=true]{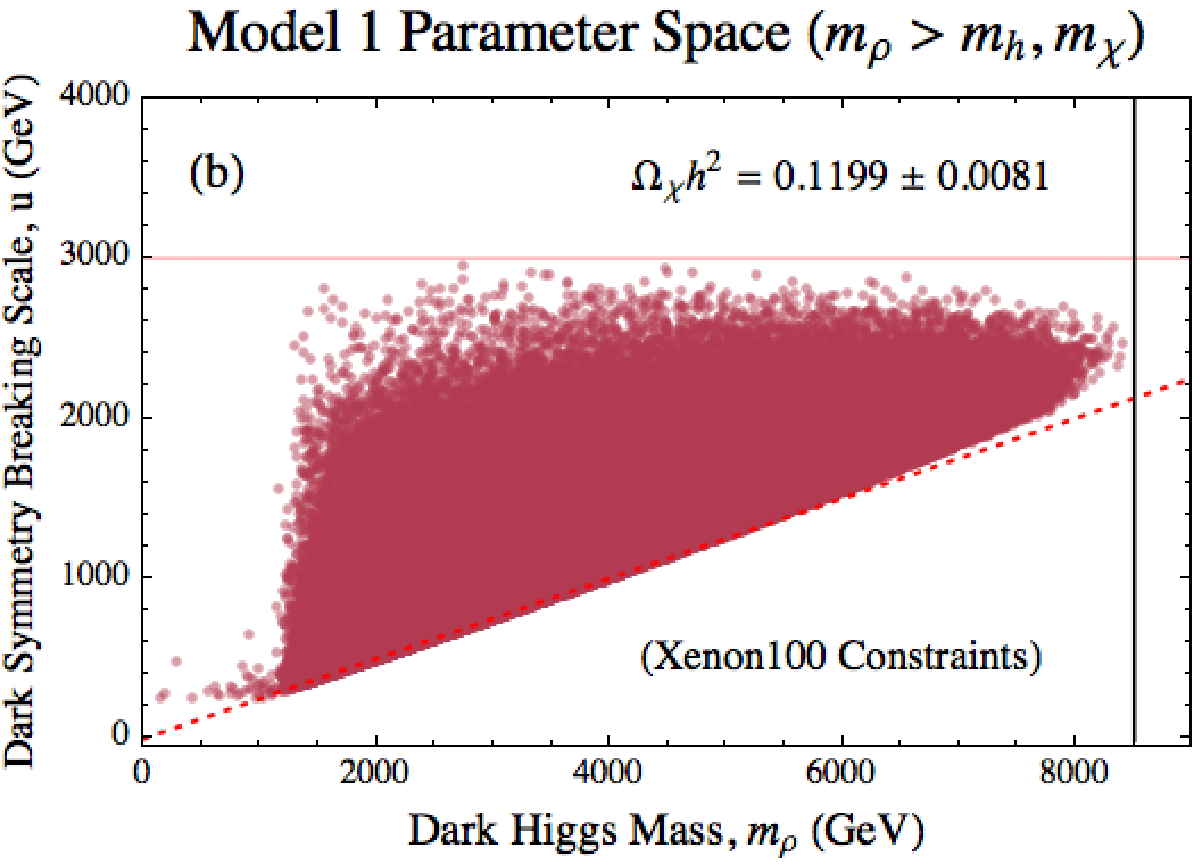}} 
	\caption{Higgs portal parameter space for Model 1 ($\lambda_{\chi_A}$ = 0) without (a) and with Xenon100 (b) constraints.  The Higgs portal dark matter is assumed to satisfy the measured dark matter relic abundance in equation~\leqn{eq:relic}.  The solid vertical line is the bound on the dark Higgs mass. The horizontal line is the bound on the dark symmetry breaking vev.  The dashed diagonal line is the unitarity constraint given by equation~\leqn{eq:firstunitconstr}.  The slope is consistent with the bound in~\cite{Lee:1977eg} and explicitly recovered when the dark vev is set to the electroweak vev.}
\end{figure*} 
\begin{figure*}[t]
\centering
{\label{fig:parameterspace3}	
	\includegraphics[width=10truecm,height=7.0truecm,clip=true]{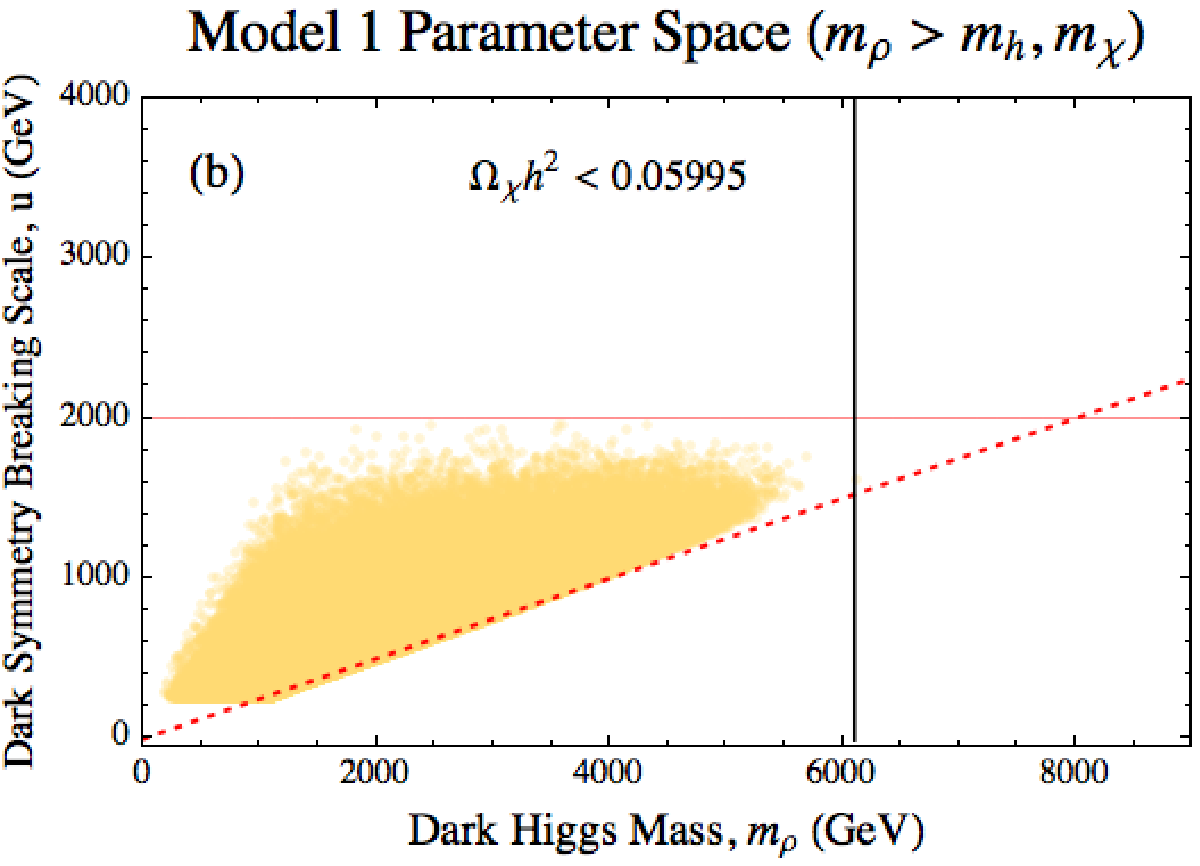} \vspace{1.5cm}}	 
	{\includegraphics[width=10truecm,height=7.0truecm,clip=true]{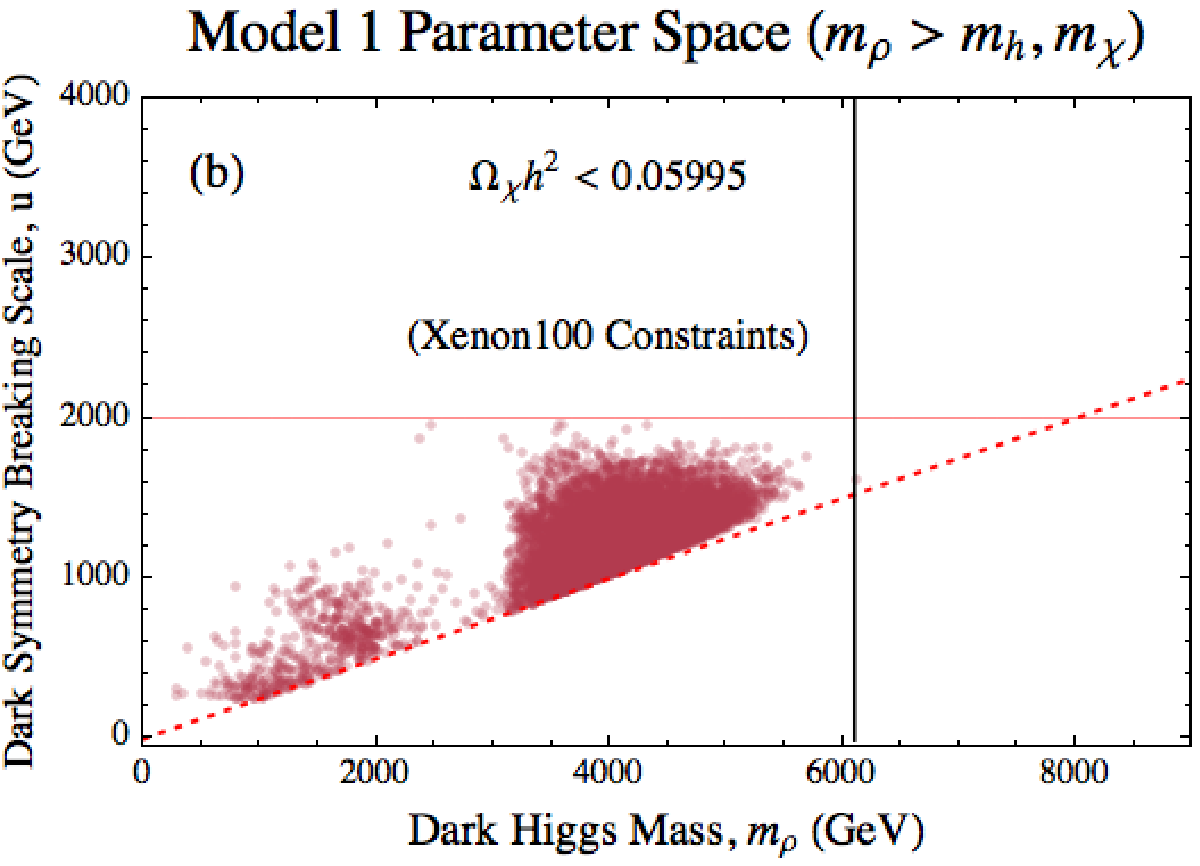}} 
	\caption{Higgs portal parameter space for Model 1 ($\lambda_{\chi_A}$ = 0) without (a) and with Xenon100 (b) constraints.  The Higgs portal dark matter is assumed to satisfy the only half of the measured dark matter relic abundance.  See equation~\leqn{eq:partialrelic}.    The vertical line is the bound on the dark Higgs mass. The horizontal line is the bound on the dark symmetry breaking vev.  The dashed diagonal line is the same as in Figure (1a) and (1b).  Notice the bounds \textbf{\textit{improve}} if the Higgs portal only satisfies only a fraction of the measured relic abundance.  In Figure (2b), the local density of points around 1.7 TeV has a fine-tuned dark matter mass so the low-velocity cross section features a small denominator.  The small denominators allow for the couplings and mixing angles to also be relatively small and therefore evade Xenon100.} 
\end{figure*} 
\begin{figure*}[t]
\centering
{\label{fig:parameterspace5}	
	\includegraphics[width=10truecm,height=7.0truecm,clip=true]{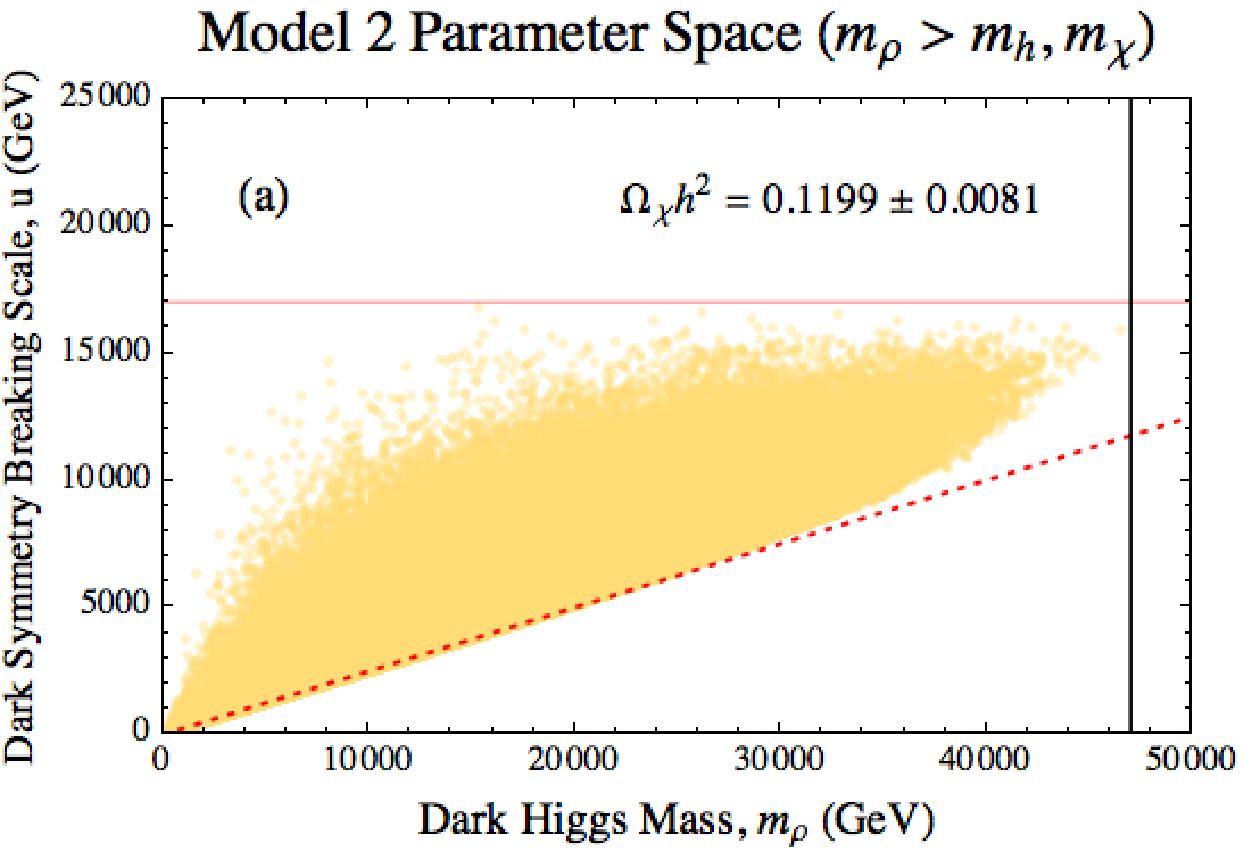} \vspace{1.5cm}}	 
	{\includegraphics[width=10truecm,height=7.0truecm,clip=true]{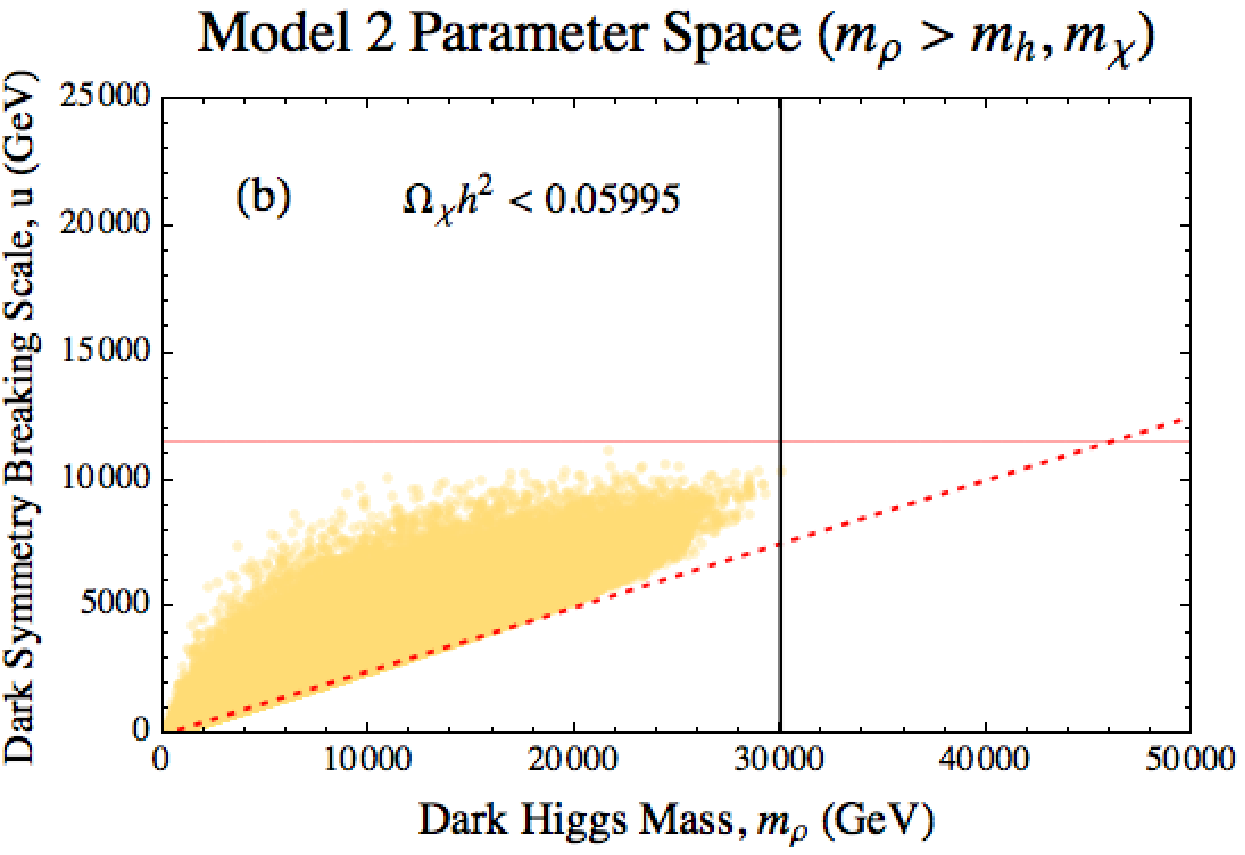}} 	
	\caption{Higgs portal parameter space for the full model.  Please see the caption in Figure (1) for the explanation of the various boundary lines.  The parameter space outlined in Figure (3a) satisfies the measured relic abundance.  Figure (3b) has the parameter space points which satisfy half of the observed relic abundance.  See equation~\leqn{eq:partialrelic}.   Like before, the bounds \textbf{\textit{improve}} if dark matter is multi-component and the Higgs portal only satisfies only a fraction of the measured relic abundance. Xenon100 only removes 1\% of the parameter space points which are clustered in the lower-left corner of the plot.  Since the difference is not visible, we have not included the plot with the Xenon100 constrained parameter space.}
\end{figure*} 
\newline
\newline
In terms of the physics, recall there is only one unsuppressed channel, equation~\leqn{eq:tchannelannhil}, for the dark matter to annihilate in Model~1.  For Model~2, there are many more available channels, equations~\leqn{eq:tchannelannhil}-\leqn{eq:schannelhh}.  (See Section III.2. for the definitions of Model 1~and~2.)  This means, given a dark matter mass, on average the couplings and mixing angles for Model 1 are going to be larger those in Model 2 in order to generate the measured relic abundance.  Thus, Model 1 has much stronger bounds than Model 2.  As mentioned above, we also consider the case where Higgs portal dark matter is only one component of the overall measured relic abundance.   For both models, satisfying a smaller relic abundance means the dark matter has to annihilate more and leads to larger couplings and mixing angles.  Therefore more stringent bounds appear.

\subsection{VIII.2. Unitarity Bounds for Model 1}
\noindent
For Model 1 ($\lambda_{\chi_A} = 0$), the dark matter scatters with nucleons with a spin-independent cross section.  In Figure (1a), we plot the results of the parameter scan in terms of the dark Higgs mass versus the dark symmetry breaking vev.  Figure (1b) has the same parameter space as Figure (1a) except the points ruled out by Xenon100 are removed.  The bounds on the parameter space that satisfies the measured relic abundance in equation~\leqn{eq:relic} at the $3\,\sigma$ level are 
\begin{eqnarray}
m_\rho &\lesssim& 8.5\,\,\mathrm{TeV}, \\
u &\lesssim& 3\,\,\mathrm{TeV}.
\end{eqnarray} 
In general the dark matter unitarity bound is of order the bound on the dark Higgs mass.  We find
\begin{eqnarray}
m_\chi &\lesssim& 8.4\,\,\mathrm{TeV}. 
\end{eqnarray}
In the Appendix, we plot the dark matter mass versus~$\cos\theta$ with and without the Xenon100 constraints for Model 1.  For dark matter masses less than 3 TeV, $\cos\theta$ can range from $0.75$ to almost $1$.  Above this mass, the parameter space asymptotes to $\cos\theta \sim 1$.  Xenon100 eliminates many of the points for dark matter masses below about $1.6$ TeV.  Notably, this bound on the dark matter mass is a significant improvement over the constraint given in~\cite{Griest:1989wd}.
\newline
\newline 
It is certainly possible for the Higgs portal dark matter in Model 1 to be only a fraction of the overall measured dark matter relic abundance.  Non-thermal relics like axions can serve as a large fraction of the abundance.  In Figure (2a) and (2b) we show consider a scenario where the Higgs portal dark matter is responsible for only half of the measured relic abundance,
\begin{equation}
\Omega_\chi h^2 \leq 0.1192/2 = 0.05995. \label{eq:partialrelic}
\end{equation} 
Figure (2b) shows the valid parameter with the Xenon100 constraints.  Figure (2a) has the parameter space without these constraints.  The bounds on the parameter space are now
\begin{eqnarray}
m_\rho &\lesssim& 6.1\,\,\mathrm{TeV},  \\
u &\lesssim& 2\,\,\mathrm{TeV} .
\end{eqnarray}
It is noteworthy that the bounds \textbf{\textit{improve}} in this multi-component limit.  The dark matter bound is now
\begin{eqnarray}
m_\chi  &\lesssim& 5.7\,\,\mathrm{TeV}.
\end{eqnarray}
As discussed in the previous sections, the reason for this improvement  is not a mystery.  The smaller relic abundance requires the couplings and mixing angles must be larger on average for a given dark matter mass.  The larger couplings translate to lower bounds.
%
%
%
%
\newline
\newline
Finally by inspection, almost all the unbounded regions of parameter space have $\lambda_i > 3$ for at least one coupling in the Higgs potential.  This implies the possibility that the couplings in the Higgs potential will become strongly coupled before reaching the unitarity bound.  This in turn implies a new scale of physics.  We address this possibility in Section XI.  

\subsection{VIII.3. Unitarity Bounds for Model 2}
\noindent
For Model 2, the dark matter scatters with nucleons with both a spin-independent and spin-dependent cross section.  We repeat exercise of the previous section and plot the dark Higgs mass versus the dark symmetry breaking scale in Figure (2).  The points that satisfy the measured dark matter relic abundance (equation~\leqn{eq:relic})  at the $3\sigma$ level generate the following bounds
\begin{eqnarray}
m_\rho &\lesssim& 45.5\,\,\mathrm{TeV}, \label{eq:unitaritybound}   \\
u &\lesssim& 17\,\,\mathrm{TeV}.
\end{eqnarray}
We did not include the parameter space plot when the Xenon100 constraints are included.  Xenon100 excludes less than 1\% of the overall parameter space points mostly concentrated in the lower, left-handed corner of Figure (3).  The difference between Figure (3) and the plot with the points excluded by Xenon100 removed is that they are not visually distinguishable.  Xenon100 excludes points where the dark Higgs mass and dark symmetry breaking vevs are less than $500$ GeV and $3$ TeV, respectively.  In the Appendix, we plot $m_\chi$ versus~$cos\theta$.  Again, we do not include a plot with the Xenon100 exclusion points as the difference is visually indistinguishable.  The bound on the dark matter mass is 
\begin{eqnarray}
m_\chi &\lesssim& 45.5\,\,\mathrm{TeV}.
\end{eqnarray}
Most of the parameter space points asymptote quickly to $\cos\theta \sim 1$ as the dark matter mass is increased.  The bounds on the dark matter mass are still an improvement over~\cite{Griest:1989wd}.  Notably like the Model in the previous section, much of the parameter space points have couplings in the Higgs potential which satisfy $\lambda_i > 3$.  Thus requiring a perturbativity in the Higgs sector means stronger bounds. 
\newline
\newline
Higgs portal dark matter can only a fraction of the overall measured dark matter relic abundance.  The parameter points that satisfy the relic abundance in equation~\leqn{eq:partialrelic} are in Figure 4. The bounds are
\begin{eqnarray}
 m_\chi &\lesssim& 30\,\,\mathrm{TeV}, \label{eq:unitaritybound2} \\  
 u &\lesssim& 11.5\,\,\mathrm{TeV}.
\end{eqnarray}
which are an improvement.  See Section VIII.1.~for an explanation for why the bounds improve.  Xenon100 excludes about 30\% of the points in Figure (3b).  The points are concentrated in the area where the dark Higgs mass and dark symmetry breaking vevs are less than $620$ GeV and $3$ TeV, respectively.  Like before, we do not include a plot with the Xenon100 exclusion points as the difference is visually indistinguishable.  As an additional check (without reference plots), we considered the parameter points which satisfy,
\begin{equation}
\Omega_\chi h^2 < 0.03997,
\end{equation}
which is roughly a third of the measured relic abundance.  For the Higgs portal dark matter to satisfy this fraction of the relic abundance, the bounds are now
\begin{eqnarray}
m_\chi, m_\rho &\lesssim& 26\, \mathrm{TeV}, \\ 
u &\lesssim& 9.2\, \mathrm{TeV}.
\end{eqnarray}
By inspection of the data, a larger portion of the remaining of the parameter space remains perturbative in comparison to Model 1. 
%
\section{IX.  Perturbativity Parameter Scans}
\noindent
In the limit where the dark matter Yukawa coupling is zero, the quartic couplings in the Higgs potential (equation~\leqn{eq:Higgspotential}) monotonically increase as a function of the momentum scale.  Eventually theory becomes strongly coupled and generates a Landau pole.  In the past, this argument was used to suggest when new physics would appear for a given SM Higgs mass.  For example, a SM Higgs mass larger than 560 GeV reaches a Landau pole before than the 1~TeV unitarity bound~\cite{Kolda:2000wi}.  Such a large SM Higgs mass implies a large quartic coupling in equation~\leqn{eq:ewHiggspotential}, $\lambda_\mathrm{SM} > 2.6$.\newline
\newline
In general, perturbativity arguments are not as robust like the unitarity arguments described in the previous section.  However if the new physics is expected to be perturbative, new particles will appear well before theory becomes strongly coupled.  As an example, the dimension six operators in four-fermi theory violate unitarity around 350 GeV.  theory starts becoming strongly coupled at scales above 292 GeV.  However, the W boson perturbatively prevents the four-fermi scattering amplitudes from growing to large by appearing at 80 GeV.  In this section, we aim to improve the bounds in Section VI by applying the one-loop renormalization group equations (RGEs).  In previous section, we considered many different bounds.  See Section VIII.1.,  In this Section we apply a peturbativity condition on the bounds given in the previous Section.

\subsection{IX.1. Perturbativity Conditions}
\noindent
To construct the perturbativity conditions, we adapt the arguments made in~\cite{Barbieri:2006dq}.  The relevant one-loop renormalization group equations~\cite{Cheng:1973nv} for Higgs portal couplings are, 
\begin{widetext}
 \begin{eqnarray}
16\pi^2\,{d\lambda_1 \over d t} &=&  24 \lambda_1^2 +  {1 \over 2} \lambda_3^2 + 12\lambda_t^2\,\lambda_1 - 6\lambda_t^4, \label{eq:lambda1rge} \\
16\pi^2\,{d\lambda_2 \over d t} &=& 18\, \lambda_2^2 +  2\, \lambda_3^2 + 4 \lambda_2\,\bigl(\lambda_{\chi_V}^2 + \lambda_{\chi_A}^2 \bigr) - {1 \over 2}\,\bigl(\lambda_{\chi_V}^2 + \lambda_{\chi_A}^2 \bigr)^2,  \label{eq:lambda2rge}\\
16\pi^2\,{d\lambda_3 \over d t} &=& 36 \,\lambda_1 \lambda_3 + 18 \,\lambda_2 \lambda_3 + 4 \bigl(\lambda_{\chi_V}^2 + \lambda_{\chi_A}^2 \bigr)\lambda_3, \\
16\pi^2\,{d\lambda_{\chi_V} \over d t} &=& {5 \over 2}\,\bigl(\lambda_{\chi_V}^2 + \lambda_{\chi_A}^2 \bigr)\lambda_{\chi_V} + {1 \over 24\pi^2}\bigl(\lambda_3^2 + 3 \lambda_2^2 \bigr)\,\lambda_{\chi_V},  \\ 
16\pi^2\,{d\lambda_{\chi_A} \over d t} &=& {5 \over 2}\,\bigl(\lambda_{\chi_V}^2 + \lambda_{\chi_A}^2 \bigr)\lambda_{\chi_A} + {1 \over 24\pi^2}\bigl(\lambda_3^2 + 3 \lambda_2^2 \bigr)\,\lambda_{\chi_A},  \label{eq:lambda5rge}
\end{eqnarray}
\end{widetext}
along with the relevant SM beta functions,
\begin{eqnarray} 
16\pi^2\,{d\lambda_t \over d t} &=& {9 \over 2} \lambda_t^3 - 8\, g_s^2\,\lambda_t \\
 16\pi^2\,{dg_s \over d t} &=& -{31 \over 3} g_s^3.
\end{eqnarray}
Here $t = \ln(\Lambda/Q)$.  $\Lambda$ and $Q$ are the ultraviolet (UV) and infrared (IR) scales, respectively.   We have neglected the contributions from the weak SM gauge sector and the light SM fermions for simplicity.  In the limit where $\lambda_3 = \lambda_{\chi_V} = \lambda_{\chi_A} = \lambda_t = 0$, the RGEs decouple; and RGEs for the Higgs couplings are simple to solve.  The Landau poles in this limit are,
\begin{eqnarray}
\Lambda_{1\,\,\mathrm{Landau}} &=& Q_1 \exp\biggl({2\, \pi^2 \over 3\, \lambda_1(Q_1)} \biggr) \\
\Lambda_{2\,\,\mathrm{Landau}} &=& Q_2 \exp\biggl({8 \,\pi^2 \over 9 \,\lambda_2(Q_2)} \biggr),
\end{eqnarray}
where $Q_1 = 1.36\,m_h$ and $Q_2 = 1.36\,m_\rho$.  Perturbativity will break down well before the Landau scale.  In~\cite{Barbieri:2006dq}, the Barbieri, Hall and Rychkov define a perturbativity scale in which the one-loop corrections to $\lambda_1$ and $\lambda_2$ reaches 30\% of the tree-level value.  Thus, 
\begin{eqnarray}
\Lambda_{p_1} &=& Q_1 \exp\biggl(0.3\,{2\, \pi^2 \over 3\, \lambda_1(Q_1)} \biggr) \\
\Lambda_{p_2} &=& Q_2 \exp\biggl(0.3\,{8 \,\pi^2 \over 9 \,\lambda_2(Q_2)} \biggr).
\end{eqnarray}
Since we require the couplings in the Higgs potential to remain perturbative up to the unitarity bound, our first perturbativity condition is 
\begin{enumerate}
\newcounter{enum_saved3}
\item We require $\Lambda_{p_i}$ to be larger than the unitarity bounds under consideration.  
\setcounter{enum_saved3}{\value{enumi}}
\end{enumerate}
If $\Lambda_{p_i}$ is smaller than the unitarity bounds, then the Higgs sector will become strongly coupled which often signals new physics.  
Requiring $\Lambda_{p_i}$ to be much larger than the unitarity bound requires smaller couplings in the IR.  Thus, the condition above is conservative.
\newline
\begin{figure*}[t]
\centering
{\label{fig:parameterspace7}	
	\includegraphics[width=11truecm,height=7.81truecm,clip=true]{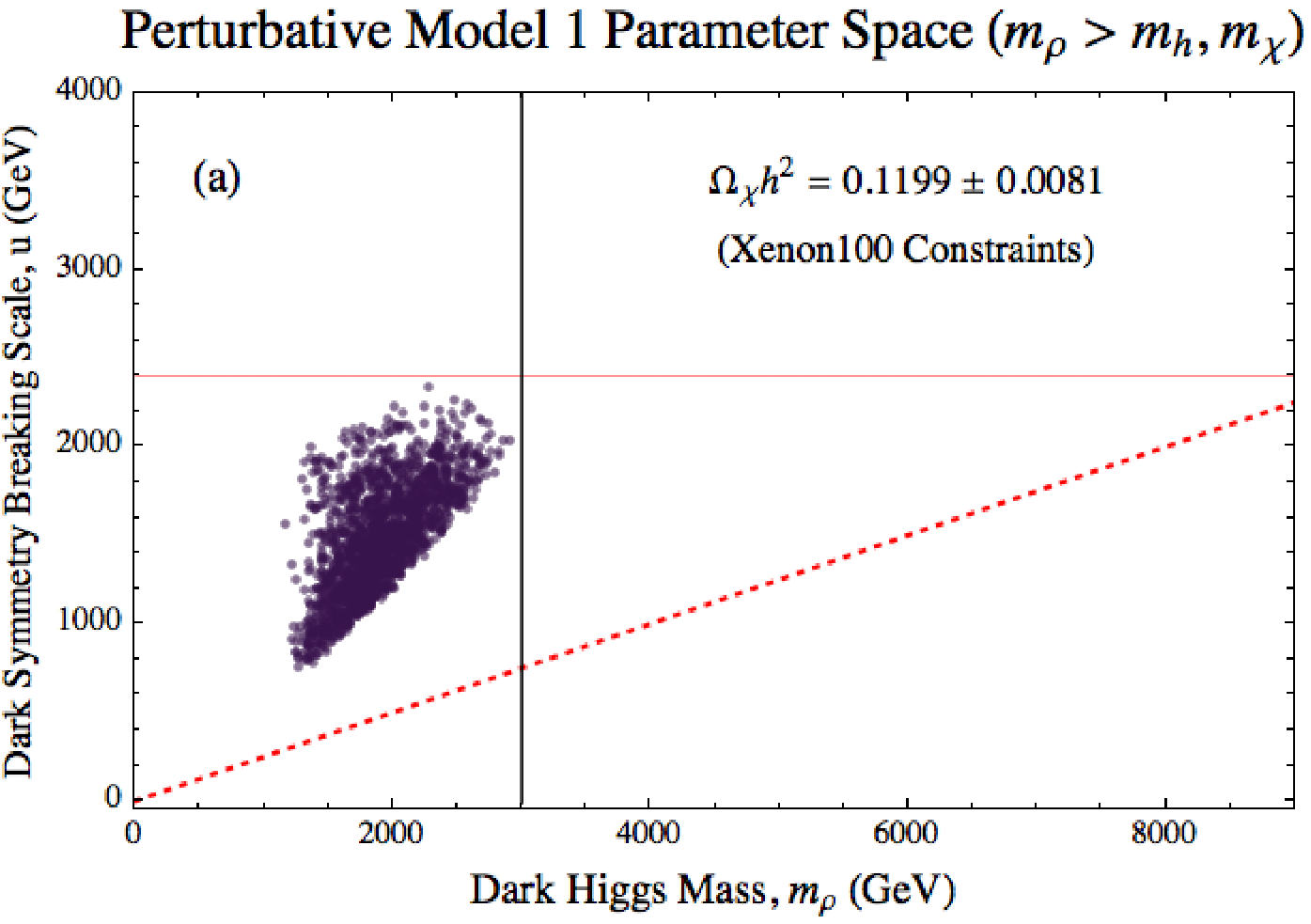} \vspace{1.0cm}}	
	{\includegraphics[width=11truecm,height=7.8truecm,clip=true]{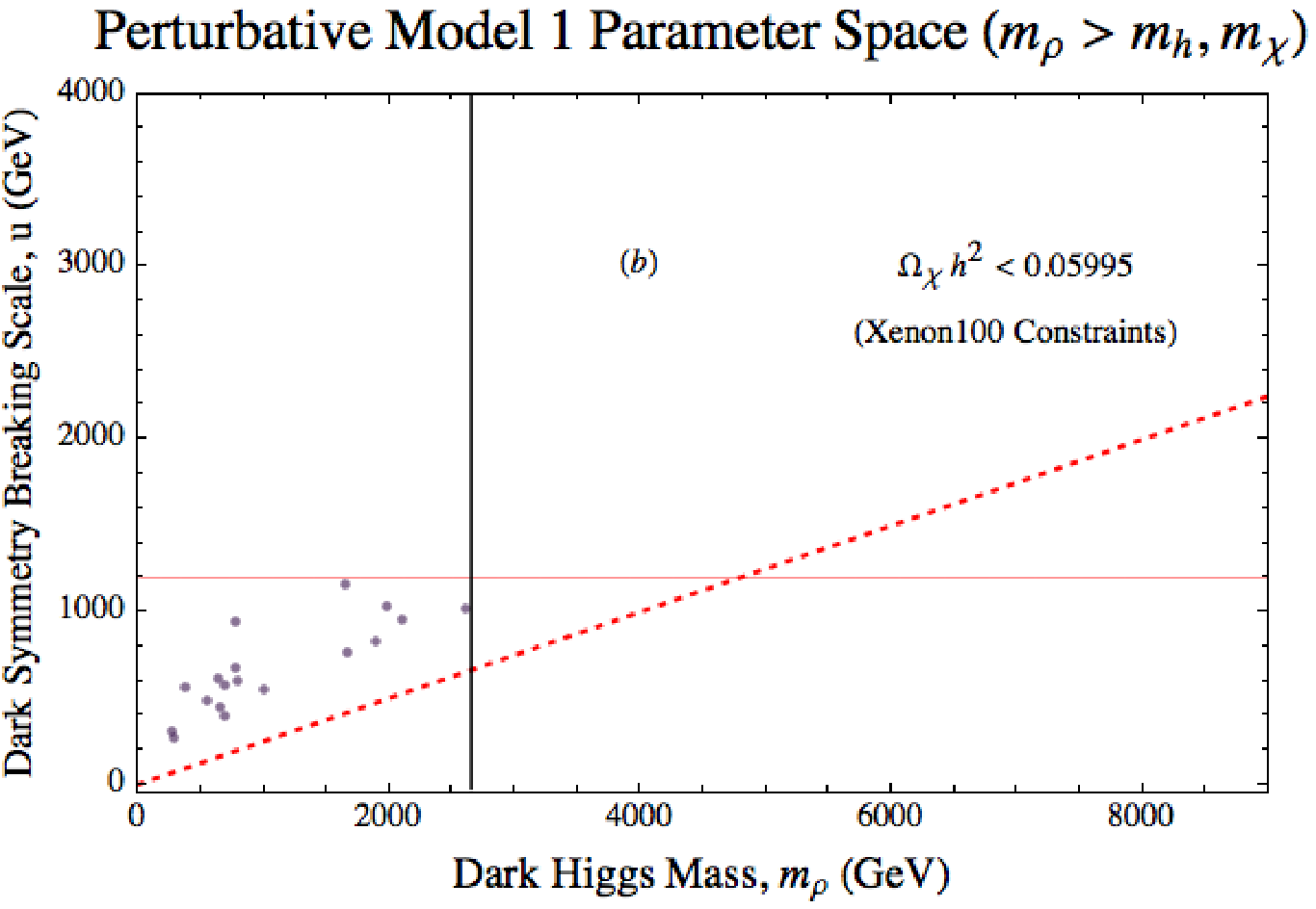}} 
	\caption{Perturbativity constraints applied to Figures (1b) and (2b).  Figure(4a) has the parameter space which satisfies the measured relic abundance.  Figure (4b) has the parameter space which satisfies $\Omega_\chi h^2 < 0.05995$.  The bounds on the various masses improve dramatically.  For comparison, we have kept the red dotted line which indicates the traditional unitarity bound in equation~\leqn{eq:firstunitconstr}.  The slope is consistent with the bound in~\cite{Lee:1977eg}.  The other lines are adjusted in comparison to the unitarity bounds given by Figures (1) and (2).  The sparse points in Figure (4b) are due to the points in parameter space where dark matter mass is tuned so the low-velocity cross section features a small denominator.  The small denominators allow for the couplings and mixing angles to also be relatively small and therefore pass the perturbativity constraints.}
\end{figure*} 
\begin{figure*}[t]
\centering
{\label{fig:parameterspace7}	
	\includegraphics[width=11truecm,height=7.81truecm,clip=true]{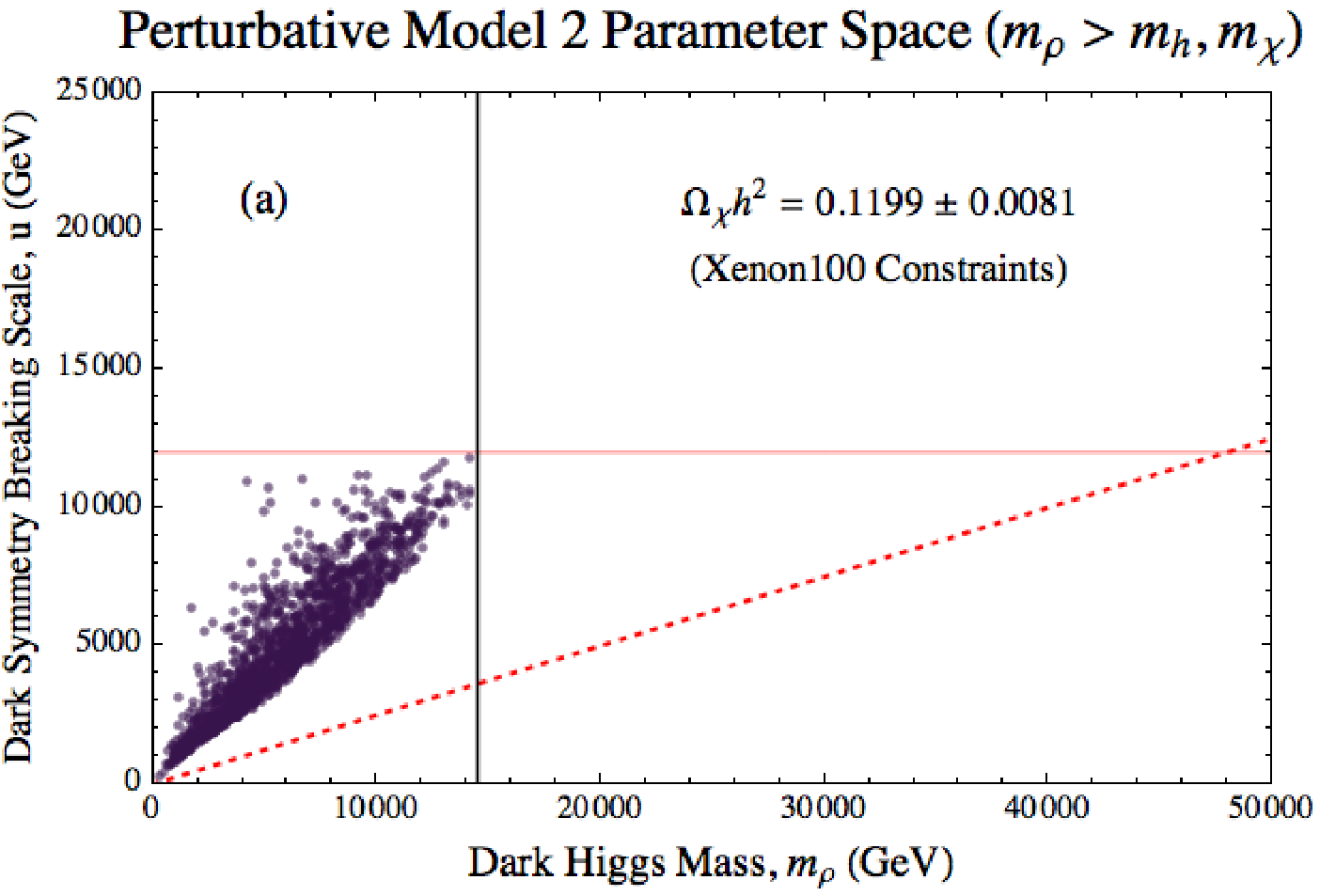} \vspace{1.0cm}}	
	{\includegraphics[width=11truecm,height=7.8truecm,clip=true]{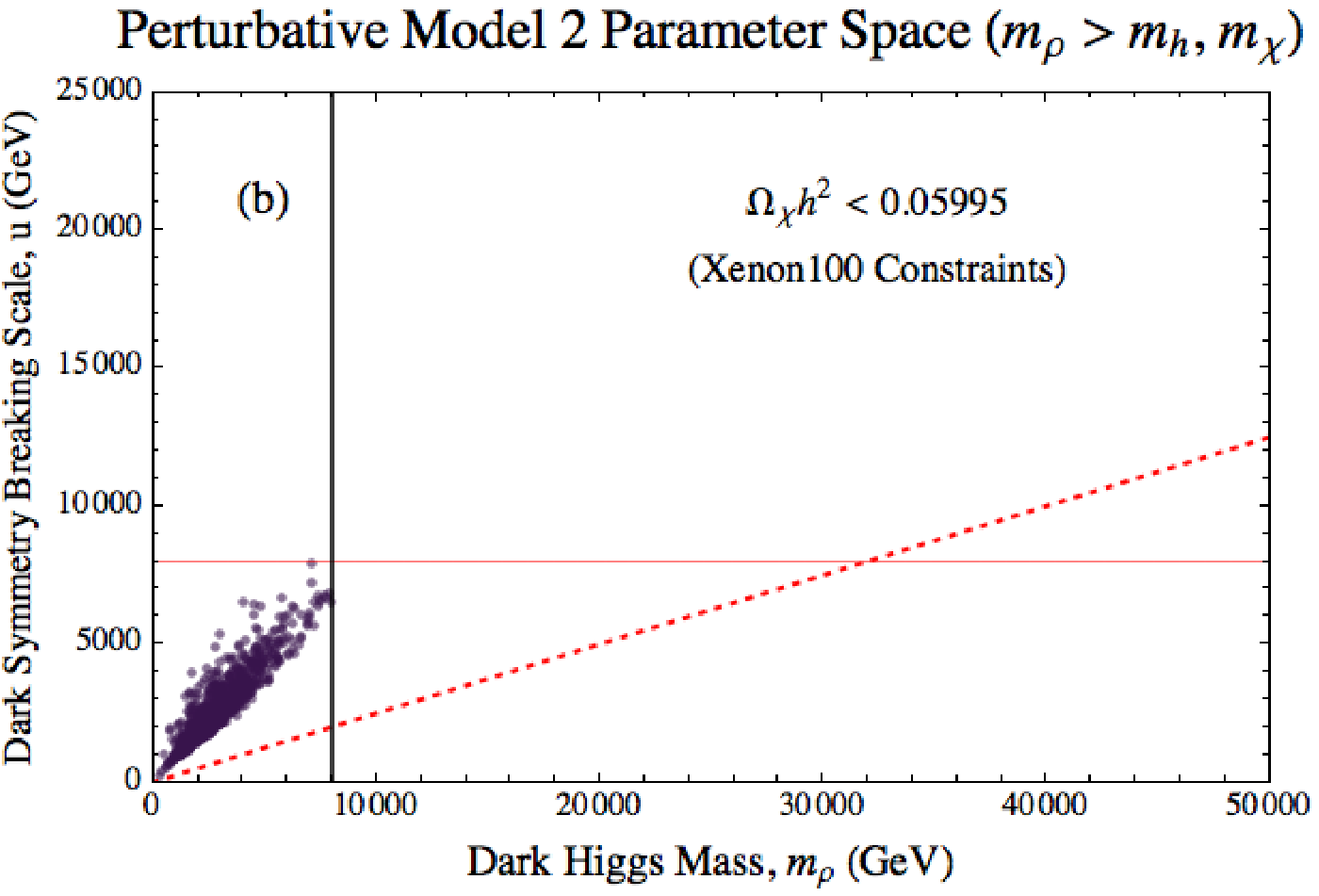}} 
	\caption{Perturbativity constraints applied to Figure (3).  Plot (a) satisfies the measured relic abundance while (b) satisfies $\Omega_\chi h^2 < 0.05995$.  Again, the bounds on the various masses improve dramatically.   For comparison, we have kept the red dotted line which indicates the traditional unitarity bound in equation~\leqn{eq:firstunitconstr}.  The slope is consistent with the bound in~\cite{Lee:1977eg}.  The other lines are adjusted in comparison to the unitarity bounds given in Figure (3).}
\end{figure*} 
\newline
The evolution of the RGEs (equations~\leqn{eq:lambda1rge}-\leqn{eq:lambda5rge}) can force $\lambda_1$ and $\lambda_2$ to be large.  To prevent this Barbieri, Hall and Rychkov~\cite{Barbieri:2006dq} suggest 
the following constraints,
\begin{eqnarray}
50 &\gtrsim&  \biggl|{1 \over 2} \lambda_3^2 + 12\lambda_t^2\,\lambda_1 - 6\lambda_t^4 \biggr| \label{eq:perturbcondtion1} \\ 
40  &\gtrsim& \biggl| 2\, \lambda_3^2 + 4 \lambda_2\,\bigl(\lambda_{\chi_V}^2 + \lambda_{\chi_A}^2 \bigr) \label{eq:perturbcondtion2}\\ 
&-& {1 \over 2}\,\bigl(\lambda_{\chi_V}^2 + \lambda_{\chi_A}^2 \bigr)^2 \biggr|, \nonumber
\end{eqnarray}
should be adopted for our Higgs potential.  Notice that the ``interaction" terms on the RHS of equations~\leqn{eq:lambda1rge} and \leqn{eq:lambda2rge} do not exceed half the value of the $\lambda_1^2$ and $\lambda_2^2$ terms.  Here $\lambda_1 \sim \lambda_2 \sim 2$.  Thus, our second perturbativity condition is simply,
 \begin{enumerate}
\setcounter{enumi}{\value{enum_saved3}}
\item We require equations~\leqn{eq:perturbcondtion1} and \leqn{eq:perturbcondtion2}.
\setcounter{enum_saved3}{\value{enumi}}
\end{enumerate}
Any point in parameter space that fails these conditions is removed.  The results are in Figures 4 and 5.  Additional plots are in the Appendix.  All the points in these plots pass the Xenon100 bounds.  

\subsection{IX.2. Perturbativity Bounds}
\noindent
In this section, we simply reproduce the plots in Section VIII.~but implement the perturbativity conditions above.  For the Model 1, the upper bounds~are 
\begin{eqnarray}
m_\chi, m_\rho &<& 3\,\,\mathrm{TeV} \\
 u &< &2.4\,\,\mathrm{TeV}.
\end{eqnarray}
The parameter space is plotted in Figure (4a) with the Xenon100 constraints.  In the Appendix we place plots of the dark matter mass versus the mass mixing, $\cos\theta$.  There the dark matter masses range from $1.1$ TeV to $2.9$ TeV with $\cos\theta$ ranging from $0.9$ to $0.99$.  
\newline
\newline
Like before, we consider the bounds on the Higgs portal parameter space when the dark matter satisfies only a fraction of the measured relic abundance.  As shown in Figure (4b), the available parameter space Model 1 with the Xenon100 constants is effectively decimated.  What remains in Figure (4b) are the points from the velocity-suppressed s-channel processes.  Recall, we considered the velocity suppressed s-channel terms for Model 1 in our analysis.  We did not write these terms out in equation~\leqn{eq:schannelff}-\leqn{eq:schannelhh}.  The points shown in Figure (4b) have fine-tuned dark matter masses so that the low-velocity scattering amplitudes have small denominators.  Thus, the $\sin^2\theta\,\cos^\theta$ prefactor can be small enough to evade the Xenon100 bounds.   For these remaining points, the upper bound on Model 1 when dark matter satisfies only \textit{half} of the abundance, equation~\leqn{eq:partialrelic}, is
\begin{eqnarray}
m_\chi, m_\rho &<& 2.65\,\,\mathrm{TeV} \\
 u &<& 1.2\,\,\mathrm{TeV}.
\end{eqnarray}
In the Appendix, because of this fine-tuned parameter space, we omit the $m_\chi$ versus~$\cos\theta$ plot.
In the next section, it will be clear the next generation of direct detection and accelerator experiments can exclude this fine-tuned parameter space.
\newline
\newline
Figure (5a) and Figure (5b) shows the perturbative parameter space for Model 2 with the Xenon100 constraints. The upper bounds are now
\begin{eqnarray}
m_\chi, m_\rho &<& 14.5\,\,\mathrm{TeV} \\
 u &< &12\,\,\mathrm{TeV}.
\end{eqnarray}
The upper bounds on Model 2 when dark matter satisfies the abundance $\Omega_\chi h^2 < 0.05995$ is
\begin{eqnarray}
m_\chi, m_\rho &<& 8\,\,\mathrm{TeV} \\
 u &<& 8\,\,\mathrm{TeV}.
\end{eqnarray}
In the next section, we show the reach of the next generation of experiments on these scales.

\section{X.  Experimental Signatures}
\noindent
In this Section, we sketch how to identity the dark Higgs when it is produced directly at accelerators.  We also discuss on how direct detection experiments can constrain the Higgs portal parameter space.  For the accelerator searches, we focus on the LHC14 with 300 fb$^{-1}$ of data, the 500 GeV ILC at 500 fb$^{-1}$ of data and an upgraded ILC at 1 TeV at 1000 fb$^{-1}$.  For the direct detection experiments, we focus on Xenon1T projections.  We note direct detection, with experiments planned for this decade, will almost cover all the parameter space associated with the Higgs portal.  However, accelerator searches are needed to understand whether the dark matter is thermal and annihilates via the Higgs or other portals/mechanisms.  If nothing is observed the dark matter is likely a non-thermal candidate (e.g., axions).  Answering the question of how the dark matter annihilates (if at all) is key to understanding the nature of dark matter.

\subsection{X.1. Direct Detection Signatures}
\noindent
As discussed in Section V.2, direct detection experiments can also constrain the Higgs mass mixing angle.  The proposed Xenon1T experiment proposes to measure the spin-independent dark matter-nucleon cross section to about $10^{-47}$ cm$^2$.  This is an improvement of two orders of magnitude over Xenon100.  We re-plotted previous Figures to determine the reach of Xenon1T.  Only Model~2 without the perturbativity constraints has a parameter space that escapes Xenon1T. See Figure 6.  This raises the distinct possibility that if dark matter has anything to do with the Higgs portal and the dark higgs is the largest scale in the effective theory, then a discovery will be made within the next decade. 
\begin{figure*}[t]
\centering
{\label{fig:parameterspace7}	
	\includegraphics[width=11truecm,height=7.81truecm,clip=true]{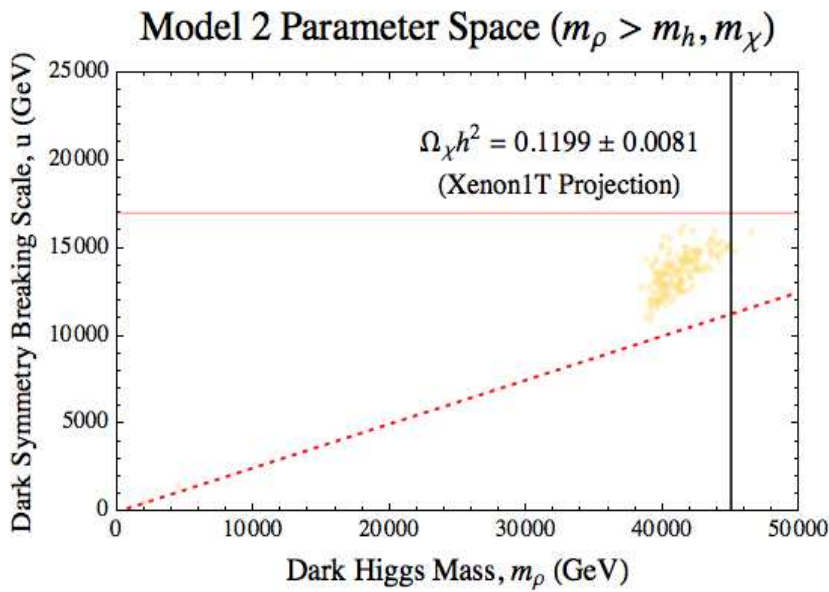} \vspace{1.0cm}}	
	\caption{Constraints on the Higgs portal parameter space of Model 2 from Xenon 1T.  The plots corresponds to Figure (3a) and has the most expansive parameter space out of all the scenarios presented.  Almost all the parameter space is potentially eliminated.  Because of this, we did not include Model 1 plots or plots for with the perturbativity constraints as the parameter space is effectively eliminated.}
\end{figure*} 
As we discuss in the next section, accelerators are needed to search for the dark Higgs and identify the Higgs portal mechanism.  After finding the dark Higgs' with direct detection, the arguments in the previous sections and Section XI. can be recycled to give definitive predictions on the mass and coupling of the dark Higgs.

\subsection{X.2. Accelerator Signatures}
\begin{figure*}[t]
\centering
{\label{fig:parameterspace7}	
	\includegraphics[width=11truecm,height=7.81truecm,clip=true]{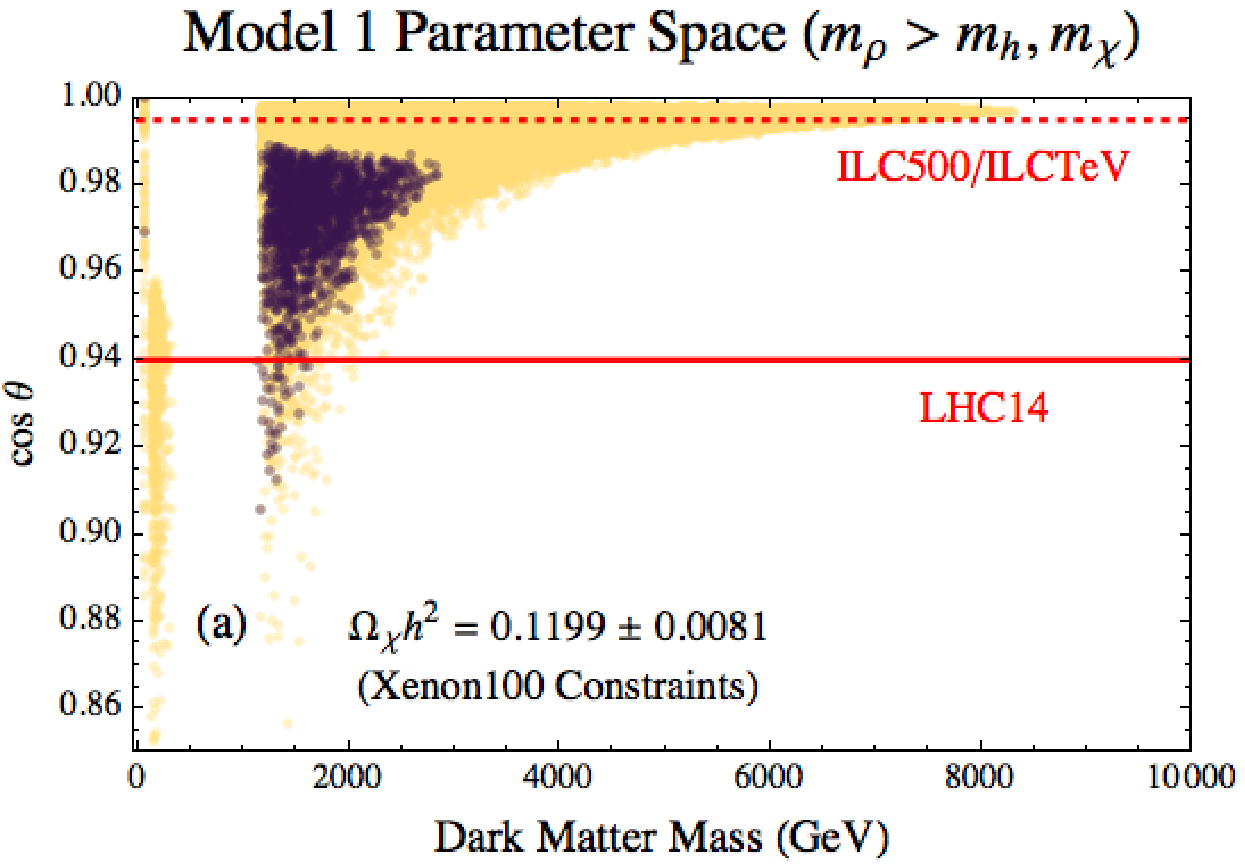} \vspace{1.0cm}}	
	{\includegraphics[width=11truecm,height=7.8truecm,clip=true]{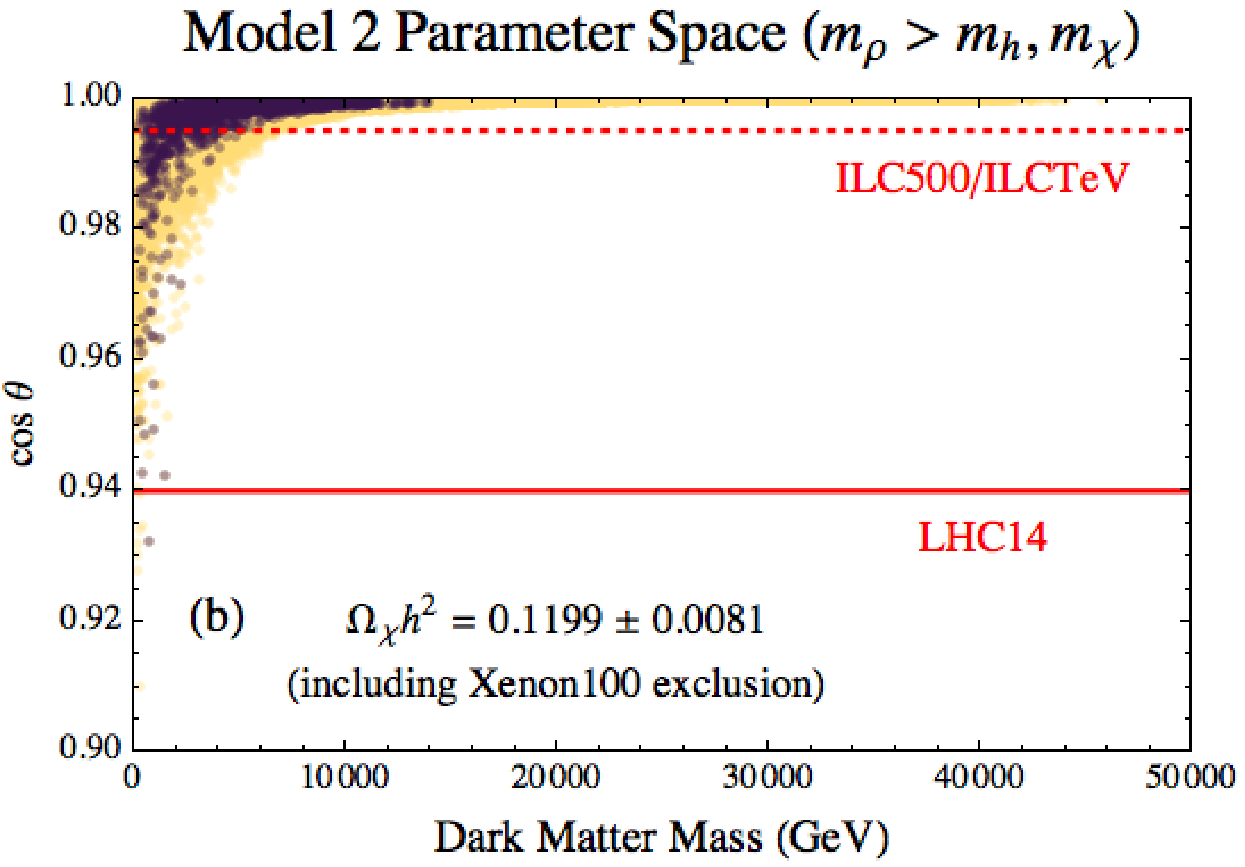}} 
	\caption{Constraints on the Higgs portal parameter space.   The lighter points are the parameter space points that satisfy the unitarity bounds.  The darker points satisfy the unitarity and perturbativity bounds. The parameter space points below the solid horizontal line are excluded by the LHC14 at 300 fb$^{-1}$ of data.  The parameter space points below the dashed horizontal line are excluded by the ILC at 500 fb$^{-1}$ and/or ILCTeV at 1000  fb$^{-1}$. Figure (7a) is for Model 1.  Figure (7b) is for Model 2.}
\end{figure*} 
\noindent
As noted in Section V.2., previous accelerator searches (LEP, Tevatron and the LHC) did not discover any Higgs-like particles.  This implies that if a nontrivial coupling exists between the dark Higgs and the SM, then the dark Higgs is likely to be heavier than the SM Higgs.  Otherwise, the dark Higgs is light and the mass mixing angle is sufficiently small that the dark Higgs evaded detection.  To trigger the unitarity bounds, we have taken the dark Higgs' mass to be the largest scale  in the effective theory.  This implies the mass hierarchy, 
\begin{equation}
m_\rho > m_\chi, m_h.
\end{equation}
Therefore for this mass hierarchy, a strategy for searching for the dark Higgs with accelerators includes 
\begin{enumerate}
\item Invisible dark Higgs decays.
\item Bump hunting on an invariant mass spectrum.        
\item Precisely measuring the Higgs mass mixing angle via the SM Higgs' coupling to SM particles.
\end{enumerate}
This last point indirectly constrains Higgs portal dark matter annihilation.  We did not emphasize invisible SM Higgs decays with Item 2.  This is possible if $m_h > m_\chi$ and only if the dark matter decays to the SM through loop-suppressed processes.  In this work, we focused on tree-level annihilation processes and therefore only consider invisible dark Higgs decays.  We do discuss invisible SM Higgs decays for other mass hierarchies in Section X.

 \subsubsection{X.2.1.  (Invisible) Dark Higgs Decays}
 \label{sec:invHiggs}
 \noindent
A dramatic signature for dark matter at accelerators is the observation of large amounts of missing energy produced from the decay of a dark Higgs.  The dark Higgs has a sizable invisible branching fraction.  
To see this, at leading order the decay width is,
\begin{equation}
\Gamma_{\rho \to \chi \chi} =  {m_\rho \over 4 \pi} \,{\lambda_\chi^2 \cos^2\theta \over 2}\,\biggl(1 - {4\,m_\chi^2 \over m_\rho^2} \biggr)^{3/2}, \label{eq:darkHiggsinvdecay}
\end{equation}
where we have assume the dark matter is fermonic.  Now, consider dark Higgs decays to SM particles.  The contribution to the total width is
\begin{eqnarray}
\Gamma_\mathrm{SM} &=& \sum_q \Gamma_{\overline{q}q} + \sum_l \Gamma_{\overline{l}l}  + \Gamma_{gg} +  \Gamma_{\gamma\gamma} + \Gamma_{Z\gamma} \nonumber\\
&+& \Gamma_{WW}  +  \Gamma_{ZZ}. \nonumber
\end{eqnarray}
These widths are proportional to 
\begin{eqnarray}
\Gamma_{\overline{f}f} &\sim&  m_\rho\,m_f^2\,G_F\,\sin^2\theta \\
\Gamma_{VV} \sim \Gamma_{gg}  &\sim& m_\rho^3\,G_F \,\sin^2\theta.
\end{eqnarray}
\newline
It is clear out of all the SM fermions the dark Higgs will dominantly decay into top quarks. 
However, because 
\begin{equation}
\lambda_\chi^2\cos^2\theta \gtrsim m_\rho^2 \,\sin^2\theta/ 4\,v^2 \gtrsim 3 \lambda_t^2\,\sin^2\theta, 
\end{equation}
holds for almost all the parameter space, then
\begin{equation}
\Gamma_{\rho \to \chi \chi} \gtrsim \Gamma_{\rho \to VV} \gtrsim \Gamma_{\rho \to \overline{t}t}
\end{equation}
and the dark Higgs often decays invisibly.  Because of the $m_\rho^3$ enhancement, the dark Higgs can appreciably decay into easily tagged electroweak gauge bosons.  In this channel, reconstruction of the dark straightforward with a sufficient number of events.
\newline
\newline
\subsubsection{X.2.2.  Invariant Mass Spectrum} 
\noindent
The dark and SM Higgses have similar production and decay channels.  Additionally, the experimentalist can ``bump hunt" on a suitable invariant mass distribution to isolate the dark Higgs cross section.  Important final states include $WW$, $ZZ$, $\gamma\,\gamma$, $\overline{t} t$ as well as $Z h$ for larger dark Higgs mass.  However, the height of the dark Higgs bump will be down by a factor of $\sin^2\theta$ which will likely make the search difficult for hadronic colliders.  To see this at the LHC, the production %
cross section is, 
\begin{widetext}
\begin{eqnarray}
\Gamma(\phi_i \to g g) &=& {\alpha_s^2\, g^2 \,m_{\phi_i}^3 \over 64 \,\pi^3\,m_W^2}\,\beta_i^2 \,\biggl| \sum_i {4 m_f^2 \over m_{\phi_i}^2} \biggl(1 + \biggl(1 - {4 m_f^2 \over m_{\phi_i}^2}\biggr)F\biggl[{4 m_f^2 \over m_{\phi_i}^2}\biggr]\biggr) \biggr| \\  
&=& \beta_i^2 \, \Gamma_0(\phi_i \to g g) \nonumber  \\  \nonumber  \\
{d\sigma \over dy}(p p \to \phi_i + X) &=& {\pi^2 \over8 m_{\phi_i}^3}\,g_A(x_A,m_{\phi_i}^2)\,g_B(x_B,m_{\phi_i}^2) \,\Gamma(\phi_i \to g g), \label{eq:Higgsprodcution}\\ \nonumber 
\end{eqnarray}
\end{widetext}
where $\sigma_0$ is defined as equation~\leqn{eq:Higgsprodcution} with $\Gamma \to \Gamma_0$.  This equation modified from~\cite{Gunion:1989we}. The notation is defined therein.
To keep the equations concise, we defined $\phi_i = (h, \rho)$ and $\beta_i^2 = (\cos^2\theta, \sin^2\theta)$, respectively.  Here we focused on dark Higgs production from initial state gluons.  $WW$ fusion,
\begin{equation}
p\,p \to W\,W \to \rho,
\end{equation}
is a dominant production mechanism for heavy dark Higgses whose production amplitude also has a prefactor of $\sin^2\theta$.
\newline
\newline
Because the production cross section for SM Higgs and dark Higgs are multiplied by a prefactor of %
%
$\cos^2\theta$ and $\sin^2\theta$, respectively, the dark and SM Higgses should satisfy 
\begin{eqnarray}
1 = \sum_i \Biggl({ \sigma(p\,p \to \phi_i)_\mathrm{measured} \over \sigma_0(p\,p \to \phi_i) } \Biggr).
\end{eqnarray}
Also, models with more complicated dark Higgs and SM Higgs mixings should satisfy a similar relation.  Satisfying this relation (or the equivalent) and the production of large amounts of missing energy via dark Higgs decays gives strong evidence of a Higgs portal. 

\subsubsection{IX.2.3 Higgs Coupling Strength}
\noindent
Precise measurement of the Higgs couplings for, 
\begin{align}
h \to W\,W &&  h \to Z\,Z, \label{eq:Higgscouplingstrength}
\end{align}
probes the mass mixing angle, $\cos\theta$.  See, for example, the coupling in equation~\leqn{eq:WWh}.  \cite{Peskin:2012we} describes how the LHC14 for 300 fb$^{-1}$ of data can measure the couplings in equation~\leqn{eq:Higgscouplingstrength} to around the 7-8\% level.  The 500 GeV ILC for 500 fb$^{-1}$ of data can measure these couplings to an accuracy of 0.5\%.  Finally, \cite{Peskin:2012we} shows the 1 TeV ILC at 1000 fb$^{-1}$ can measure this SM Higgs coupling well below the 0.3\% level.  In Figure 7, we re-plot the figures in the Appendix and overlay the potential constraints.  Notably, the ILC will be able to fully probe the Higgs portal parameter space leftover after perturbativity is required of the Higgs sector.

\section{XI.  Additional Matter Hierarchies}
\noindent
We have focused the mass hierarchy where dark Higgs is the largest scale in the effective theory,
\begin{equation}
m_\rho > m_\chi, m_h. \label{eq:origmasshierarchy}
\end{equation}
We have showed the effective theory generates the bounds on the dark Higgs mass in this limit.  Other mass hiercharcies are also possible,
\begin{eqnarray}
m_h >  m_\chi, \,m_\rho, \label{eq:mhmasshierarchy}\\
m_\chi > m_h, \,m_\rho. \label{eq:mchimasshierarchy}
\end{eqnarray}  
In this section, we outline how one can implement bounds on these hierarchies.

\subsection{XI.1. Bounds on the $m_h  > m_\chi, \,m_\rho$ Hierarchy}
\noindent
In this scenario, the dark Higgs is lighter than the SM Higgs.   Because of  LHC, Tevatron and LEP bounds on light SM Higgses, the dark Higgs must have a small mass mixing angle.  The small mixing angle ensures the dark Higgs production cross section is sufficiently small to evade the accelerator bounds.   For dark matter that annihilates into the dark Higgs, the low velocity cross section goes 
\begin{equation}
\langle \sigma v \rangle \sim {\cos^4\theta \over m_\chi^2}.
\end{equation}
In the limit where $\theta \to 0$, the low velocity cross section does not vanish.  Thus, the arguments made for the hierarchy in equation~\leqn{eq:origmasshierarchy} do not apply.  In this case, new constraints on the mass mixing angle are needed.  If the LHC or ILC determines the higgs coupling strength is nontrivially smaller than the SM expectation (see Section IX.2.3.), then an upper bound on $\cos\theta$ can be made.  The arguments made in the previous section can proceed as normal.  %
If the dark matter is lighter than the dark Higgs then the dark matter can annihilate to the SM via effective operators generated from the integrated out SM Higgs. For dark matter annihilating into SM fermions, $f$, the low velocity cross section goes as,
\begin{equation}
\langle \sigma v \rangle \sim {\sin^2\theta\cos^2\theta \over m_\chi^2}\,{m_f^4 \over m_h^4}.
\end{equation}
Likely the dark matter coupling to the dark Higgs is  very strong in order to generate the right relic abundance.  Should this be a viable scenario, the arguments from the previous sections would proceed as normal.
\newline
\newline
In the limit where $m_\chi  >m_\rho$ and the mixing angle is vanishingly small, the best bounds on this scenario are from observational and experimental constraints.  If the dark Higgs is long-lived, then for a given epoch in the universe's history, this mediator behaves like dark matter.  Thus, the same constraints for decaying dark matter apply to the mediator.  For example, decaying dark matter is severely constrained from decaying during the CMB epoch.  The injection of entropy and ionized particles produces strong constraints~\cite{Finkbeiner:2011dx}.  Moreover, if the production cross section is large enough to allow for production at colliders, the dark Higgs could generate easily tagged signatures such as displaced vertices.  Importantly, the SM higgs can decay invisibly.  The ILC would provide a strong platform for searching for this physics.  See~\cite{Peskin:2012we} and the projections for measuring the invisible SM Higgs width.  

\subsection{XI.2. Bounds on the $m_\chi  > m_h, \,m_\rho$ Hierarchy}
\noindent
In this scenario the dark matter can annihilate to both the SM and dark sectors, 
\begin{eqnarray}
\chi + \chi &\to& \rho + \rho \label{eq:DMDM-rhorho} \\
\chi + \chi &\to& h + h,  \label{eq:DMDM-hh}
\end{eqnarray}
as well as,
\begin{align}
\chi + \chi \to \overline{f} + f && \chi + \chi \to V + V.  \label{eq:DMDM-SMSM}
\end{align}
Here $f$ and $V$ are the SM fermions and gauge bosons, respectively.  The low-velocity annihilation cross sections go as 
\begin{align}
\langle \sigma v \rangle \sim {\sin^4\theta \over m_\chi^2} &&\mathrm{or}&&  \langle \sigma v \rangle \sim {\sin^2\theta\cos^2\theta \over m_\chi^2}.
\end{align}
For the processes in equations~\leqn{eq:DMDM-hh} and~\leqn{eq:DMDM-SMSM}.  The process in equation~\leqn{eq:DMDM-rhorho} goes as 
\begin{equation}
\langle \sigma v \rangle \sim {\cos^4\theta \over m_\chi^2}.
\end{equation}
In the limit of $\theta \to 0$, a bound is not generated as the relic abundance does not vanish.  In this limit, the dark matter annihilates into the hidden sector.  Please note:  The unitarity bound in Section VII.2.~is still valid and should provide a much stronger limit on the dark matter mass than the bounds generated in~\cite{Griest:1989wd}.  
\newline
\newline
If LHC or ILC measures the SM Higgs coupling strength to be smaller than the expected SM value (see Section IX.2.3.), then a bound on $\cos\theta$ can be generated.  Given this bound, the arguments in the previous sections can be adapted.  As an example, suppose the LHC14 measures the mass mixing angle to be,
\begin{equation}
\cos\theta_\mathrm{measured} \sim 0.8 \pm 0.01, 
\end{equation}
we can raise the dark Higgs mass to be much larger than the SM Higgs mass while maintaining $m_\chi > m_\rho$.  Consider $WW$ scattering.  The tree-level scattering amplitudes below the dark Higgs mass go as
\begin{eqnarray}
\mathcal{M}_\mathrm{gauge} &=& {g^2 \over 4\,m_W^2}(s + t) \label{eq:gaugecontrib2} \\
\mathcal{M}_\mathrm{SM\,\,Higgs} &=& -{g^2 \over 4\,m_W^2}(s + t)\, \cos^2\theta.  \label{eq:smhcontrib2}
\end{eqnarray}
Because of the incomplete cancellation between the two terms, the dark Higgs mass must be
\begin{equation}
m_\rho \lesssim 1.6\,\,\mathrm{TeV}.
\end{equation}
Here we used equation~\leqn{eq:unitconstraint}.  The relic abundance and direct detection constraints would shape the available parameter space for this scenario and improve this bound.  

\section{XII. Conclusions}
\noindent
In this paper, we considered unitarity constraints on models where the dark matter annihilates via the Higgs portal.  Higgs portal models feature a mediating (or ``dark") Higgs that mixes with the SM Higgs in order to facilitate dark matter annihilations.  This mixing upsets electroweak unitarity constraints by forcing SM Higgs amplitude to incompletely cancel the pure gauge contribution. 
We exploited this fact to place an upper bound on the dark Higgs and dark matter masses.  We also placed constraints on the dark symmetry breaking vev and bounded the basic parameter space.  We considered two basic models.  The dark matter in Model~1 annihilates via femion exchange.  Model~2 has dark matter that annihilates via both femion and Higgs boson exchange.  We find the upper bound on the dark Higgs and dark matter masses to be 
\begin{align}
\mathrm{\textbf{Model\,\,1:}}  &&  &m_\rho < 8.5\,\,\mathrm{TeV}\,\,\,\,\,\,  \mathrm{(unitarity)} \nonumber \\
\mathrm{\textbf{Model\,\,2:}}  &&  &m_\rho < 45.5\,\,\mathrm{TeV}.  \nonumber
\nonumber
\end{align}
The bounds on the dark matter masses are roughly the same as the bounds on the dark Higgs.  For each Model, the bound on the dark symmetry breaking vev is roughly a factor of $1/2.75$ less than the bound on the dark Higgs mass.  The difference in bounds for Model~1 and Model~2 can be traced back to the mixing angle.  The unsuppressed annihilation channel for Model~1 depends sensitively on the mixing and requires a larger mixing angle to get the right relic abundance.  Model~2 has more annihilation channels and depends less sensitively on the mixing.  A smaller mixing angle is needed to get the right relic abundance and a larger unitarity bound is generated.
\newline
\newline
We also considered scenarios where the dark matter is only one component of the overall measured relic abundance.  We showed the bounds on the dark matter, dark Higgs and dark symmetry breaking vevs drastically improved.  A smaller relic abundance implies more dark matter annihilations and therefore a larger mass mixing angle.  With a larger mixing angle, stronger bounds on the dark Higgs and dark matter masses as well as the dark symmetry breaking vevs are generated.
We also improved our unitarity bounds by requiring a perturbativity in the Higgs sector up to the scales in which unitarity breaks down.  We find 
\begin{align}
\mathrm{\textbf{Model\,\,1:}}  &&  &m_\rho < 3\,\,\mathrm{TeV}\,\,\,\,\,\, \,\,\,\, \mathrm{(pertubativity)}  \nonumber\\
\mathrm{\textbf{Model\,\,2:}}  &&  &m_\rho < 14.5\,\,\mathrm{TeV}.  \nonumber
\nonumber
\end{align}
The dark symmetry breaking vevs in this limit are the same order of the dark Higgs masses.  The dark matter has similar bounds as the dark Higgses.  Also, when requiring the Higgs portal dark matter to satisfy only a fraction of the measured relic abundance, these bounds become even stronger. 
\newline
\newline
We considered the impact of current and planned experiments on these bounds.  Notably, proposed direct detection experiments, such as Xenon1T, can cover almost all the Higgs portal parameter space.  Since these direct detection experiments are scheduled to run by the end of this decade, this raises the likely possibility that a signature will be discovered in the next decade if the dark matter annihilates via the Higgs portal.  To verify that the Higgs portal is responsible, accelerator experiments such as the LHC and/or ILC are needed.  Given the discovery of dark matter, the arguments made within can be used to generate a prediction for the dark Higgs mass and symmetry breaking vev assuming Higgs portal dark matter.
\newline
\newline
In all, the Higgs portal is an interesting mechanism to facilitate dark matter annihilation.  By providing bounds on generic models the goal is to determine a new scale of physics.  This new scale of physics also translates to bounds in the parameter space that future experiments can probe.  Historically, new scales of physics are used as scaffolding in order to design new models of physics.  With the increasing sensitivity of data expected with in the next decade, a very vibrant period for understanding dark matter is upon us.

\vskip 0.2cm
\noindent
{\it Acknowledgments:}
We thank J.~Berger, H.~Davoudiasl, H.-C.~Fang, J.~Hewett,   A.~Ismail, M.~Peskin, T.~Rizzo, W.~Shepard, T.~Tait and J.~Virzi for useful discussions.  Special thanks to J.~Virzi for early motivation.  Special thanks also goes to M.~Peskin for critical readings of the draft.  This work was supported by the US National Science Foundation, grant NSF-PHY-0705682, the LHC Theory Initiative and a grant from the US National Academies of Science.  SLAC is operated by Stanford University for the US Department of Energy under contract DE-AC02-76SF00515.

\appendix
\section{The Appendix} 

\section{A. Expanded Higgs Potential}
\noindent
The Higgs potential before going to unitarity gauge is given by equation~\leqn{eq:Higgspotentialext}.  After mixing the neutral Higgses, the potential is now
\begin{widetext}
\begin{eqnarray}
V &=& {1 \over 2}\,m_h^2\,h^2  + v\,\lambda_{hw^+ w^-}h \,w^+ w^- + {1 \over 6}\,v\,\lambda_{h^3} h^3 + {1 \over 2}\,v\,\lambda_{hz^2}\,h \,z^2 + {1 \over 4}\,\lambda_{1}\,(w^+ w^-)^2   \\
&+& {1 \over 24}\,\lambda_{h^4}\,h^4 + {1 \over 24}\,\lambda_{1}\,z^4 +  {1 \over 2} \,\lambda_{h^2w^+w^-}\,h^2\,w^+w^- + {1 \over 4} \, \lambda_{h^2z^2}\,h^2\,z^2 + {1 \over 2} \,\lambda_{z^2w^+w^-}\,z^2\,w^+w^- \nonumber \\
&+& {1 \over 2} \, m_\rho^2\,\rho^2 + {1 \over 6}\,u \, \lambda_{\rho^3}\,\rho^3 + {1 \over 24} \, \lambda_{\rho^4}\rho^4 + {1 \over 2} \, \lambda_{h\rho^2}\,v\,h\,\rho^2 + u\,\lambda_{\rho w^+ w^-}\rho \,w^+ w^-  \nonumber \\
&+& {1 \over 2}\, v\,\lambda_{\rho h^2}\rho \,h^2  + {1 \over 2}\, v\,\lambda_{\rho h^2}\rho \,z^2   + {1 \over 2} \,\lambda_{\rho^2w^+w^-} \rho^2 \,w^+ w^- +  {1 \over 4} \,\lambda_{\rho^2h^2} \,\rho^2\,h^2 +  {1 \over 4} \,\lambda_{\rho^2z^2} \,\rho^2\,z^2 \nonumber  
\end{eqnarray}
\end{widetext}
where the couplings are 
\begin{eqnarray} 
\lambda_{hw^+ w^-} &=& 2 \lambda_1 \cos\theta- \lambda_3 u \sin\theta/v \\
\lambda_{h^3} &=&  6\,\bigl(-\lambda_2 u \sin\theta^3 + \lambda_1 v \cos\theta^3  \label{eq:lambdah3}\\ 
&-&\lambda_3 u \sin\theta \cos\theta^2/2  \nonumber \\
&+&\lambda_3 v \sin\theta^2 \cos\theta/2\bigr) \nonumber \\
\lambda_{hz^2} &=& 2 \lambda_1 v \cos\theta-\lambda_3 u \sin\theta \\
\lambda_{h^4} &=& \lambda_1 \cos\theta^4+\lambda_2 \sin\theta^4  \\
&+&\lambda_3 \sin\theta^2 \cos\theta^2  \nonumber \\
\lambda_{h^2 w^+ w^-} &=& 2 \lambda_1 \cos\theta^2+\lambda_3 \sin\theta^2 \\
\lambda_{h^2 z^2} &=& 2 \lambda_1 \cos\theta^2+\lambda_3 \sin\theta^2  \\
\lambda_{z^2w^+w^-} &=& 2 \lambda_1 \\
\lambda_{\rho^3} &=& 6 \bigl(\lambda_2 \cos\theta^3+\lambda_3 \sin\theta^2 \cos\theta/2 \,\,\,\,\,\,\, \\
&+& \lambda_1 v \sin\theta^3/(2 u) \nonumber \\
&+&\lambda_3 v \sin\theta \cos\theta^2/(2 u)\bigr) \nonumber \\
\lambda_{\rho^4} &=& 6 \bigl(\lambda_1 \sin\theta^4+\lambda_2 \cos\theta^4 \\
&+&\lambda_3 \sin\theta^2 \cos\theta^2 \bigr) \nonumber\\
\lambda_{h\rho^2} &=& -2 \sin\theta^2 \cos\theta (\lambda_3-3 \lambda_1) \\
&+& \lambda_3 \cos\theta^3 \nonumber \\
&+&2 u \sin\theta \cos\theta^2 (\lambda_3-3 \lambda_2)/v \nonumber \\
&-& \lambda_3 u \sin\theta^3/v \nonumber \\
\lambda_{\rho h^2} &=& 2 u \sin^2\theta \cos\theta (3 \lambda_2-\lambda_3) \label{eq:hhp}\\
&+&  \lambda_3 u \cos^3\theta \nonumber \\
&+& v \sin\theta \cos^2\theta (6 \lambda_1-2 \lambda_3) \nonumber \\
&+& \lambda_3 v \sin^3\theta. \nonumber
\end{eqnarray}

\section{B.  High-Energy Scattering Diagrams}
\noindent
Here we use the Goldstone boson equivalence theorem to compute all the high-energy Goldstone-Goldstone, Goldstone-Higgs and Higgs-Higgs scattering diagrams.  We define the following couplings,
\begin{eqnarray}
\kappa_1 &=& 2 \lambda_1\, v \,\cos\theta - \lambda_3 \,u\, \sin\theta \\
\kappa_2 &=& 2 \lambda_1\, v\, \sin\theta + \lambda_3\, u\, \cos\theta \\
\kappa_3 &=& \bigl(3\, \kappa_1\,\bigl(\lambda_3 \sin\theta\cos\theta(v \sin\theta- u \cos\theta)  \\
&+&2 \lambda_1 v \cos\theta^3 -2 \lambda_2 \,u \sin^3\theta\bigr)\bigr)^{1/2} \nonumber \\
\kappa_4 &=& \bigl(\kappa_2\,\bigl(2\, u \sin^2\theta \cos\theta (3 \lambda_2-\lambda_3)  \\
&+& \lambda_3 \,u \cos^3\theta +2 \,v \sin\theta \cos^2\theta (3 \lambda_1-\lambda_3)  \nonumber \\
&+& \lambda_3\, v \sin^3\theta\bigr)\bigr)^{1/2} \nonumber \\
\kappa_5 &=&\kappa_3 \bigl(u \to -u, \cos\theta \to \sin\theta, \sin\theta \to \cos\theta\bigr)\, \,\,\,\\
\kappa_6 &=&\kappa_4(u \to -u, \cos\theta \to \sin\theta, \sin\theta \to \cos\theta)\,\,\,\, \\ 
\kappa_7 &=& \bigl((2 \lambda_1\, v \,\cos\theta - \lambda_3 \,u\, \sin\theta)\\
& &(2 \lambda_1\, v\, \sin\theta+\lambda_3\, u\, \cos\theta)\bigr)^{1/2} \nonumber \\
\kappa_8 &=& \bigl(\kappa_1\,\bigl(2\, u \sin^2\theta \cos\theta (3 \lambda_2-\lambda_3)  \\
&+& \lambda_3 \,u \cos^3\theta +2 \,v \sin\theta \cos^2\theta (3 \lambda_1-\lambda_3)  \nonumber \\
&+& \lambda_3\, v \sin^3\theta\bigr)\bigr)^{1/2} \nonumber \\ 
\kappa_9 &=& \bigl( 2 \lambda_3 \sin\theta \cos^3\theta \left(\lambda_3\, u^2+\lambda_1 v^2\right)\\
&-& 2\, u\, v\, \sin^2\theta \cos^2\theta \left(6 \lambda_1 \lambda_2-5 \lambda_1 \lambda_3+\lambda_3^2\right)\nonumber  \\
&-& 2 \lambda_1 \lambda_3 \,u \,v \,\sin^4\theta+\lambda_3^2\, u\, v\, \cos^4\theta \nonumber \\
&-& \cos\theta \bigl(\lambda_3 \sin^3\theta \left(u^2 (6 \lambda_2+\lambda_3)+4 \lambda_1\, v^2\right)  \nonumber \\
&-& 4 \lambda_1^2 \,v^2 \,\sin\theta\bigr) \bigr)^{1/2} \nonumber \\
\kappa_{10} &=& 3\, \bigl( \lambda_3 \sin\theta\cos\theta (v \sin\theta-u \cos\theta)\\
&+& 2 \lambda_1 v \cos^3\theta - 2 \lambda_2\, u\, \sin^3\theta\bigr) \nonumber \\
\kappa_{11} &=& 2 u \sin^2\theta \cos\theta (3 \lambda_2-\lambda_3)+\lambda_3 u \cos^3\theta\\
&+& 2 v \sin\theta \cos^2\theta (3 \lambda_1-\lambda_3)+\lambda_3 v \sin^3\theta \nonumber\\
\kappa_{12}  &=& 2 u \sin\theta \cos^2\theta (\lambda_3-3 \lambda_2)-\lambda_3 u \sin^3\theta\\
&+& 2 v \sin^2\theta \cos\theta (3 \lambda_1-\lambda_3)+\lambda_3 v \cos^3\theta\nonumber \\
\kappa_{13} &=& 3 \,\bigl(\lambda_3 \sin\theta \cos\theta (u \sin\theta + v \cos\theta) \\
&+& 2 \lambda_1 v \sin^3\theta + 2 \lambda_2 u \cos^3\theta\bigr). \nonumber  
\end{eqnarray}
All the amplitudes below recover the results in~\cite{Lee:1977eg} in the decoupling limit.  

\subsection{B.1.  Longitudinal Gauge Boson Scattering Amplitudes}
\noindent
\begin{eqnarray}
\mathcal{M}_{WW\to WW} &=& 4 \lambda_1 \hspace{4.1cm}\\
&+& \kappa_1^2\, \biggl({1 \over s - m_h^2} + {1 \over t - m_h^2} \biggr) \nonumber \\
&+& \kappa_2^2 \,\biggl({1 \over s - m_\rho^2} + {1 \over t - m_\rho^2} \biggr), \nonumber 
\end{eqnarray}
\begin{eqnarray}
\mathcal{M}_{WW\to ZZ} &=& 2 \lambda_1 \\
&+& \kappa_1^2\, \biggl({1 \over s - m_h^2} \biggr) + \kappa_2^2\, \biggl({1 \over s - m_\rho^2} \biggr), \nonumber
\end{eqnarray}
\begin{eqnarray}
\mathcal{M}_{ZZ\to ZZ} &=& 6 \lambda_1 \\
&+& \kappa_1^2\, \biggl({1 \over s - m_h^2} + {1 \over t - m_h^2} + {1 \over u - m_h^2}\biggr) \nonumber \\
&+& \kappa_2^2\, \biggl({1 \over s - m_\rho^2} + {1 \over t - m_\rho^2} + {1 \over u - m_h^2} \biggr).\nonumber 
\end{eqnarray}

\subsection{B.2. Longitudinal Gauge and Higgs Scattering Amplitudes}
\noindent
\begin{eqnarray}
\mathcal{M}_{hh\to WW} &=&  2 \lambda_1\cos^2\theta + \lambda_3 \sin^2\theta \\
&+& \kappa_1^2\, \biggl( {1 \over t - m_W^2} + {1 \over u - m_W^2}\biggr)\nonumber \\
&+& \kappa_3^2 \,\biggl({1 \over s - m_h^2} + {1 \over t - m_h^2} + {1 \over u - m_h^2} \biggr) \nonumber \\
&+& \kappa_4^2 \,\biggl({1 \over s - m_\rho^2} + {1 \over t - m_\rho^2} + {1 \over u - m_\rho^2} \biggr),  \nonumber \\  \nonumber \\
\mathcal{M}_{hh\to ZZ} &=&  \mathcal{M}_{hh\to WW}(m_W \to m_Z), 
\end{eqnarray}
\begin{eqnarray}
\mathcal{M}_{hZ\to hZ} &=& 2 \lambda_1 \cos^2\theta+\lambda_3 \sin^2\theta\\
&+& \kappa^2_{1}\, \biggl(  {1 \over s - m_Z^2} +{1 \over t - m_Z^2} + {1 \over u - m_Z^2}\biggr)\nonumber 
\end{eqnarray}
\begin{eqnarray}
\mathcal{M}_{h Z\to \rho Z} &=& (2 \lambda_1 - \lambda_3 )\sin\theta\cos\theta \\
&+& \kappa_{1}\kappa_2\,\biggl( {1 \over s - m_Z^2}  + {1 \over t - m_Z^2} + {1 \over u - m_Z^2}\biggr) \nonumber
\end{eqnarray}
\begin{eqnarray}
\mathcal{M}_{\rho Z\to \rho Z} &=&2 \lambda_1 \sin^2\theta+\lambda_3 \cos^2\theta \\
&+& \kappa^2_{2}\, \biggl(  {1 \over s - m_Z^2} +{1 \over t - m_Z^2} + {1 \over u - m_Z^2}\biggr)\nonumber
\end{eqnarray}

\begin{eqnarray}
\mathcal{M}_{\rho\rho\to WW} &=& 2 \lambda_1\sin^2\theta + \lambda_3 \cos^2\theta \\
&+& \kappa_2^2\, \biggl( {1 \over t - m_W^2} + {1 \over u - m_W^2}\biggr)\nonumber \\
&+& \kappa_5^2\, \biggl({1 \over s - m_h^2} + {1 \over t - m_h^2} + {1 \over u - m_h^2} \biggr) \nonumber \\
&+& \kappa_6^2\, \biggl({1 \over s - m_\rho^2} + {1 \over t - m_\rho^2} + {1 \over u - m_\rho^2} \biggr),  \nonumber \\   \nonumber \\ 
\mathcal{M}_{\rho\rho\to ZZ} &=& \mathcal{M}_{\rho\rho\to WW}(m_W \to m_Z) \nonumber \\ \nonumber \\ 
%
\mathcal{M}_{h\rho\to WW} &=& (2 \lambda_1 - \lambda_3 )\sin\theta\cos\theta \\
&+& \kappa_7^2\, \biggl( {1 \over t - m_W^2} + {1 \over u - m_W^2}\biggr)\nonumber \\
&+& \kappa_8^2\, \biggl({1 \over s - m_h^2} \biggr) + \kappa_9^2\, \biggl({1 \over s - m_\rho^2} \biggr),  \nonumber \\ \nonumber \\
\mathcal{M}_{h\rho\to ZZ} &=& \mathcal{M}_{h\rho\to WW}(m_W \to m_Z) 
%
\end{eqnarray}
%

\subsection{B.3. Higgs Scattering Amplitudes}
\noindent
\begin{eqnarray}
\mathcal{M}_{hh\to hh} &=& 6\,\bigl(\lambda_1 \cos^4\theta+ \lambda_2 \sin^4\theta + \\ 
&+&  \lambda_3 \,\sin^2\theta \cos^2\theta\bigr) \nonumber \\
&+& \kappa_{10}^2\, \biggl( {1 \over s - m_h^2}  + {1 \over t - m_h^2} + {1 \over u - m_h^2}\biggr)\nonumber \\
&+& \kappa_{11}^2\, \biggl( {1 \over s - m_\rho^2} + {1 \over t - m_\rho^2} + {1 \over u - m_\rho^2}\biggr),  \nonumber 
\end{eqnarray}
\begin{eqnarray}
\mathcal{M}_{\rho\rho\to \rho\rho} &=& 6 \bigl(\lambda_1 \sin^4\theta+ \lambda_2 \cos^4\theta \\
&+& \lambda_3 \sin^2\theta \cos^2\theta \bigr) \nonumber\\
&+& \kappa_{12}^2\, \biggl( {1 \over s - m_h^2}  + {1 \over t - m_h^2} + {1 \over u - m_h^2}\biggr)\nonumber \\
&+& \kappa_{13}^2 \biggl( {1 \over s - m_\rho^2}  + {1 \over t - m_\rho^2} + {1 \over u - m_\rho^2}\biggr),  \nonumber
\end{eqnarray}
\begin{eqnarray}
\mathcal{M}_{\rho h\to \rho \rho} 
&=& 3\sin\theta\cos\theta\bigl( \,(2\, \lambda_1-\lambda_3) \sin^2\theta  \\
&+& 3\,(\lambda_3-2 \lambda_2) \cos^2\theta \bigr)  \nonumber \\
&+& \kappa_{11}\,\kappa_{12}\, \biggl({1 \over s- m_h^2} + {1 \over t - m_h^2} + {1 \over u - m_h^2}\biggr)\nonumber \\
&+& \kappa_{12}\,\kappa_{13}\, \biggl({1 \over s - m_\rho^2} + {1 \over t - m_\rho^2} + {1 \over u - m_\rho^2} \biggr),  \nonumber 
\end{eqnarray}
\begin{eqnarray}
\mathcal{M}_{h\rho\to hh} &=& 3 \sin\theta \cos\theta \bigl( \,(2\, \lambda_1-\lambda_3)\cos^2\theta\\
&+&  (\lambda_3-2 \lambda_2)\sin^2\theta \bigr) \nonumber\\
&+& \kappa_{10}\,\kappa_{11}\,  \biggl( {1 \over s - m_h^2}  + {1 \over t - m_h^2} + {1 \over u - m_h^2}\biggr)\nonumber \\
&+& \kappa_{11}\,\kappa_{12}\, \biggl( {1 \over s - m_\rho^2}  + {1 \over t - m_\rho^2} + {1 \over u - m_\rho^2}\biggr),  \nonumber
\end{eqnarray}
\begin{eqnarray}
\mathcal{M}_{\rho\rho\to hh} &=& 2 \sin^2\theta \cos^2\theta \,\bigl(3 \,(\lambda_1+\lambda_2)-2 \,\lambda_3\bigr) \\
&+& \lambda_3 (\sin^4\theta + \cos^4\theta) \nonumber\\
&+& \kappa_{10}\kappa_{12}\, \biggl( {1 \over s - m_h^2}  + {1 \over t - m_h^2} + {1 \over u - m_h^2}\biggr)\nonumber \\
&+& \kappa_{11}\kappa_{13}\, \biggl( {1 \over s - m_\rho^2}  + {1 \over t - m_\rho^2} + {1 \over u - m_\rho^2}\biggr), \nonumber 
\end{eqnarray}
\begin{eqnarray}
\mathcal{M}_{h\rho\to h\rho} &=& 2 \sin\theta^2 \cos\theta^2 \,\bigl(3 (\lambda_1+\lambda_2)-2\, \lambda_3\bigr) \\
&+& \lambda_3 (\sin\theta^4 + \cos\theta^4) \nonumber \\
&+& \kappa_{11}^2\, \biggl( {1 \over s - m_h^2}  + {1 \over t - m_h^2} + {1 \over u - m_h^2}\biggr)\nonumber \\
&+& \kappa_{12}^2\, \biggl( {1 \over s - m_\rho^2}  + {1 \over t - m_\rho^2} + {1 \over u - m_\rho^2}\biggr). \nonumber
%
\end{eqnarray}

\subsection{B.4. Dark Matter Scattering Amplitudes}
\noindent
Here we calculate the high-energy scattering diagrams associated with the dark matter.  In the decoupling limit and when the pseudoscalar coupling is set to zero, these amplitudes reduce to the results in~\cite{Chanowitz:1978mv}.  Following~\cite{Chanowitz:1978mv}, we define $\lambda$, $\bar{\lambda}$ to be the helicities of the fermions and anti-fermions, respectively.  We define $\lambda'$ and  $\bar{\lambda}'$ to be the helicities of the outgoing fermions and anti-fermions, respectively.  The spin-up and spin-down spinors are represented by $+$ and $-$, respectively.  In the following, $\alpha$ is the center-of-mass scattering angle.
\newline
\newline
The self-scattering amplitude for s- and t-channel exchange is
\begin{widetext}
\begin{eqnarray}
\mathcal{M}_{\chi\, \chi \to \chi\, \chi} &=& -{s \over 2} \,\biggl( {\cos^2\theta \over s - m_\rho^2}  +  {\sin^2\theta \over s - m_h^2}  \biggr) 
\,\biggl(\lambda_{\chi_V}^2 \lambda\lambda'  + i \, \lambda_{\chi_V}\lambda_{\chi_A} \lambda'  - i \, \lambda_{\chi_V}\lambda_{\chi_A} \lambda  + \lambda_{\chi_A}^2 \biggr) \delta_{\lambda \bar{\lambda}}\,\delta_{\lambda' \bar{\lambda}'},\nonumber \\
\mathcal{M}_{\chi\, \chi \to \chi\, \chi} &=& -{1 \over 2} \,\biggl( {\cos^2\theta \over t - m_\rho^2}  +  {\sin^2\theta \over t - m_h^2}  \biggr)\,E^2\sin^2{\alpha \over 2} \\
&\times&\biggl( (\lambda_{\chi_V} - i \lambda_{\chi_A})^2\,\delta_{\lambda,-}\delta_{\bar{\lambda},-}  + (\lambda_{\chi_V} + i \lambda_{\chi_A})^2\,\delta_{\lambda,+}\delta_{\bar{\lambda},+}  \nonumber \\ 
&+& (\lambda_{\chi_V}^2 + \lambda_{\chi_A}^2)\,\delta_{\lambda,+}\delta_{\bar{\lambda},-}  + (\lambda_{\chi_V}^2 + \lambda_{\chi_A}^2)\,\delta_{\lambda,-}\delta_{\bar{\lambda},+}   \biggr)\,\lambda\bar{\lambda}\,\delta_{\lambda,-\lambda'}\,\delta_{\bar{\lambda},-\bar{\lambda}'}, \nonumber
\end{eqnarray}
respectively.
\end{widetext}

\subsection{C.  Dark Matter Mass versus Mixing Angle Plots}
\begin{figure*}[t]
\centering
{\label{fig:parameterspace4}	
	\includegraphics[width=10truecm,height=7.0truecm,clip=true]{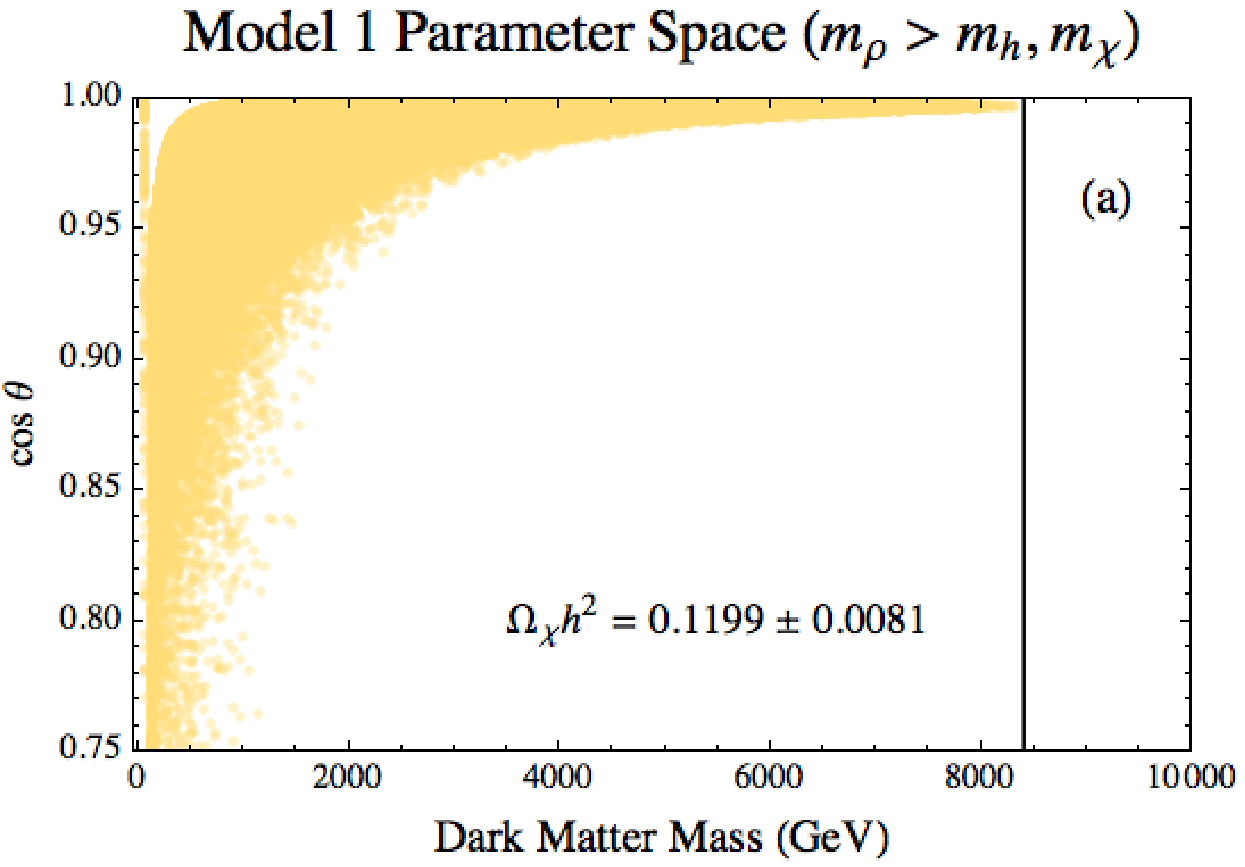} \vspace{1.5cm}}	
	{\includegraphics[width=10.1truecm,height=7.2truecm,clip=true]{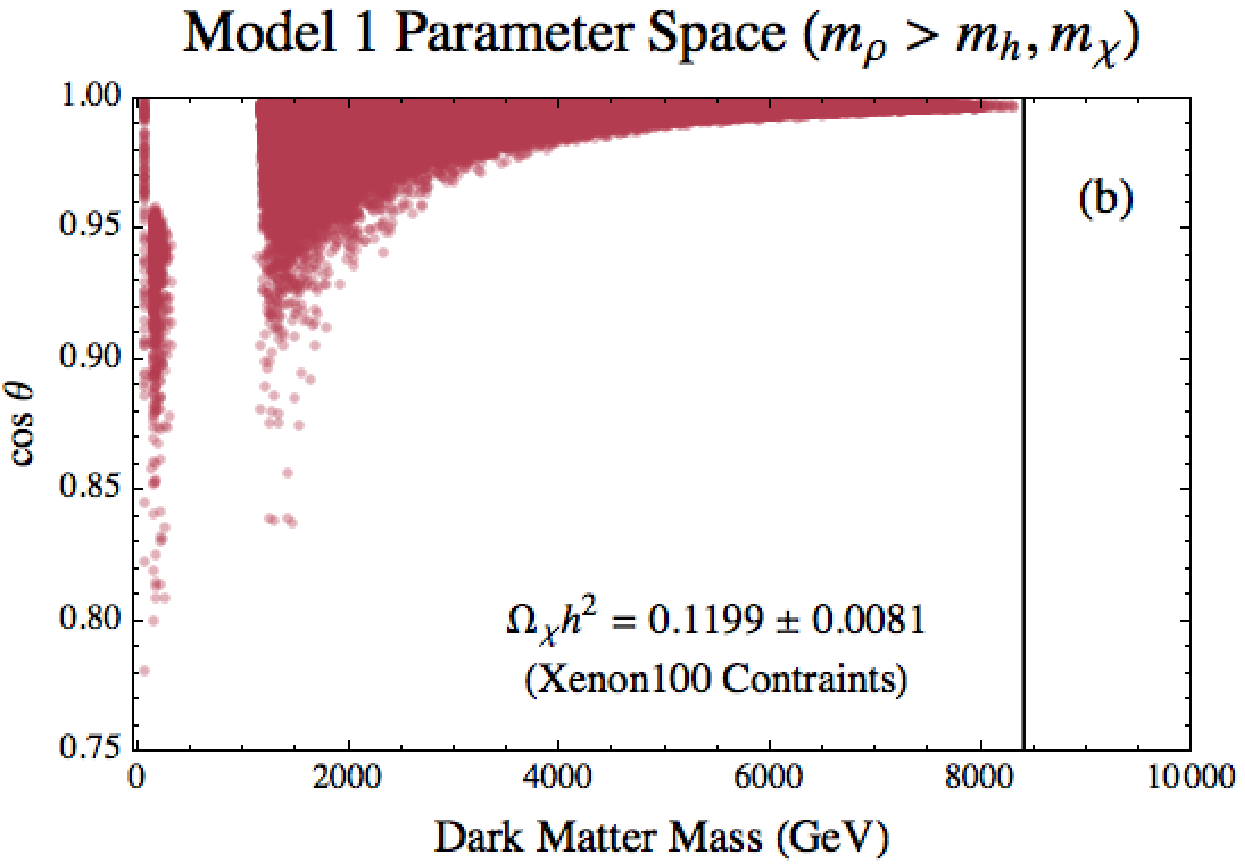}} 
	\caption{Higgs portal parameter space for Model 1 ($\lambda_{\chi_A}$ = 0).  The parameter space without the Xenon100 constraints are shown in Figure (8a).  The parameter space with the Xenon100 constraints are shown in Figure (8b).  The parameter space also satisfies the measured relic abundance.}
\end{figure*} 
\begin{figure*}[t]
\centering
{\label{fig:parameterspace4}	
	\includegraphics[width=10truecm,height=7.0truecm,clip=true]{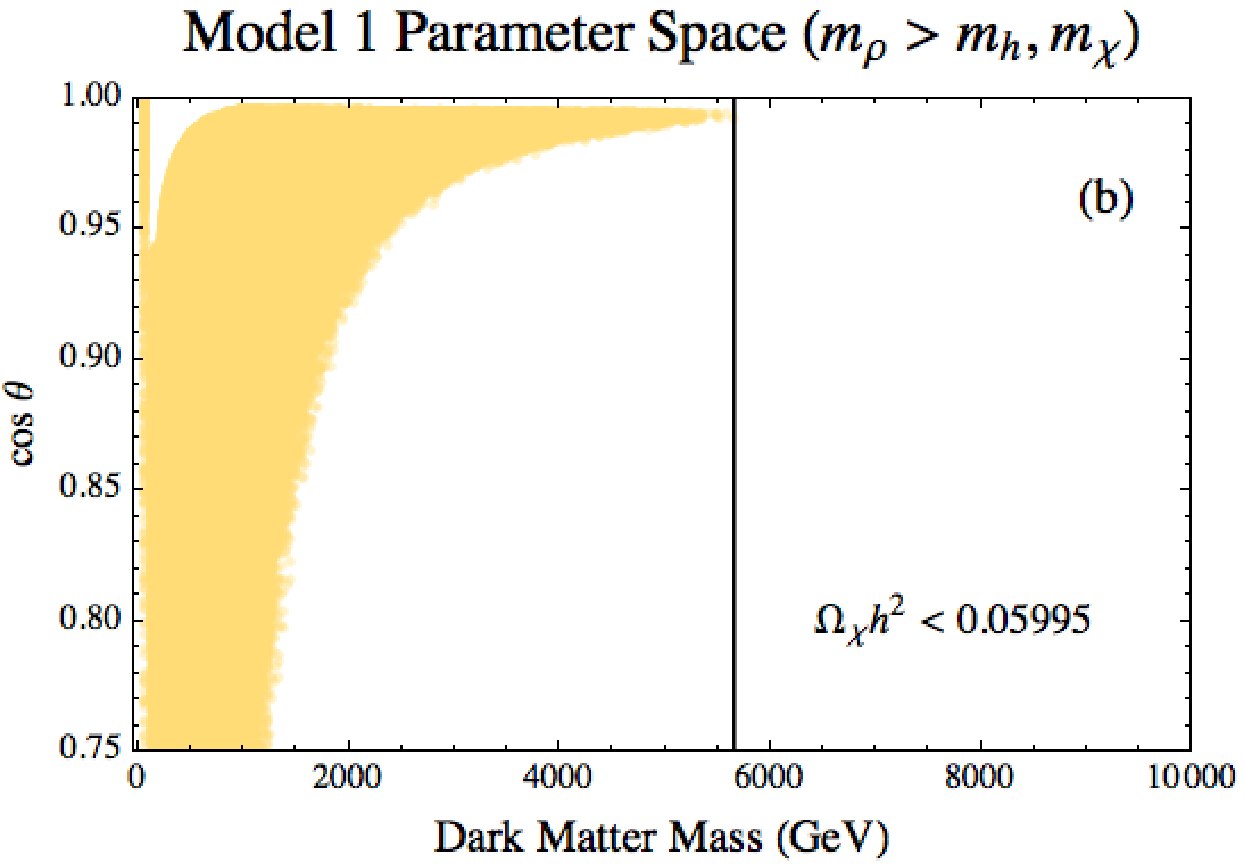} \vspace{1.5cm}}	
	{\includegraphics[width=10.1truecm,height=7.2truecm,clip=true]{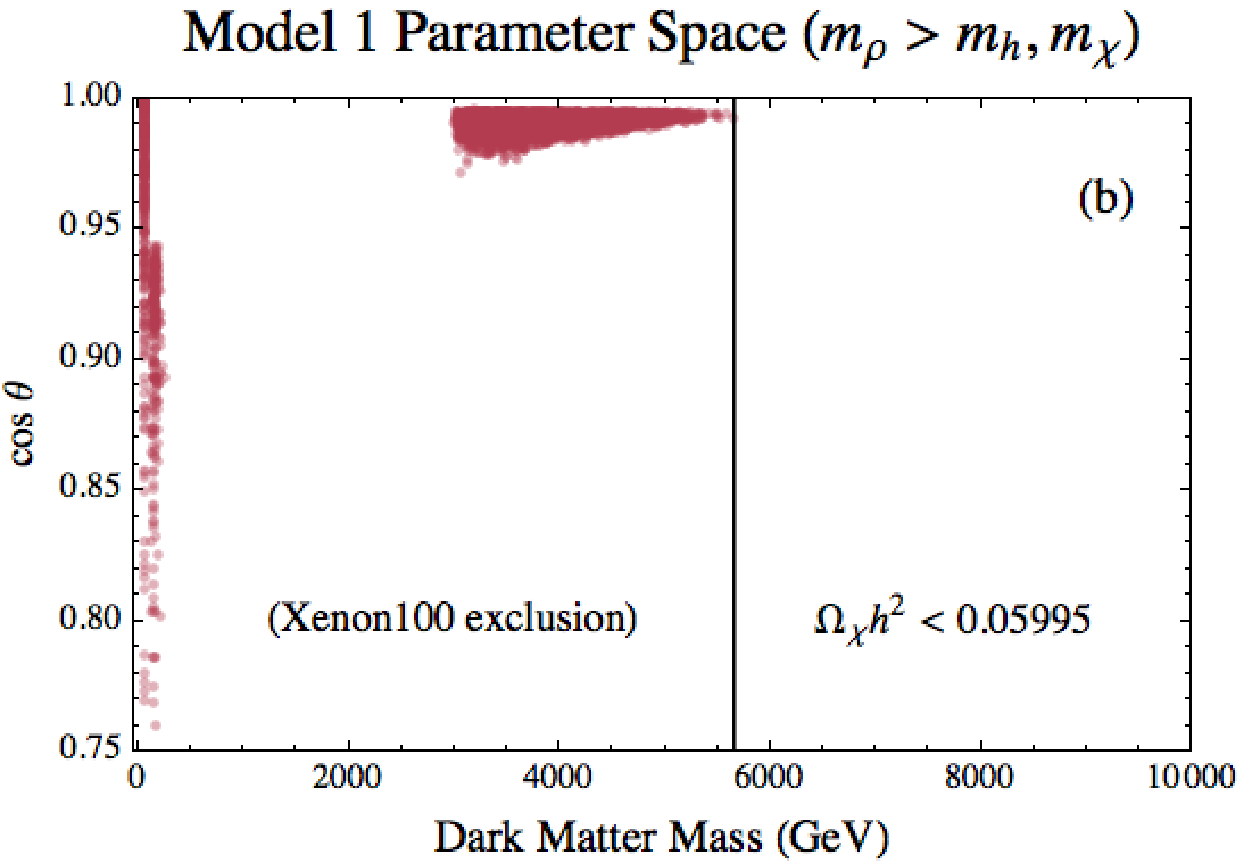}} 
	\caption{Higgs portal parameter space for Model 1.  The parameter space without the Xenon100 constraints are shown in Figure (8a).  The parameter space with the Xenon100 constraints are shown in Figure (8b).  The parameter space is for the points that satisfy at least half of the measured relic abundance.}
\end{figure*} 
\begin{figure*}[t]
\centering
{\label{fig:parameterspace4}	
	\includegraphics[width=10truecm,height=7.0truecm,clip=true]{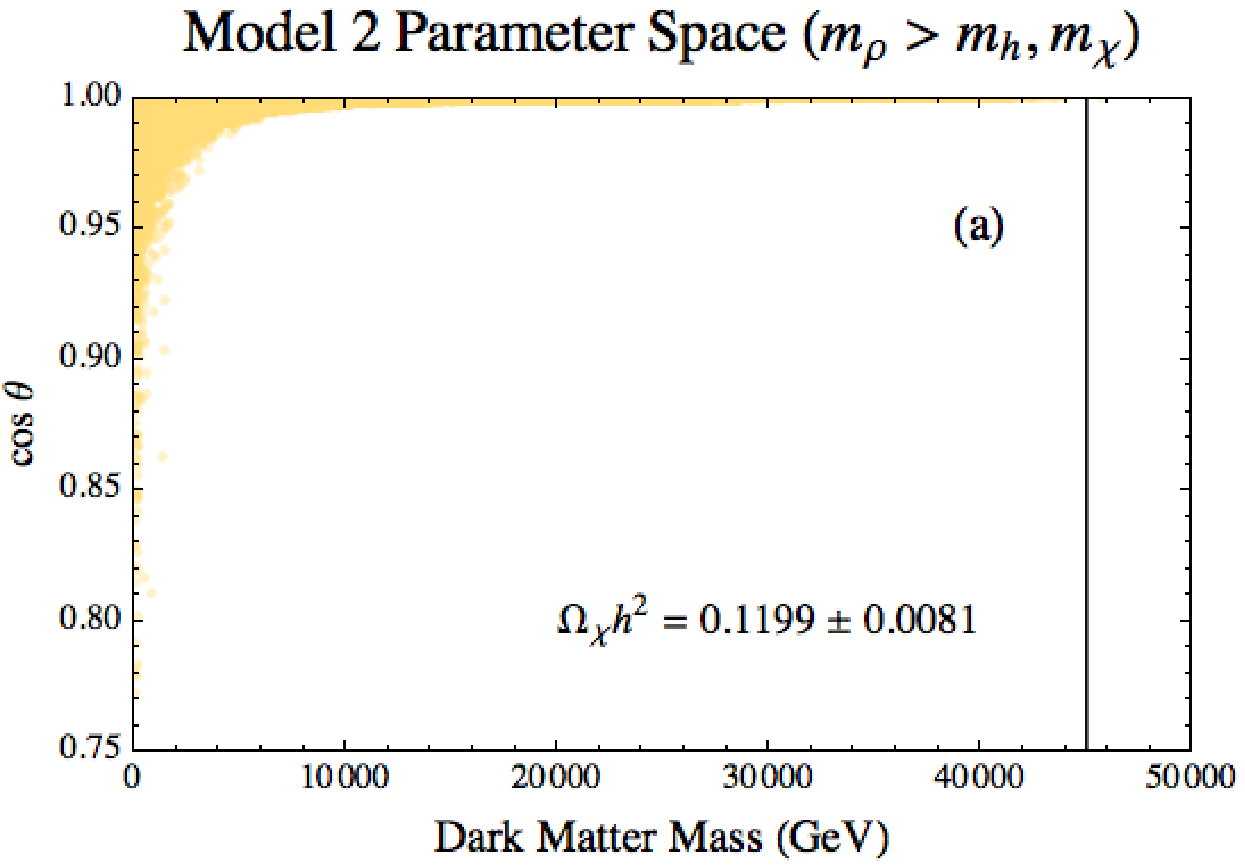} \vspace{1.0cm}}	
	{\includegraphics[width=10truecm,height=7.2truecm,clip=true]{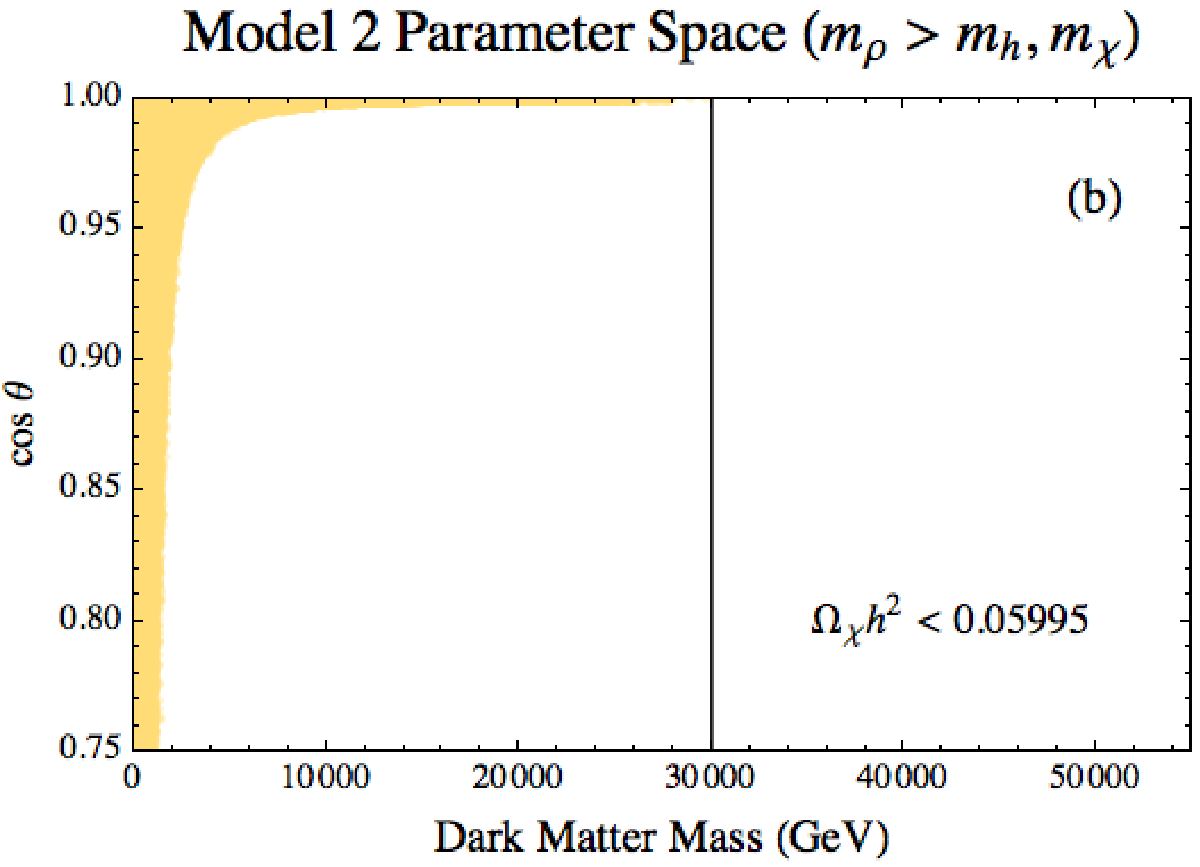}} 
	\caption{Higgs portal parameter space for Model 2.  The vertical lines are the bounds on the dark matter mass.  Figure (10a) has the points that satisfy the measured relic abundance.  Figure (10b) has  the parameter space points which satisfy at least half of the measured relic abundance.  Both plots include the Xenon100 constraints.  Visually, these constraints are no different from the parameter space points without the Xenon100 constraints.}
\end{figure*} 
\begin{figure*}[t]
\centering
{\label{fig:parameterspace7}	
	\includegraphics[width=10truecm,height=7.0truecm,clip=true]{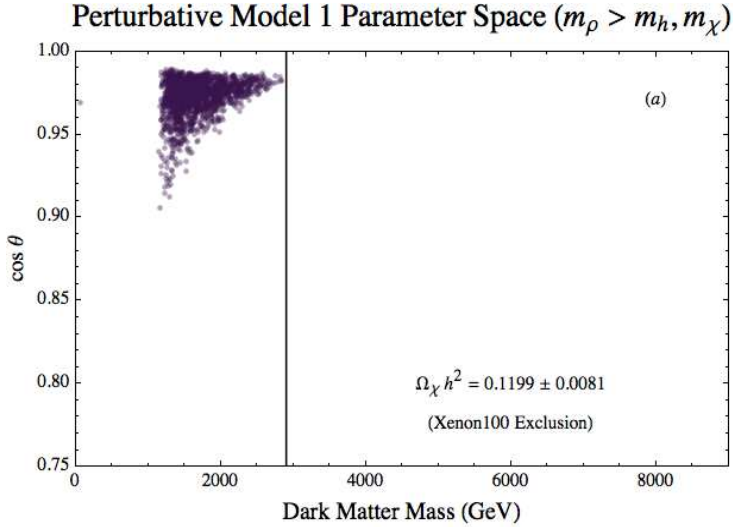}}	%
	\caption{Perturbativity constraints applied to Figure (8a).  The dark matter mass bound (vertical line) is significantly lower than the bound in Figure (8a).  The available parameter space was decimated when the perturbativity constraints were applied to Figure (8b).}
\end{figure*} 
\begin{figure*}[t]
\centering
{\label{fig:parameterspace7}	
	\includegraphics[width=10truecm,height=7.0truecm,clip=true]{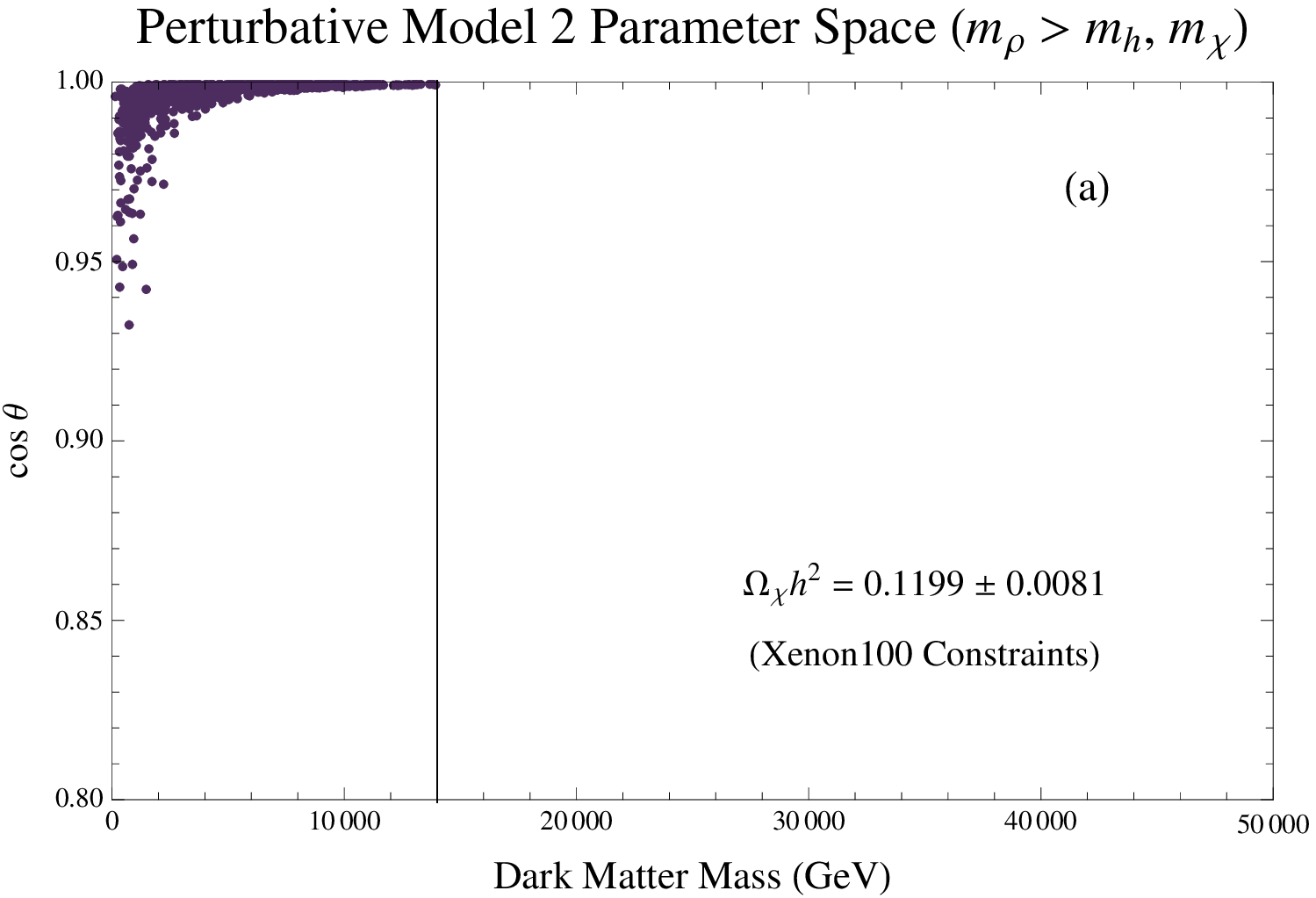} \vspace{1.0cm}}	
	{\includegraphics[width=10truecm,height=7.2truecm,clip=true]{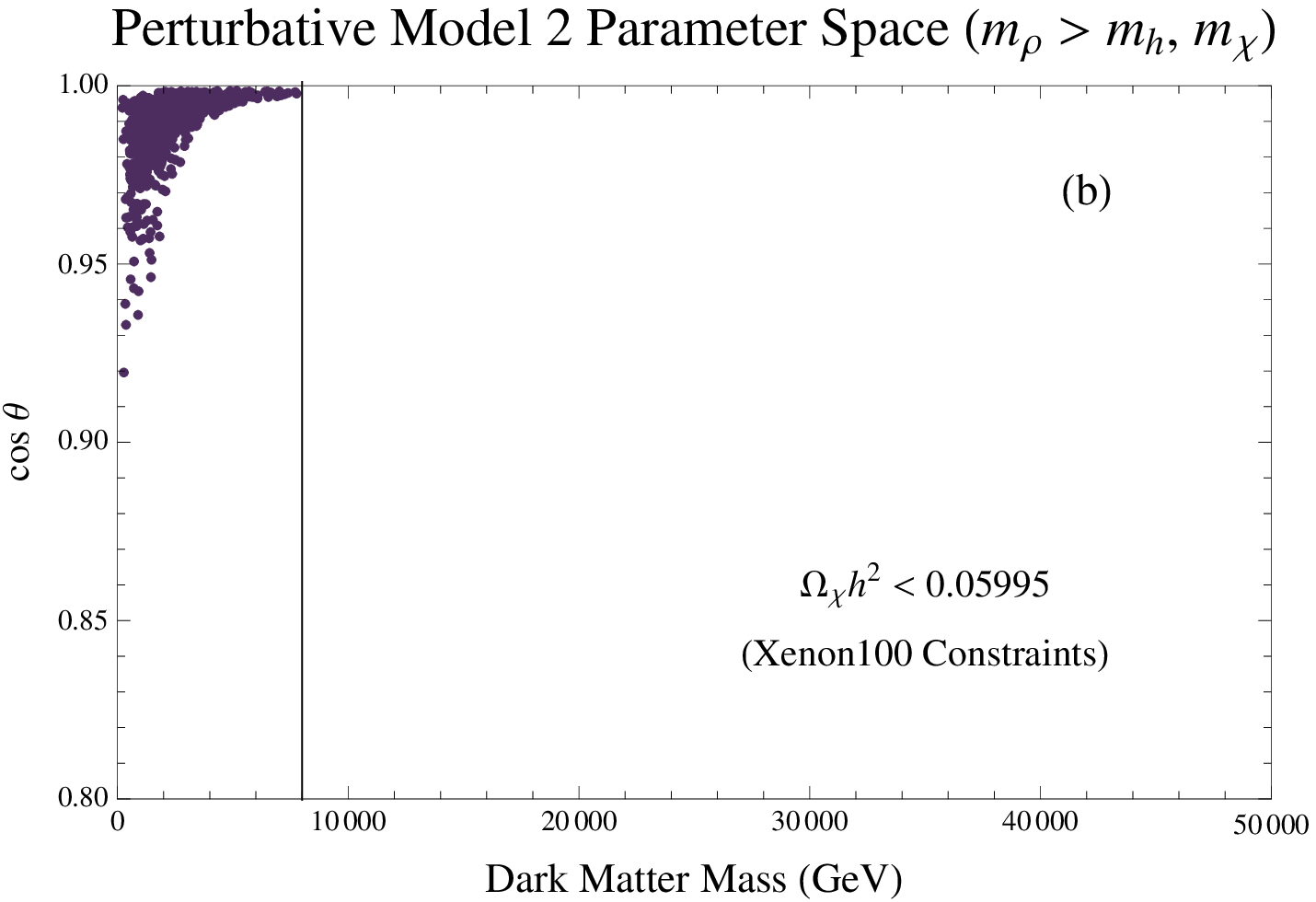}} 
	\caption{Perturbativity constraints applied to Figure (10).  Figure (12a) shows the available parameter space for the measured relic abundance.  Figure (12b) shows the available parameter space for half the measured relic abundance.  The dark matter bounds (vertical lines) are much stronger than the bounds in Figure (10).}
\end{figure*} 
\noindent
In the main text, we described the bounds on the dark matter mass and mass mixing angle.  In this part of the Appendix, we show the plots with these bounds.  For Model~1 in Figure (8), the bound on the dark matter mass is 
\begin{align}
m_\chi \leq 8.4\,\,\mathrm{TeV} &\hspace{0.4cm}& (\Omega_\chi\,h^2 = 0.1199 \pm 0.0081)
\end{align}
assuming the Higgs portal dark matter satisfies the full relic abundance $\pm 3\,\sigma$.  Figures~(8a) and (8b) show the parameter space with and without the Xenon100 constraints. For dark matter that satisfies only half of the measured relic abundance, $\Omega_\chi\,h^2 < 0.05995$, the bound is lowered to
\begin{align}
m_\chi \leq  6.2\,\, \mathrm{TeV} &\hspace{0.4cm}& (\Omega_\chi\,h^2 < 0.05995).
 \end{align}
Figure~(9b) shows the available parameter space is largely decimated by Xenon100.  For Model~2, the bound on the dark matter mass is 
\begin{align}
m_\chi \leq 45.5\,\,\mathrm{TeV} &\hspace{0.4cm}& (\Omega_\chi\,h^2 = 0.1199 \pm 0.0081).
\end{align}
Here we assume the Higgs portal dark matter satisfies the full relic abundance $\pm 3\,\sigma$.  In the limit where the dark matter satisfies only half of the measured relic abundance, $\Omega_\chi\,h^2 < 0.05995$, the bound is lowered to
\begin{align}
m_\chi \leq  30\,\, \mathrm{TeV} &\hspace{0.4cm}& (\Omega_\chi\,h^2 < 0.05995).
 \end{align}

\section{D. Scalar Dark Matter Considerations}
\noindent
Throughout this paper, we have placed unitarity constraints on a Higgs portal model with fermonic dark matter.  In the minimal scenario, there is a total of five degrees of freedom which are constrained with relic abundance and unitarity constraints.  In this section, we sketch the same argument for bosonic dark matter.  
By definition, Higgs portal scalar dark matter couples to the dark Higgs directly.  The mass mixing between the SM and dark Higgs facilitates the Higgs portal mechanism.  The potential is,
\begin{equation}
V = \lambda_\chi\,\phi^*\phi\,\chi^*\,\chi + \lambda_{\chi'}(\chi^*\chi)^2, 
\end{equation}
where $\chi$ can be a real or complex scalar.  We choose a complex scalar; but, the arguments hold for either case.  We assume a $Z_2$ symmetry to stabilize the dark matter.   By the conventions in this paper, we take $\phi$ to be the dark Higgs.  Thus for the scalar case, we have two couplings to constrain.  For the fermion case there is only one.  Please note the operator,
\begin{equation}
V' =  \lambda_{\chi h}\,h^\dagger h\,\chi^*\,\chi,
\end{equation}
is allowed.  However we take $\lambda_{\chi h} \to 0$ for this discussion.  We discuss this operator in~\cite{devinjoannetim}. 
%
%
\newline
\newline  
For the scalar case, the Goldstone-Higgs scattering matrix in Section VII.1.1.~is now expanded to have an additional column and row for dark matter.  The new matrix must account for the additional scattering processes,
\begin{eqnarray}
\chi + \chi  &\leftrightarrow& \chi + \chi \\
\chi + \chi  &\leftrightarrow& \rho + \rho \\
\chi + \chi &\leftrightarrow& h + h \\
\chi + \chi &\leftrightarrow& w^+ + w^- \\
\chi + \chi &\leftrightarrow& z + z \\
\chi + \chi&\leftrightarrow& h + z \\ 
h + \chi &\leftrightarrow& h + \chi \\
\rho + \chi &\leftrightarrow& \rho + \chi. 
\end{eqnarray}
Here $z$ and $w^\pm$ are the Goldstone bosons eaten by the $W$ and $Z$ bosons.  These processes directly constrain $\lambda_\chi$ and $\lambda_{\chi'}$.  For the fermionic case, dark matter scattering decoupled from the Goldstone-Higgs scattering diagram.  However, with a larger matrix~\cite{Lee:1977eg} the bounds on dark matter-Higgs-Goldstone couplings are likely stronger than the fermionic case with only Goldstone-Higgs couplings.

\subsection{E. A Dark Higgs Sector without Symmetry Breaking}
\noindent
Consider the case where the dark Higgs is simply a scalar that mixes with the SM Higgs but does not under spontaneous symmetry breaking.  A generic potential is therefore
\begin{eqnarray}
V' &=&  \lambda_1\,\biggl(h^\dagger h -  {v \over 2} \biggr)^2 + m_\rho^2\,\rho^2 + \lambda_\rho\,\rho^4  \label{eq:Higgspotential2} \\
 &+&{1 \over 2}\,m_\rho''\,\rho\,h^2 +  v\,m_\rho'\,\rho\,h  + {1 \over 2}\,\lambda_\rho''\,v\,h\,\rho^2 \nonumber \\
 &+& {1 \over 4}\,\lambda_\rho'\,h^2\rho^2. \nonumber
\end{eqnarray}
Like Appendix D., this potential just adds couplings to be constrained by the Goldstone-Higgs scattering matrix in Section VII.1.1.

\subsection{F. Precision Electroweak Parameters}
\noindent
The discovery of the SM Higgs has tightened precision electroweak constraints on the S and T parameters~\cite{Peskin:1991sw,Baak:2012kk}.  The overall contribution to the S and T parameters is given by~\cite{Grimus:2007if,Grimus:2008nb},
\begin{widetext}
\begin{eqnarray}
T &=& {g^2 \over 64\pi^2\alpha}\,{1 \over m_W^2}\Biggl[ -3\sin^2\theta \biggl(F(m_Z^2,m_h^2) -  F(m_W^2,m_h^2) \biggr) + 3\sin^2\theta \biggl(F(m_Z^2,m_\rho^2) -  F(m_W^2,m_\rho^2) \biggr) \Biggr] \\
S &=& {g^2 \over 382\pi^2\cos^2\theta_W}\,{1 \over m_W^2}\Biggl[- \sin^2\theta \,\ln m_h^2 + \sin^2\theta\, \ln m_\rho^2 - \sin^2\theta\, \hat{G}(m_h^2, m_Z^2) + \sin^2\theta\, \hat{G}(m_\rho^2, m_Z^2) \Biggr] 
\end{eqnarray}
where
\begin{eqnarray}
F(m_1^2, m_2^2) &=& 0  \hspace{5.6cm} (m_1 = m_2) \\
F(m_1^2, m_2^2) &=& {m_1^2 + m_2^2 \over 2} - {m_1^2\,m_2^2 \over m_1^2 - m_2^2} \ln{m_1^2 \over m_2^2} \hspace{1.5cm} (m_1 \neq m_2) \\ \nonumber \\
\hat{G}(m_1^2,m_2^2) &=& - {79 \over 3} + {9 \,m_1^2 \over m_2^2} -  {2 \,m_1^4 \over m_2^4}  + \Biggl(-10 + {18\,m_1^2 \over m_2^2} - {6\,m_1^4 \over m_2^4} + {m_1^6 \over m_2^6} - {9(m_1^2 + m_2^2) \over m_1^2 - m_2^2} \Biggr) \ln{m_1^2 \over m_2^2} \nonumber \\
&+& \biggl( 12 - {4\,m_1^2 \over m_2^2} + {m_1^4 \over m_2^4} \biggr) \,{f (m_1^2, m_1^4 - 4\,m_1^2\,m_2^2)  \over m_2^2} \nonumber \\
f (m_1^2, m_1^4 - 4\,m_1^2\,m_2^2)  &=& \sqrt{m_1^4 - 4\,m_1^2\,m_2^2} \,\log\biggl[\mathrm{Abs}\Biggl[{m_1^2 - \sqrt{m_1^4 - 4\,m_1^2\,m_2^2}  \over m_1^2 + \sqrt{m_1^4 - 4\,m_1^2\,m_2^2}} \Biggr]\biggr].
\end{eqnarray}
\end{widetext}
Here the $U$ parameter is set to zero; and, $\cos^2 \theta_W$ is the cosine of the Weinberg angle.  We checked that Higgs portal parameter space survives this test.
\newline
\newline
We performed a parameter scan over this Higgs portal parameter space.  To do so, we took the SM central value~\cite{Baak:2012kk}, 
\begin{align}
S\bigl|_{U = 0} = 0.05\pm 0.09 && T\bigl|_{U = 0} =  0.08 \pm 0.07,
\end{align}
and added the logarithmically enhanced pieces shown above.  We kept all points that satisfied the 95\% c.l.~constraint ellipse which were none.  This constraint is model dependent.  New physics can push the Higgs portal parameter space out of the ellipse. 
%

\subsection{G.  Current Higgs Mixing Constraints}
\noindent
To leading order~\cite{LHCHiggsCrossSectionWorkingGroup:2012nn,ATLAS:2012wma}, the Higgs signal cross section can be expressed as
 \begin{eqnarray}
 n_\mathrm{signal} &=& \biggl(\sum_i\,\mu_i\,\sigma_{i\,\mathrm{SM}} \times A_i \times \epsilon_{i} \biggr) \\
 &\times& \mu_f \,B_{f\,\mathrm{SM}} \times \mathcal{L}.\,\, \nonumber
  \end{eqnarray}
  \noindent
Here $A$ is the detector acceptance, $\epsilon$ the reconstruction efficiency and $\mathcal{L}$ the integrated luminosity.  $B$ denotes the branching fraction The signal strength factor is defined by $\mu_i = \sigma_i/\sigma_{i,\mathrm{SM}}$.  
Similarly, the decay strength factor is defined by $\mu_f = B_f/B_{f,\mathrm{SM}}$.  The dominant way\footnote{In some models, corrections by new fermions may enhance the signal strength.  The dark matter, by definition, does not couple at tree-level to the SM fermions or gauge bosons; therefore any correction involving these particles is at best two-loops.  Corrections involving the dark Higgs is one additional loop suppressed in comparison the SM leading order contribution.  We therefore do not consider these suppressed contributions from beyond the SM physics.} the Higgs portal modifies this signal strength is through the SM Higgs couplings which are reduced by the mixing parameter, $\cos\theta$.  (See Section II.3 for example couplings.)  
Thus, we make the simple assumption that only SM particles 
contribute to the cross section.  This implies
\begin{align}
\mu_i \to \cos^4\theta &&  \mu_f \to 1.  \label{eq:cosHiggsstrength}
\end{align}  
The ATLAS fit to the global signal strength (for all Higgs decay channels) is~\cite{ATLAS:2012wma} is 
\begin{equation}
\sqrt{\mu} = 1.19\, \pm\, 0.11\, (\mathrm{stat}) \,\pm\, 0.03\, (\mathrm{sys}). \label{eq:ATLASHiggsgit}
\end{equation}
The number of events exceeds what one expects with the SM.  $\cos\theta$ cannot be greater than one.  However, the statistics are small; and the high value for this fit is primarily due to the excess of $h \to \gamma\gamma$ events in comparison to $h \to ZZ$, $h \to WW$, $h \to \bar{\tau}\tau$ and $h \to \bar{b}b$.  The CMS collaboration~\cite{CMSHiggsgammagamma} has also seen the same excess of $h \to \gamma \gamma$ events.  
However, if new physics is responsible for the $h \to \gamma\gamma$ signal, then one would also expect $h \to Z Z$ to have an equivalent excess.  Both collaborations report $h \to Z Z$ (along with $h \to WW$, $h \to \bar{\tau}\tau$ and $h \to \bar{b}b$) can be consistent with the SM and $\cos\theta$ less than one.  More data is needed to shrink the error bars and determine associated branching fractions as well as the Higgs mass.  Since $h \to \gamma \gamma$ and $h \to ZZ$ are the most constraining for our naive scenario and new physics would theoretically impact both equally, we use 
a combination of those measurements to set the range of $\cos\theta$.  CMS~\cite{CMSHiggsgammagamma} reports a signal strength for
\begin{eqnarray}
\mu_{\gamma\gamma+ZZ} &\in& [0.58, 2.15]\,\,\, \,\,\,\,95\%\,\,\mathrm{c.l.}
\end{eqnarray}
However, it should be noted that CMS has also shown~\cite{CMSHiggsgammagamma} their $h \to \gamma \gamma$ results vary depending on whether a vector-boson tag was applied or not.  The $h \to \gamma \gamma$ events are consistent with maximal dark/SM Higgs mixing ($\cos\theta \sim 1/\sqrt{2}$) at the $2\sigma$ level.  Again, more data is needed to resolve these ambiguities.  We therefore take the range of $\cos\theta$ to be
\begin{equation}
\cos\theta \in [1/\sqrt{2},\, \lesssim 1].
\end{equation}

\end{document}